\RequirePackage{fix-cm}
\documentclass[twocolumn,smallextended,hidelinks]{svjour3}       
\smartqed  

\usepackage{hyperref}
\hypersetup{colorlinks=true,urlcolor=black,linkcolor=black,citecolor=black}
\usepackage{tabulary}
\usepackage{tabularx}
\usepackage{booktabs}
\usepackage{relsize,xspace}
\usepackage{multirow}
\usepackage{microtype}
\usepackage{enumitem}
\usepackage{balance}
\usepackage{array}
\usepackage[table,xcdraw]{xcolor}
\usepackage{multicol}
\usepackage{makecell}
\usepackage{graphicx}
\usepackage{natbib}
\usepackage{tikz}
\usetikzlibrary{arrows}
\usetikzlibrary{arrows.meta}

\usepackage{listings}
\usepackage{subcaption}
\AtBeginDocument{\DeclareCaptionSubType{lstlisting}}

\definecolor{codegreen}{HTML}{567A0D}
\definecolor{codegray}{HTML}{999999}
\definecolor{codeblue}{HTML}{015493}
\definecolor{codeorange}{HTML}{B75301}
\definecolor{backcolour}{rgb}{1,1,1}

\lstdefinestyle{mystyle}{
    backgroundcolor=\color{backcolour},
    commentstyle=\color{codegray},
    keywordstyle=\color{codeblue},
    numberstyle=\tiny\color{codegray},
    stringstyle=\color{codegreen},
    identifierstyle=\color{codeorange},
    basicstyle=\ttfamily\smaller,
    breakatwhitespace=false,
    breaklines=true,
    postbreak=\mbox{\textcolor{red}{$\hookrightarrow$}\space},
    captionpos=b,
    keepspaces=true,
    numbers=left,
    numbersep=5pt,
    showspaces=false,
    showstringspaces=false,
    showtabs=false,
    tabsize=1,
    keywordsprefix={@},
    moredelim=[is][\color{black}]{[*}{*]},
    xleftmargin=9pt
}

\newcommand\parhead[1]{\vspace{.5mm}\noindent\textbf{{#1}}}

\usepackage{ifthen}
\usepackage{amssymb}
\usepackage{xcolor}
\usepackage{url}

\usepackage[flushleft]{threeparttable}
\usepackage{caption}

\lstdefinelanguage{Config}
{
    basicstyle=\ttfamily\smaller,
    columns=fullflexible,
    morecomment=[s][\color{black}\bfseries]{[}{]},
    morecomment=[l]{\#},
    morecomment=[l]{;},
    commentstyle=\color{codegray}\ttfamily,
    morekeywords={},
    otherkeywords={=,:},
    keywordstyle={\color{black}\bfseries},
    identifierstyle=\color{black},
    tabsize=1
}

\newboolean{showcomments}
\setboolean{showcomments}{true} 
\ifthenelse{\boolean{showcomments}}
{\newcommand{\nb}[2]{
		\fcolorbox{gray}{yellow}{\bfseries\sffamily\scriptsize#1}
		{\sf\small$\blacktriangleright$\textit{#2}$\blacktriangleleft$}
	}
	
}
{\newcommand{\nb}[2]{}
	
}

\newcommand{\add}[1]{#1}

\newcommand{\orcid}[1]{\href{https://orcid.org/#1}{\includegraphics[width=7pt]{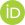}}}

\begin{document}

\title{A Taxonomy of Functional Security Features and How They Can Be Located}
\subtitle{}

\author{
	    Kevin Hermann\textsuperscript{\textsection\orcid{0009-0004-6207-4045}}
        \and
		Simon Schneider\textsuperscript{\textsection\orcid{0000-0001-8605-615X}}
		\and
		Catherine Tony\textsuperscript{\textsection\orcid{0000-0002-9916-4456}}
        \and
		Asli Yardim\textsuperscript{\textsection\orcid{0009-0005-6117-6581}}
        \and
		Sven Peldszus\textsuperscript{\orcid{0000-0002-2604-0487}}
        \and
		Thorsten Berger\textsuperscript{\orcid{0000-0002-3870-5167}}
        \and
		Riccardo Scandariato\textsuperscript{\orcid{0000-0003-3591-7671}}
        \and
		M. Angela Sasse\textsuperscript{\orcid{0000-0003-1823-5505}}
        \and
		Alena Naiakshina\textsuperscript{\orcid{0009-0008-1843-2027}}
	}

\institute{
        K. Hermann \at
			Ruhr University Bochum, Germany\\
			\email{kevin.hermann@rub.de}
        \and
		S. Schneider \at
		    Hamburg University of Technology\\
		    \email{simon.schneider@tuhh.de}
		\and
		C. Tony \at
		    Hamburg University of Technology\\
		    \email{catherine.tony@tuhh.de}
		\and
		A. Yardim \at
			Ruhr University Bochum, Germany\\
			\email{asli.yardim@rub.de}
		\and
		S. Peldszus \at
			Ruhr University Bochum, Germany\\
			\email{sven.peldszus@rub.de}
		\and
		T. Berger \at
			Ruhr University Bochum, Germany\\
			Chalmers University of Technology and the University of Gothenburg, Sweden\\
			\email{thorsten.berger@rub.de}
		\and
		R. Scandariato \at
		    Hamburg University of Technology\\
		    \email{riccardo.scandariato@tuhh.de}
		\and
		M. A. Sasse \at
			Ruhr University Bochum, Germany\\
			\email{martina.sasse@rub.de}
		\and
		A. Naiakshina \at
			Ruhr University Bochum, Germany\\
			\email{alena.naiakshina@rub.de}
}
\authorrunning{Kevin Hermann et al.}

\date{}

\maketitle

\begingroup\renewcommand\thefootnote{\textsection}
\footnotetext{These four authors contributed equally to the paper.}

\begin{abstract}
	\looseness=-1
Security must be considered in almost every software system.
Unfortunately, selecting and implementing security features remains a challenge due to the wide variety of security threats and possible countermeasures. While security standards are intended to help developers, they are usually too abstract and vague to help implementing security features, or they merely help configuring such.
A resource that describes security features at an abstraction level that lies between high-level (i.e., rather too general) and low-level (i.e., rather too specific) security standards could facilitate secure systems development.
This resource should support the selection of appropriate security features to achieve high-level security goals, allow easy retrieval of relevant low-level details, and provide pointers to suitable ways to realize the security features.
To realize security features, developers typically use external security libraries or frameworks, to minimize implementation mistakes.
Even when using libraries, developers still make mistakes when writing code to integrate them, often resulting in security vulnerabilities.
When security incidents occur or the system needs to be audited or maintained, it is essential to know what security features have been implemented and, more importantly, where they are located.
This task, commonly referred to as feature location, is often tedious and error-prone. While dedicated feature location techniques exist, they require significant manual effort or adherence to strict development processes, preventing their use.
Therefore, we have to support long-term tracking of implemented security features.

	\looseness=-1
We present a study of security features in the literature and their coverage in popular security frameworks.
We contribute (1) a taxonomy of 68 functional implementation-level security features including a mapping to widely used security standards, (2) an examination of 21 popular security frameworks concerning which of these security features they provide, and (3) a discussion on the representation of security features in source code.
Our taxonomy aims to aid developers in selecting appropriate security features and security frameworks, as well as relating them to security standards when they need to choose and implement security features for a software system.
\end{abstract}

%


\section{Introduction}
\label{sec:intro}

\noindent
Considering security when developing software is crucial. Software vulnerabilities pose a major threat to the operation of software systems\,\citep{bau2012vulnerability, egele2013empirical, lazar2014does, nadi2016jumping, fahl2013rethinking, krombholz2017have, roth202112}. 
\add{In 2020, an entire hospital had to be shut down due to a successful attack on its IT systems, preventing access to patient data\,\citep{hospital_example1}.}
Unfortunately, considering the wide variety of threats and implementing appropriate countermeasures to create a secure design for a software system requires special expertise\,\citep{Oyetoyan2016, Oyetoyan2019}.

\looseness=-1
Security standards were created to help selecting appropriate security measures to protect software systems from threats. 
Unfortunately, their support for realizing \emph{security features}---the concrete implementations of security measures in code---is limited. 
Such standards are often too abstract \add{and rather focus on the development process, on non-functional security requirements (e.g., the criticality of data), or on low-level details, such as specific implementation aspects of cryptography. 
While security design patterns exist to help implementing non-functional security features (e.g., secure logging pattern), developers lack guidance selecting and implementing \textit{functional security features} (e.g., authentication or encryption) to achieve security goals. 
Specifically, a functional security feature as a label representing code that aims to mitigate an attack, the impact of one, or to protect an asset.}

\looseness=-1
Engineering functional security features is challenging. \textbf{First, developers lack an overview of functional security features}. Such an overview should facilitate selecting security features from both the high-level security goals considered by many security standards and from the many low-level details of how to implement specific features securely.
\textbf{Second, after selecting suitable functional security features 
developers either need to implement them} either from scratch or by incorporating them from a security library or frameworks. 
Unfortunately, a systematic overview of what security feature is offered by which library or framework is missing. 
Developers, therefore, often fall back on the ones they already know. 
However, depending on the project, choosing a different security framework would allow using libraries that might provide better-suited implementations of security features.
Even when using security libraries, security issues often arise in the manually implemented parts of applications, e.g., due to the insecure use of libraries\,\citep{Acar17} or bad usability\,\citep{Patnaik2019}.
Fixing new vulnerabilities requires developers to review and fix them quickly once they are discovered\,\citep{RDVC2019, PBKJ2021}.
\textbf{Third, it is important to know what security features are implemented in a system at hand, and where they are located. }
Many security standards, such as the \emph{Common Criteria} (CC)\,\citep{CC} or the \emph{ISO/SAE 21434} for road vehicles\,\citep{ISO21434}, require maintaining and tracing security features.
Unfortunately, today's traceability techniques require significant manual effort, even when using tools, such as DOORS\,\citep{DOORS}. 
Others depend on strict development processes 
and impose high overhead\,\citep{Peldszus2022}.
When feature are not re{-}corded or maintained properly, recovering them is laborious and error-prone\,\citep{biggerstaff1994program,dit2013, krueger2019, Rubin2013ASO}.
Recording features during development
, when the feature is still fresh in the mind of the developer\,\citep{seiler2017,Ji2015,Martinson.2021,Bergel.2021,Schwarz.2020,Entekhabi.2019,Andam.2017,Mukelabai2023}, is rarely done in practice.
While automated feature location techniques exist, they are difficult to use and produce too many false positives\,\citep{Rubin2013ASO,benOthmane.2015, cornell2012, hewett2009, benOthmane.2017, Abukwaik.2018} to be relevant in practice.
Improving our empirical understanding of how security features are represented security frameworks, using what mechanisms (e.g., configuration options, code annotations, or APIs), would help to build better methods and tools to locate security features in code.

\looseness=-1
In summary, supporting the development of secure software systems requires effective methods and tools for selecting, implementing, and locating security features in code bases.
A problem are the different granularities at which security features can be considered\,\citep{Peldszus2022}, as well as their scattering over the code base and cross-cutting nature.
While high-level security features are often hard to locate, as they are implemented across the codebase, locating fine-grained security features requires intricate knowledge that many developers lack..  
It is unclear yet, at which level of granularity security features manifest in implementations, preventing the development of lightweight support.
Even security standards do not provide an adequate level of abstraction to be effectively used by developers for selecting which security features must be implemented to reach desired security goals.
What is missing is a systematic representation of what functional security features exist, accompanied by a description suitable for developers, and a mapping to relevant security standards. 
We aim to improve the understanding of implementation-level security features and explore the following research questions:  
\begin{description}
    \item[\textbf{RQ1:}] What functional implementation-level security features are considered in the literature? 
    \item[\textbf{RQ2:}] What functional implementation-level security features are provided by security frameworks in practice?
    \item[\textbf{RQ3:}] Which functional implementation-level security features can be located by leveraging information from security frameworks?
\end{description}



We addressed these research questions as follows.
First, we established a taxonomy of functional implemen\-tation-level security features by reviewing literature that systematically describes security features.
Second, we mapped the taxonomy to four generally recognized and well-established security standards: the \textit{ISO/IEC 27000} series, the \textit{Common Criteria (CC)}, the \textit{NIST SP800-53}, and the \textit{NIST Cybersecurity Framework}.
Third, we investigated state-of-the-art security frameworks \add{as discussed by developers on platforms such as Stack Overflow and Reddit}. 
Finally, we explored the mechanisms used by the frameworks for providing security features to developers and how these can be used for locating security features within applications.
We demonstrate that our taxonomy can be related to all functional security features from popular security standards.
\looseness=-1
Further, we demonstrate that security features from security frameworks target all security aspects covered in our taxonomy but do not capture the more detailed concepts considered in the literature.
Finally, we show that security frameworks offer an entry point for locating security features through their API, configuration file and annotations, but still require considering additional code, that is required to integrate them.
\vspace{-1em}

\section{Background and Related Work}
\label{sec:related_work}
\vspace{-0.5em}

We now discuss the notion of security features to motivate our work and introduce the necessary background.

\begin{figure}
    \centering
    \includegraphics[width=\columnwidth]{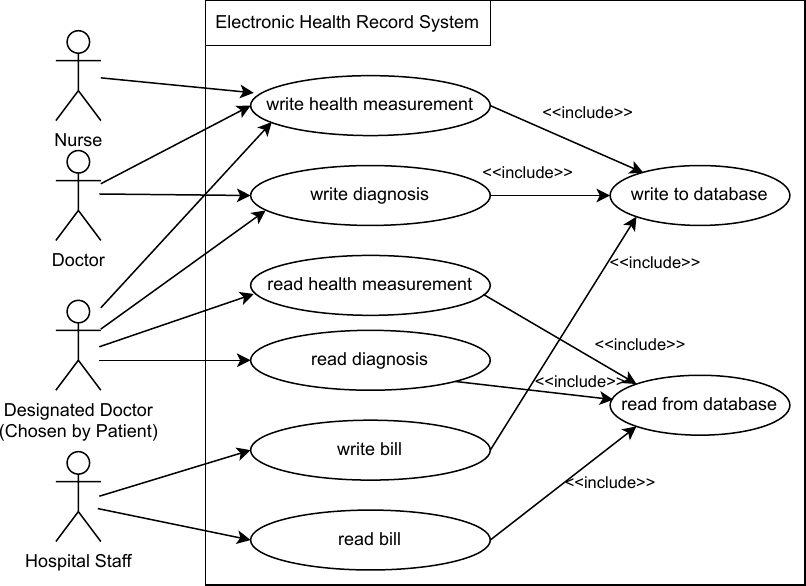}
    \caption{\add{Use case diagram illustrating how different users store and retrieve data from our simplified exemplary EHRS}}
    \label{fig:use_case}
    \vspace{-4mm}
\end{figure}

\add{
\subsection{Running Example}
\label{subsec:running_eg}

As a motivating example for this work, we consider a simplified electronic health record system (EHRS) for a hospital.

In hospitals, many different groups of people are involved in treating a patient.
Treatment requires data from a variety of sources, such as a diagnosis from a physician, health measurements collected by nurses, or data from specialists such as radiologists.
An EHRS enables the capture and analysis of medical data, but often includes additional supporting functionality such as appointment management, medical data analysis, and administrative support such as billing.
%
%
As illustrated in Figure~\ref{fig:use_case}, our EHRS assumes several groups of users, which interact differently with the system: Patients, Doctors, Nurses, Administrative Hospital Staff, and many more.
A doctor can store a diagnosis or an examination report for a patient within the EHRS.
Similarly, nurses store measurements such as data about body temperature or blood pressure within the system.
While doctors and nurses can store data for a patient, only a designated doctor, chosen by the patient, can retrieve the data to e.g., plan further treatment.
Administrative hospital staff can store and retrieve data related to billing within and from the system.

Since an EHRS manages sensitive data such as patient data, hospital data, or diagnosis data that is used to decide about the treatment of patients, the system must preserve data integrity and confidentiality at all times.
Also, availability must be ensured to allow the hospital to operate---in 2020 a hospital was shut down due to an attack, and new emergency patients had to be relocated to other hospitals further away\,\citep{hospital_example1}.

In summary, our simplified exemplary EHRS allows doctors to plan the treatment of patients by storing all related data in a central data storage.
Due to the sensitivity of this data, it must be securely stored, and access must be controlled to ensure that only authorized personnel is allowed to view sensitive data of patients.
This requires considering a wide range of security features to be implemented into the system.
}

\subsection{Security Features}
\label{sub:secfeats}
\looseness=-1
A feature is a distinct label representing the capabilities or behaviors of software systems\,\citep{berger2015feature}.
A feature can be seen as \textit{``a logical unit of behavior specified by a set of functional and non-functional requirements"}\,\citep{bosch2000}.
\add{For example, every use case in Fig.~\ref{fig:use_case} can be seen as a required feature of the EHRS.}
A feature can also be defined as a characteristic, that distinguishes a system from other systems within a family of related systems\,\citep{Batory:2004bw}.
Other definitions describe a feature as a user-visible aspect of a system\,\citep{chen2005,kang1990} or an aspect that increases value for a customer\,\citep{riebisch2003}.

In this work, we use the notion of \textit{security features}, which provide functionalities that address security issues by preventing a security attack or realizing a security requirement\,\citep{mcgraw2004}.
\add{A typical example of a security feature is the authentication of the different users of the outlined EHRS.}
Security features must be carefully planned, even at the architecture level, since missing security features can lead to severe weaknesses in software systems\,\citep{santos2017, santos2019}.
While security features could realize, e.g., non-functional requirements\,\citep{Potter2004}, we focus on functional security features, which are security measures that manifest in the code base and address a functional requirement of a software system.
On the other hand, an example of a non-functional security feature that is not in the scope of our work is the secure design pattern of distrustful decomposition\,\citep{dougherty2009}.


\subsection{\add{Security Feature Taxonomies and Ontologies}}
\add{There are several works that organize software security-related concepts into constructs including taxonomies and ontologies to show how they are interrelated.
For instance, a work by Tsipenyuk et al. \citep{Tsipenyuk2005} presents a taxonomy of coding errors and configuration issues that lead to security vulnerabilities. The main aspects covered in this taxonomy include \textit{input validation and representation, API abuse, security features, time and state, errors, code quality, encapsulation, and environment}. Even though this taxonomy has a dedicated section for security features, it only covers 9 features, 5 of which are related to password management.
A taxonomy for cloud systems security\, \citep{HabibaMSN14} organized the security features into categories such as \textit{authentication, authorization, identity federation, privacy, user-centricity, logging, and editing} that are essential for cloud-based identity management systems.
Security aspects of the Internet of Things (IoT) domain are also discussed in the literature such as by \cite{KhanamAIJS20}, who presented a taxonomy of IoT security attacks in physical, network, and application layers along with their corresponding countermeasures. Another work\, \citep{Blythe2019} analyzed the user manuals and support pages of IoT devices to collect security features such as \textit{two-factor authentication, product lock, and local communication encryption} provided by consumer IoT products.}
\add{Similarly, there are also several other works that organizes security aspects related to cloud security\, \citep{HendreJ15, BhatiaV17}, web services\, \citep{DenkerKFPS03, KimLK07, BuschW15}, information security\, \citep{VenterE03, HerzogSD07, VorobievB10}, and IoT\, \citep{Abbas2005ASO, HerzogSD07} into taxonomies and ontologies.

Although these studies provide valuable insights into security across various fields, they are often domain-specific and largely focus on attacks and vulnerabilities, offering limited comprehensive lists of security features for developers to reference during software development.
Therefore, extracting and consolidating the security features discussed in these works into a single, accessible resource would benefit developers by providing a centralized reference during software development.
}


\subsection{Feature Location}
To maintain and evolve features, developers need to know their location in the code base at hand\,\citep{Ji2015}.
Feature location is the process of identifying the code that implements a particular feature\,\citep{RevelleBC05}.
As such, it is one of the most common activities of developers.
Unfortunately, feature location is laborious and error-prone, especially for long-living software systems with many developers and features that are scattered over the code base.
Documenting features would help, but requires upfront effort and is often avoided, requiring recovery of features and their locations\,\citep{Rubin2013ASO}.


\looseness=-1
Feature location classifies into eager and lazy strategies\,\citep{Ji2015}.
The \textit{eager strategy} refers to recording information on feature locations during their development, either directly within the software assets or in external trace databases.
Different methods exist, such as using embedded code annotations for recording features, together with tools for browsing/visualizing features\,\citep{seiler2017,Martinson.2021,Andam.2017,Bergel.2021,Entekhabi.2019}, as well as feature traceability databases, such as FEAT\,\citep{Robillard2007}.
In contrast, the \textit{lazy strategy} recovers feature locations when needed.
Both, manual\,\citep{krueger2019} and automated\,\citep{Rubin2013ASO} techniques have been explored in research.
However, manual recovery is laborious and error-prone, and automated techniques (often relying on natural-language processing or machine-learning methods) yield too many false positives to be usable in practice. 
As such, our long-term goal is to establish methods and tools to record security features eagerly.
However, to construct effective techniques, we need to improve our empirical understanding of what security features are and how they manifest in source code.
In other words, developers need to know what security features are traceworthy and on which level of abstraction they should be captured---the goal of our study.
In addition, shedding light on what security features can be located easily in the implementation can also help improve manual and automated feature-location methods that try to retroactively recover features from software assets.







\subsection{Security Feature Tracing}
The interrelation of features and their implementation in code throughout the development process is called tracing.
It is often required by security standards such as the Common Criteria\,\citep{CC}.
To this end, previous work proposes techniques to enable the traceability of security features.
The technique SecSTAR by \citet{FangMK12} traces a software system's security structure and properties and generates diagrams to support security analysis.
Enterprise Architect\,\citep{enterprisearchitect} provides commercial tool support for strictly coupling UML models to code to facilitate the synchronization between them, which could also be used for UML models describing security features.
SecReq\,\citep{Houmb2010} is a methodology for eliciting security requirements as well as the early detection and refinement of security issues with traceability support for UML design models.
\citet{Islam2011} propose a framework for obtaining security requirements from laws and regulations and tracing them to security requirements throughout the whole development life cycle to enable checking compliance with laws and regulations.
The GRaViTY\,\citep{Peldszus2020,Peldszus2022} framework maintains traceability between different artifacts, such as UML models, Java source code, and program models.
It uses trace links to propagate security requirements into the implementation.
Strong coupling between the source code and the models is required to enable the traceability of security features using these approaches.
In summary, these approaches do not yet provide enough flexibility for a vast practical application.


\subsection{Security Standards and Guidelines}
\label{sub:background_standards}
Security standards and guidelines provide developers with security features that need to be realized to secure a software system.
Many product requirements in the industry are formulated around security standards, for example, a system should adhere to all certification requirements of a specific standard.
\add{In fact, standards compliance is mandatory for systems like the EHRS~\citep{MDR2017,HIPAA1996}.}
An organization's information security management system or a single software system can be certified according to a certain security standard if it can be proven that the required security controls are implemented.
Such proofs are usually in the form of documentation of carried-out activities, e.g., the identification of security threats and the specification and realization of mitigating security features.
Due to the procedural nature of the standards, the requirements, for the most part, describe actions that have to be performed or high-level security functionality that has to be achieved.
The few implementation-level security features that are mentioned are mostly in terms of specific technologies that are given as examples of how to realize some security control and are often lost in a huge body of text.
\begin{figure}
    \centering
    \includegraphics[width =\columnwidth]{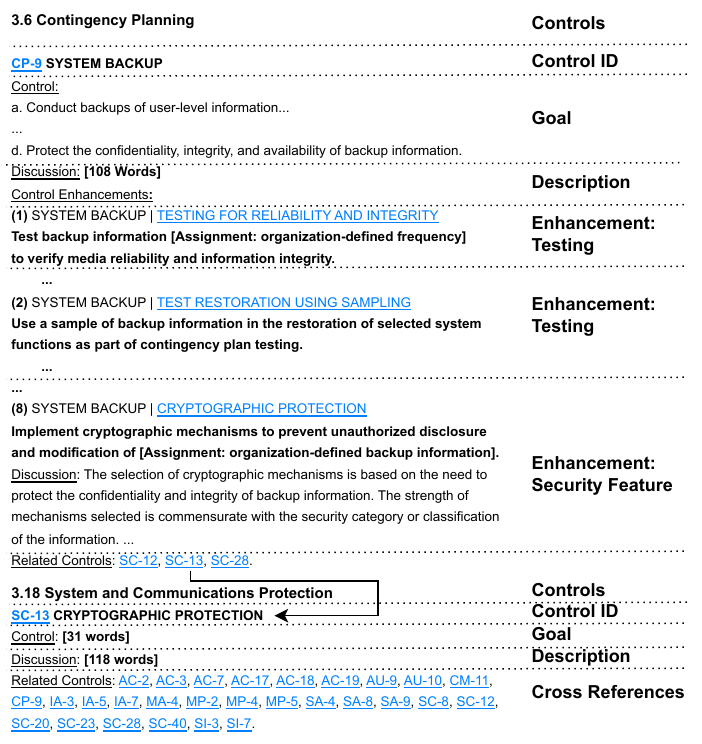}
    \caption{\add{Excerpt of the NIST SP 800-53 standard for security and privacy controls for information systems and organizations}}
    \label{fig:encryption-standard}
    \vspace{-4mm}
\end{figure}
\add{For illustration, Fig.~\ref{fig:encryption-standard} shows an excerpt from the NIST SP 800-53 standard which provides security and privacy controls for information systems and organizations. The excerpt focuses on security controls for \textit{contingency planning}, such as \textit{system backup} and presents associated control enhancements that add functionality or specificity to this base control. It can be seen that the functional-level security features such as \textit{cryptographic protection} are hidden among several other security-related information such as \textit{testing for reliability and integrity, test restoration using sampling} and so on. The figure also shows multiple cross-references (e.g., SC-12, SC-13, SC-28) meant to provide additional details on the control obscuring specific functional-level security features in an extensive and interconnected array of information.
Additionally, the descriptions for such security features such as \textit{“implement cryptographic mechanisms to prevent unauthorized disclosure,”} as in the figure, are often broad and abstract providing little concrete guidance for its practical implementation.}
Therefore, we see the need for a comprehensive overview of implementation-level functional security features.
A taxonomy of such features, together with a mapping to the standards and guidelines, could assist developers by giving actionable advice for how to realize required security controls.

\section{Methodology}
\noindent
We conducted a \emph{systematic review}~\citep{Ralph21_empirical_standards} of literature and security frameworks to elicit functional security features and how they are provided in security frameworks. Figure~\ref{fig:methodology} shows our research methodology. To identify implementation-level security features, we reviewed the literature that presents structured collections of security features~(\textbf{RQ1}). 
To ensure the applicability of the taxonomy in practice and to validate it, we created a mapping between our taxonomy and security features described in widely used security standards and potentially adapted the taxonomy. Additionally, we collected and inspected existing security frameworks discussed by developers to understand which functional implementation-level security features are provided to developers~(\textbf{RQ2}) and investigated their representation in source code through different mechanisms (e.g., code annotations)~(\textbf{RQ3}). 

\add{All steps in the creation of the taxonomy, the mapping to the security standards, and the analysis of the security frameworks followed the same general process, considering the recommendations by~\cite{McDonald2019Reliability}. Two authors performed the initial analysis (of the literature, the standards, and the frameworks) and discussed the results. Discrepancies were discussed in group meetings with the first five authors.
In more than 30 meetings lasting about one hour each, we further regularly discussed the resulting taxonomy to reach full agreement, i.e., each decision was discussed until all involved authors agreed on the solution. Since we reached full agreement after our discussion rounds, we did not calculate an inter-coder agreement~\citep{McDonald2019Reliability}.
}

\begin{figure*}
    \centering
    \includegraphics[width = 0.8\linewidth]{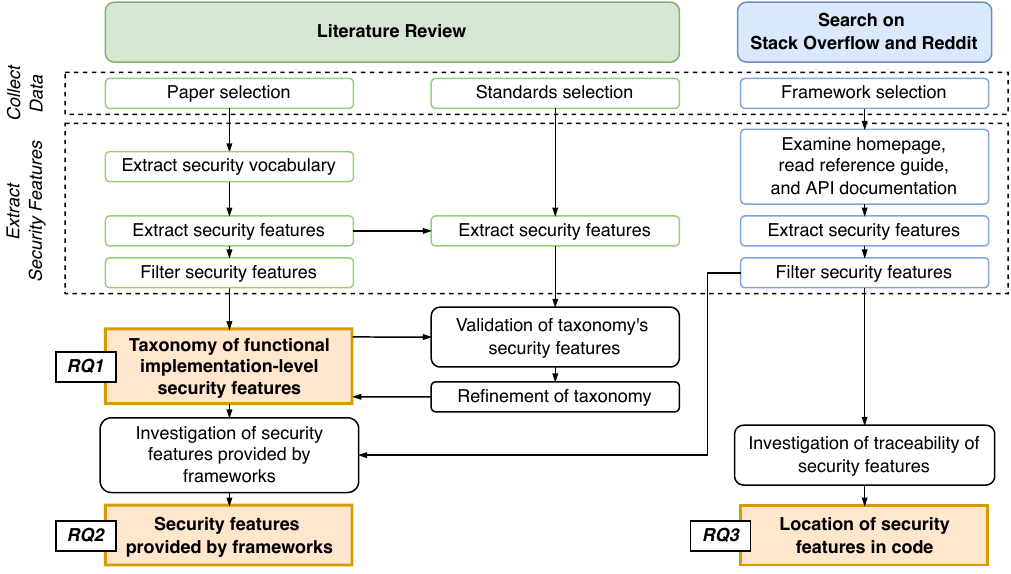}
    \caption{Overview of the applied research methodology}
    \label{fig:methodology}
\end{figure*}

\subsection{Systematic Literature Review}
\label{sec:methodology:slr}

To establish an empirical understanding of functional implementation-level security features, we reviewed structured collections of such in the literature.

\subsubsection{Paper Selection}
\label{subsec:selecting_ontologies}
\begin{figure}
    \centering
    \includegraphics[width = \linewidth]{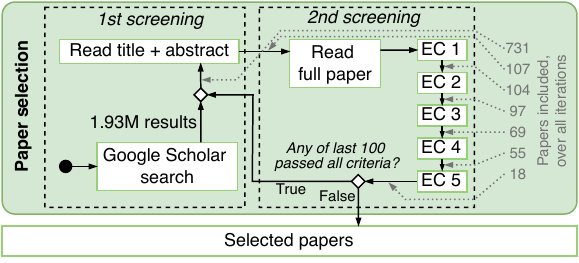}
    \caption{Paper selection process of the SLR with sequential application of exclusion criteria}
    \label{fig:methodology_slr}
    \vspace{-4mm}
\end{figure}
\noindent We conducted a manual two-step screening process to select relevant papers, as shown in Figure \ref{fig:methodology_slr}.
We searched for relevant publications on Google Scholar using the tool \textit{Publish or Perish} \citep{Harzing07}.
\add{Google Scholar covers the typical major data sources for literature reviews such as IEEExplore or the ACM Digital Library.
In a study by \cite{Valente22_academic_databases_for_slrs}, the authors observed that Google Scholar provides the most comprehensive search results for literature reviews in the computer science domain.
By using a single data source, we could directly apply a saturation criterion without having to merge search rankings of different sources.
}
To this end, \add{five authors} collected keywords to tailor the search \add{in a group meeting}: (1) terms related to the considered implementation level, (2) synonyms of ``security,'' and (3) terms describing a systematic representation of aspects.
Thus, we ended up with the following compound search term: 

\vspace{2pt}
\begin{center}
\texttt{(implementation OR code OR program) AND\\ (security OR secure) AND (ontology OR taxonomy OR "body of knowledge" OR\\ "system of knowledge" OR "conceptual model")}
\end{center}
\vspace{2pt}


    

\noindent We performed a query with this search term and examined each result in their ranked order in a two-step screening process. 
In the first step, we read the title and abstract of the paper to verify that it includes a structured collection of security features. 
Whenever the title and abstract were not enough to make this decision, we read other parts of the paper. 
We considered 111 papers for further review in the next step.
%
When a paper passed the first step, we read the full paper in the second step. 
We filtered the papers according to five criteria, that we chose to fit our scope of functional security features and applied all of them one after another to each resulting paper.
Furthermore, we excluded papers that are only applicable in specific domains, as our goal is to provide an overview of security features in general software systems.

Through the investigation of the literature, we mostly encountered ontologies and taxonomies for structuring such collections.
While a taxonomy represents a ``general categorization based on a class/subclass relationships,'' an ontology is ``the formal specification of domain concepts and their relationships'' \citep{hakeem2004ontology}.
Additionally, a number of security standards describe security requirements, which indicate security features needed to fulfill the requirements.


\noindent\textbf{Exclusion criteria:}
\begin{description}

    \item[EC1:] the paper is not published in a conference, or journal
    \item[EC2:] the collection of security features is not made available
    \item[EC3:] the scope of the paper is limited to a specific application domain, e.g., only CAN bus security
    \item[EC4:] only threats, vulnerabilities, risks, and so on are considered without presenting countermeasures
    \item[EC5:] the paper is not associated with functional security features considered in software engineering
\end{description}

\add{For data analysis, we followed a process considering the recommendations by McDonald et al.~\cite{McDonald2019Reliability}. Two authors initially performed the selection and repeatedly compared their results to check whether the saturation criterion was fulfilled. Then, the results including all possible conflicts and ambiguities were discussed in regular group meetings with the first five authors.
}
\noindent We reached saturation at the mark of 731 search results, as we observed no new papers that passed the screening process within over 100 search results before that point.
Table~\ref{tab:ontologies} lists the 18 papers that passed all exclusion criteria and were considered for extracting security features. As shown in Fig. \ref{fig:methodology_slr}, we excluded most papers (37) in the second screening step based on EC5. Note that we report only the first applicable exclusion criterion per paper, as we did not check the additional criteria after exclusion. 


\subsubsection{Extraction of Security Features}
\label{subsubsec: feature_extraction}
\noindent After shortlisting the papers through the two-step screening process, we collected all security-related terms presented in them.
\add{The extraction process was adapted to each analyzed paper's structure and presentation.
Most papers contained a visual representation of the presented features as a graph, some presented them as a table.
We analyzed these representations and their corresponding descriptions of the contained features to identify all security features.
The process and its results were performed and discussed by the first five authors in recurring meetings.}
During the extraction, we considered implementation-level security features related to software engineering. 
To this end, we excluded security features that are only limited to specific application domains of software systems such as automotive systems to keep the resulting set of security features as widely applicable as possible.
In particular, we removed all terms meeting any of the following properties:

\begin{itemize}
    \item Specific to hardware (e.g., ID card or credit card)
    \item Limited to a single platform, such as operating systems, libraries, or other technologies (e.g., $\mu$C/OS) 
    \item Not related to security (e.g., supplier or memory)
    \item Associated with security attacks or vulnerabilities (e.g., DoS attack, sniffing attack or P2P attack)
    \item Restricted to a single application domain (e.g., the automotive domain) 
\end{itemize}

\noindent We created a taxonomy containing all the collected security features. 
With a graph editor, we compared the presented security concepts and identified overlaps\add{, i.e., security features contained in multiple of the analyzed papers.}
We merged the individual sets of terms of each of the selected papers, \add{starting from one paper and iteratively adding the others by identifying security features included in the merged set and the newly added paper and adding all connected features at this place.}
Finally, we classified and grouped the security features to give them a coherent structure, following the classification and hierarchy rules from the originating papers.
Five authors discussed the taxonomy's terms to agree on a structure. 
The resulting taxonomy contains all implementation-level security features identified in the final set of papers of our SLR.

\subsubsection{Mapping to Security Standards}
\label{methodology:standard-mapping}
Security standards are generally regarded as highly reliable sources of information for securing software systems because they undergo rigorous review processes before being published.
Despite the lack of implementation details, official standards and guidelines by large, reputable organizations are a common source of information about software security.
Thus, creating a mapping of our taxonomy to established security standards increases its relevance for application in the industry.
Furthermore, a successful mapping allows reasoning about the validity of the derived taxonomy.

We expect that each functional security feature in the standards can be mapped to one or more security features in the taxonomy.
Therefore, the mapping allowed us to validate the completeness of the taxonomy we derived from the literature. 
As relevant standards for the mapping, we analyzed the \textit{ISO/IEC 27000 family}, the \textit{Common Criteria} (CC), the \textit{NIST SP800-53}, and the \textit{NIST Cybersecurity Framework}, which are widely recognized in the industry as the most important security standards and guidelines.

To create the mappings, two authors independently analyzed each part of the standards, identified functional security features, and proposed security features from the taxonomy that correspond to this description. 
In addition, for identified security features that were not yet part of the taxonomy, they also proposed adaptations to the taxonomy to support all functional security features from the standards.
Then, together with three further authors, each part was discussed, and a decision for the mapping was taken collaboratively.
To this end, for the security features that could not be immediately mapped to the taxonomy, the first five authors of this work discussed whether, where, and how to adapt the taxonomy to include the features.
This validation and adaptation process was performed for each standard, starting with the CC.

We report in Section~\ref{sub:comparison_standards} for each standard how well it could be mapped to the taxonomy, and what minor and major changes to the existing taxonomy were required to allow the mapping of all security features.
In this way, we provide guidance to developers who can use the taxonomy as an abstraction of the standards.
The granularity of security features in our taxonomy lies between the high-level descriptions of security mechanisms found in most standards and the detailed requirements for specific technologies found in others.

\subsection{Identification of Security Frameworks}
\noindent
We systematically identified popular security frameworks discussed on the popular developer platform Stack Overflow \add{and the programming community of Reddit} to compare the state-of-practices of functional security features with our derived taxonomy.
\add{We chose Stack Overflow since it is one of the largest and also most popular platforms for content related to software development amongst developers\,\citep{Xia17_developers_searches}.
On Stack Overflow, developers mainly discuss problems or seek recommendations when facing problems during their development tasks.
As a second data source, we chose Reddit's largest developer community ``r/programming'', which, from its origins, is the most popular place on the platform for exchanging programming related content.
In contrast to Stack Overflow, developers do not discuss the usage of security frameworks, but present them to other developers by sharing articles or repositories, allowing us to capture a different type of discussion.}
We investigated which implementation-level security features are provided by frameworks used in practice and their relation to the literature captured in our taxonomy.
Further, we investigated the mechanisms used to provide security features and how these could be leveraged for locating security features.
We refer to \textit{security frameworks} when they focus on providing security mechanisms and related functionality.


\subsubsection{Identifying Security Frameworks from Stack Overflow and Reddit}
\noindent To identify relevant security frameworks discussed in practice, we searched for \textit{``security framework''} on the widely used developer discussion platform \textit{stackoverflow.com}.
We used the Stack Exchange API v2.3~\citep{stackexchange} to download threads, ensuring that they remained unaltered throughout the entire analysis period when we reviewed the results. 
\add{Two authors} sorted the threads by relevance and investigated the results by \add{independently} reading the questions, answers, and comments of each thread. 
In the threads, we manually searched for mentions of security frameworks or security modules of general frameworks.
We continued the search until no new frameworks were mentioned in the last 20 threads. 
We reached this data saturation \citep{glaser1978} at 250 threads.

\add{For the search on Reddit, we employed a similar approach as we did for the search on StackOverflow.
Searching for \textit{``security framework''} resulted in 249 threads.
On Reddit, threads contain comments and either a user created discussion, or a link to an article.
As an initial filtering step, two authors exhaustively and independently read the titles of each thread, including all threads discussing security or securing applications for further investigation.
They then merged their sets of included threads, resulting in 68 threads.
Afterwards, they read all threads, including linked articles and comments, to extract all mentioned security frameworks.
Finally, we merged the results with our Stack Overflow search.}

\subsubsection{Extracting Security Features from Security Frameworks}
\label{subsubsec:methodology_extraction_frameworks}
\noindent To derive a final list of relevant frameworks, we selected all security frameworks that were mentioned in at least two threads during the identifications of security frameworks in the merged results.
We examined the selected frameworks in depth to capture the provided security features.
To this end, \add{two authors} used three different sources of information for each framework (unless not provided for a specific framework), the related \textit{homepage}, a \textit{reference guide}, and the \textit{official documentation}.
\add{Each author independently recorded security features described in each source.}
The framework's homepage usually provided a general overview of the security features included in the framework, while in the reference guide, a more detailed look at the security features was often given. Using the official documentation, we investigated the low-level components and encountered an in-depth description of the framework and its methods.
In cases where the three sources used different terminology to describe the same security feature, \add{the two authors compared the terminology and descriptions across the sources in joint sessions, and} chose the term used by at least two, or the best-fitting one if all three used different terms. Furthermore, \add{in these discussions} we categorized some specific terms, such as \textit{username} and \textit{password}, into broader security features such as \textit{credentials}.
We considered any security feature that offers a reusable functionality at the implementation level that addresses a security requirement or security issue\,\citep{mcgraw2004}.
\add{The same two authors} organized the features into a hierarchy based on the structure in the frameworks' documentation. \add{Discrepancies were discussed and resolved through collaborative sessions, ensuring that the resulting hierarchy accurately reflected the frameworks' intended structure.}
While investigating the security features, we documented in parallel information on using the individual security features offered by the selected frameworks in source code.
\add{Based on the mechanism described in the documentation,} we grouped the security features into the three realization methods \textit{annotations}, \textit{APIs}, and \textit{configuration files}. \add{Any disagreements were addressed by re-examining the documentation together to reach a consensus.}
Whenever a security feature was mentioned in combination with an API artifact, such as a method, interface, variable, or class, we grouped the security feature to a realization with an API.
Likewise, if a configuration file, such as a \texttt{.xml}, \texttt{.properties}, or \texttt{.conf} file was mentioned along the security feature, we mapped the realization to a configuration file.
Finally, we applied the same procedure for annotation mechanisms, such as Java annotations or attributes in C\#.
We collected this information for each security feature and framework to reason about their traceability (RQ3).



\section{Taxonomy of Implementation-level Security Features (\textbf{\textit{RQ1}})}\label{sec:RQ1}
\noindent
In our SLR, we identified papers that present ontologies and taxonomies of software security features from which we extracted functional code-level security features.
Thereafter, we constructed a taxonomy out of these and mapped it to four security standards to further validate and refine it.
In the following, we describe the results of our analysis.

Table~\ref{tab:ontologies} presents the 18 papers (referred to as \textit{P1} - \textit{P18} in this work) identified in our SLR for instantiating our taxonomy of functional security features.
Our taxonomy consists of 68 im\-ple\-men\-ta\-tion-level security features shown in Figure~\ref{fig:taxonomy_full}.
Note that security features are not necessarily mutually exclusive from each other.
As such, they may be combined to achieve a higher security goal or property (e.g., a system might realize both credentials and multifactor authentication to protect the confidentiality of data).
Security features with a similar goal or security property that they achieve were grouped under top-level security features, which we identified from the hierarchies of the reviewed ontologies and taxonomies.
\add{We additionally added annotations to the corresponding security features to indicate the frameworks and standards in which they were identified (see Fig.~\ref{fig:taxonomy_access_control}, and Fig.~\ref{fig:taxonomy_cryptography} to Fig.~\ref{fig:taxonomy_system_state_protection}).}




\begin{table*}
\centering\footnotesize
\setlength{\tabcolsep}{2.0pt}
\caption{Shortlisted papers presenting security features}
\label{tab:ontologies}

\begin{tabular}{c p{2.3cm} p{13cm} p{0.7cm}} \toprule

\textbf{ID} & \textbf{Authors} & \textbf{Title} & \textbf{Year} \\ \midrule

\textbf{\textit{P1}} & Venter et al. & A taxonomy for information security technologies~\citep{VenterE03} & 2003 \\ \midrule
\textbf{\textit{P2}} & Denker et al. & Security for DAML  web services: annotation and matchmaking~\citep{DenkerKFPS03} & 2003 \\ \midrule
\textbf{\textit{P3}} & Abbas et al. & A state of the art security taxonomy of internet security: threats and countermeasures~\citep{Abbas2005ASO} & 2005 \\ \midrule
\textbf{\textit{P4}} & Herzog et al. & An Ontology of Information Security~\citep{HerzogSD07} & 2007 \\ \midrule
\textbf{\textit{P5}} & Kim et al. & Security Ontology to Facilitate Web Service Description and Discovery~\citep{KimLK07} & 2007 \\ \midrule
\textbf{\textit{P6}} & Vorobiev et al. & An Ontology-Driven Approach Applied to Information Security~\citep{VorobievB10} & 2010 \\ \midrule
\textbf{\textit{P7}} & Kang et al. & A Security Ontology with MDA for Software Development~\citep{KangL13} & 2013 \\ \midrule
\textbf{\textit{P8}} & Habiba et al. & Cloud identity management security issues \& solutions: a taxonomy~\citep{HabibaMSN14} & 2014 \\ \midrule
\textbf{\textit{P9}} & Hendre et al. & A semantic approach to cloud security and compliance~\citep{HendreJ15} & 2015 \\ \midrule
\textbf{\textit{P10}} & Busch et al. & An ontology for secure web application~\citep{BuschW15} & 2015 \\ \midrule
\textbf{\textit{P11}} & Talooki et al. & Security concerns and countermeasures in network coding based communication systems: A survey~\citep{TalookiBLRFMT15} & 2015 \\ \midrule
\textbf{\textit{P12}} & Kaur et al. & Security of software-defined networks: Taxonomic modeling, key components and open research area~\citep{Kaur2016} & 2016 \\ \midrule
\textbf{\textit{P13}} & Bhatia et al. & Data security in mobile cloud computing paradigm: a survey, taxonomy, and open research issues~\citep{BhatiaV17} & 2017 \\ \midrule
\textbf{\textit{P14}} & Adat et al. & Security in Internet of Things: issues, challenges, taxonomy, and architecture~\citep{AdatG18} & 2018 \\ \midrule
\textbf{\textit{P15}} & Harbi et al. & A review of security in internet of things~\citep{HarbiAHBR19} & 2019 \\ \midrule
\textbf{\textit{P16}} & Kumar et al. & On cloud security requirements, threats, vulnerabilities, and countermeasures: A survey~\citep{KumarG19} & 2019 \\ \midrule
\textbf{\textit{P17}} & Khanam et al. & A survey of security challenges, attacks taxonomy and advanced countermeasures in the internet of things~\citep{KhanamAIJS20} & 2020\\ \midrule
\textbf{\textit{P18}} & Mahapatra et al. & A Survey on Secure Transmission in Internet of Things: Taxonomy, Recent Techniques, Research Requirements, and Challenges~\citep{Mahapatra2020} & 2020 \\ \bottomrule

\end{tabular}
\end{table*}

\subsection{Taxonomy of Functional Security Features}

\begin{figure*}
    \centering
    \includegraphics[width = 0.98\textwidth]{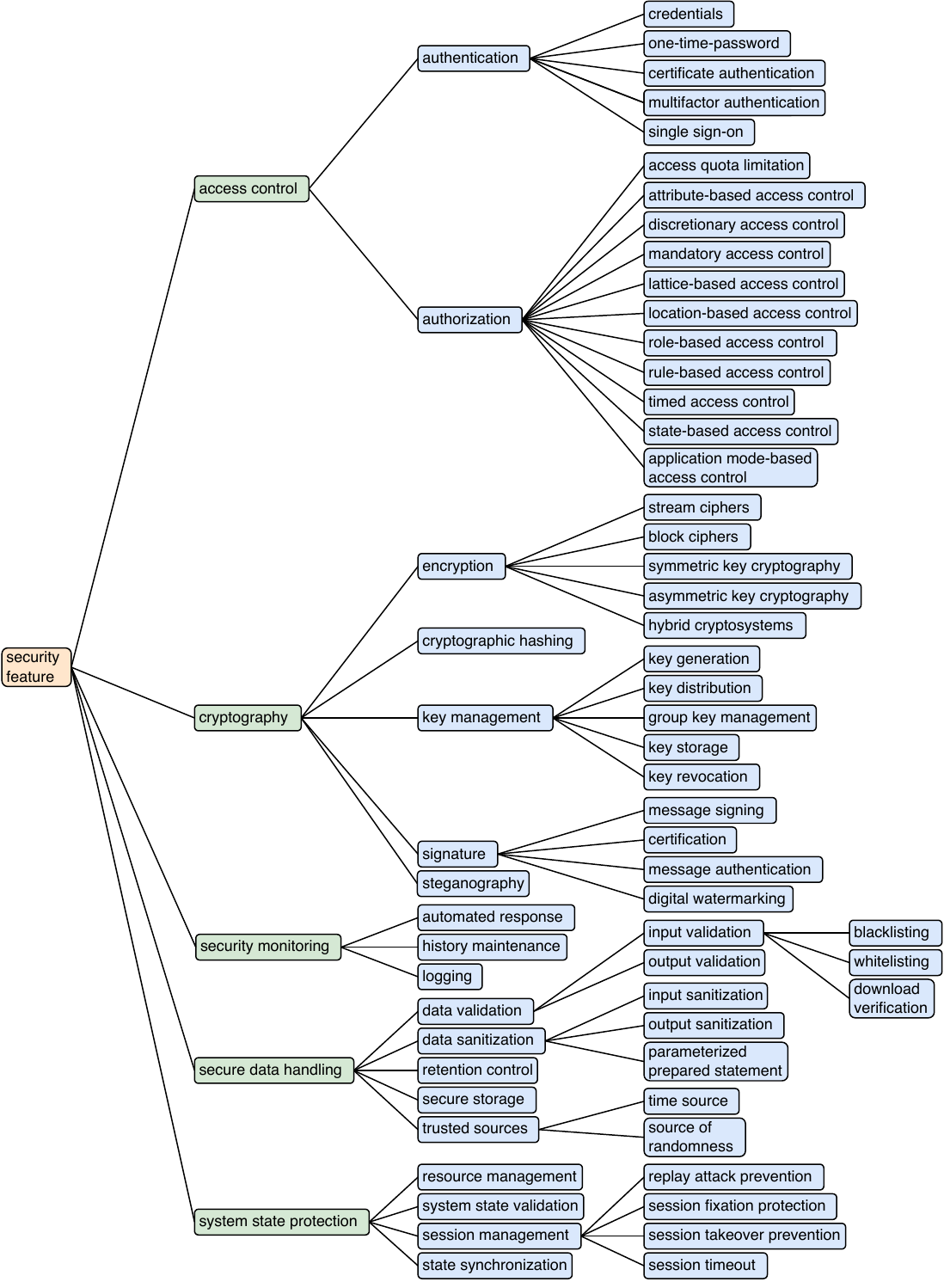}
    \caption{\add{Our full taxonomy}}
    \label{fig:taxonomy_full}
\end{figure*}

We identified five top-level security features, as shown in Figure~\ref{fig:taxonomy_full}: \textit{access control}, \textit{cryptography}, \textit{security monitoring}, \textit{secure data handling} and \textit{system state protection}.
Table~\ref{tab:ontology-feature} shows the occurrences of these security features in the papers.
Three papers included all top-level features\add{, however, our taxonomy contained more security features beyond the top-level ones for each of the papers}.
The top-level feature \textit{access control} is included in all 18 papers.
Only one paper does not include \textit{cryptography}.
This shows the importance of these two groups of features.
The detailed taxonomy of these security features is presented in Fig. \ref{fig:taxonomy_full}.
We now describe each of them in detail.


\begin{table*}[]
\centering
\caption{\add{Security feature occurrence in the 18 shortlisted papers}}
\label{tab:ontology-feature}
\resizebox{\textwidth}{!}{%
\begin{tabular}{@{}l|c>{\columncolor[gray]{0.925}}cc>{\columncolor[gray]{0.925}}cc>{\columncolor[gray]{0.925}}cc>{\columncolor[gray]{0.925}}cc>{\columncolor[gray]{0.925}}cc>{\columncolor[gray]{0.925}}cc>{\columncolor[gray]{0.925}}cc>{\columncolor[gray]{0.925}}cc>{\columncolor[gray]{0.925}}c@{}}
\toprule
\textbf{} & \textbf{P1} & \textbf{P2} & \textbf{P3} & \textbf{P4} & \textbf{P5} & \textbf{P6} & \textbf{P7} & \textbf{P8} & \textbf{P9} & \textbf{P10} & \textbf{P11} & \textbf{P12} & \textbf{P13} & \textbf{P14} & \textbf{P15} & \textbf{P16} & \textbf{P17} & \textbf{P18} \\ \midrule
\textbf{access control} &  &  &  &  &  &  &  &  &  &  &  &  &  &  &  &  &  &  \\
authentication & \checkmark & \checkmark & \checkmark & \checkmark & \checkmark & \checkmark & \checkmark & \checkmark & \checkmark & \checkmark & \checkmark & \checkmark & \checkmark & \checkmark & \checkmark & \checkmark & \checkmark & \checkmark \\
authorization &  &  & \checkmark & \checkmark & \checkmark & \checkmark & \checkmark & \checkmark & \checkmark & \checkmark & \checkmark & \checkmark & \checkmark & \checkmark & \checkmark & \checkmark & \checkmark & \checkmark \\ \midrule
\textbf{cryptography} &  &  &  &  &  &  &  &  &  &  &  &  &  &  &  &  &  &  \\
cryptographic hashing &  &  & \checkmark &  &  & \checkmark &  &  &  & \checkmark & \checkmark &  & \checkmark & \checkmark & \checkmark & \checkmark & \checkmark & \checkmark \\
encryption & \checkmark & \checkmark & \checkmark & \checkmark & \checkmark & \checkmark & \checkmark & \checkmark & \checkmark & \checkmark & \checkmark &  & \checkmark & \checkmark & \checkmark & \checkmark & \checkmark & \checkmark \\
key management &  & \checkmark & \checkmark & \checkmark & \checkmark & \checkmark &  &  & \checkmark & \checkmark & \checkmark &  & \checkmark & \checkmark & \checkmark & \checkmark & \checkmark & \checkmark \\
signature & \checkmark & \checkmark &  & \checkmark & \checkmark & \checkmark & \checkmark & \checkmark &  & \checkmark & \checkmark &  & \checkmark & \checkmark & \checkmark & \checkmark & \checkmark & \checkmark \\
steganography &  &  &  & \checkmark &  &  &  &  &  & \checkmark &  &  &  &  &  &  &  &  \\ \midrule
\textbf{security monitoring} &  &  &  &  &  &  &  &  &  &  &  &  &  &  &  &  &  &  \\
automated response &  &  &  &  &  &  &  &  & \checkmark &  &  &  &  & \checkmark &  & \checkmark &  &  \\
history maintenance &  &  &  &  &  &  &  & \checkmark &  &  &  &  &  &  &  & \checkmark &  &  \\
logging & \checkmark &  &  & \checkmark &  &  &  & \checkmark & \checkmark & \checkmark & \checkmark &  & \checkmark & \checkmark &  & \checkmark & \checkmark &  \\ \midrule
\textbf{secure data handling} &  &  &  &  &  &  &  &  &  &  &  &  &  &  &  &  &  &  \\
data validation &  &  &  &  &  &  &  &  &  & \checkmark & \checkmark &  & \checkmark &  &  & \checkmark &  &  \\
data sanitization &  &  &  &  &  &  &  &  &  & \checkmark &  &  &  &  &  & \checkmark &  &  \\
retention control &  &  &  &  &  &  &  &  &  & \checkmark &  &  &  &  &  &  &  &  \\
secure storage &  &  &  &  &  &  &  &  &  & \checkmark & \checkmark &  & \checkmark & \checkmark & \checkmark & \checkmark & \checkmark & \checkmark \\
trusted source &  &  &  &  &  &  &  &  &  & \checkmark &  &  & \checkmark &  & \checkmark & \checkmark & \checkmark &  \\ \midrule
\textbf{system state protection} &  &  &  &  &  &  &  &  &  &  &  &  &  &  &  &  &  &  \\
resource management &  &  &  &  &  &  &  &  &  &  & \checkmark &  &  & \checkmark &  & \checkmark &  &  \\
system state validation &  &  &  &  &  &  &  &  &  &  &  &  &  &  &  & \checkmark &  &  \\
session management &  &  &  & \checkmark & \checkmark &  &  &  &  & \checkmark & \checkmark &  &  &  &  & \checkmark &  &  \\
state synchronization &  &  &  &  &  &  &  &  &  &  &  &  &  &  &  & \checkmark &  &  \\ \bottomrule
\end{tabular}%
}
\end{table*}



\begin{figure*}
    \centering
    \includegraphics[width = 0.7\textwidth]{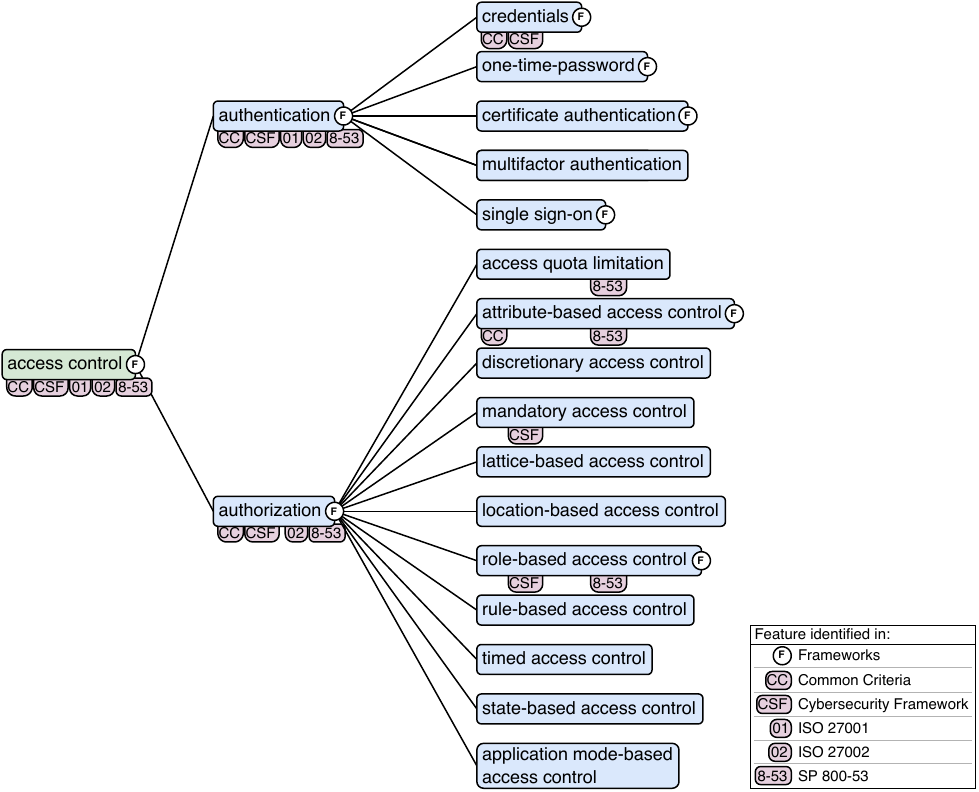}
    \caption{Sub-features of the top-level security feature \textit{access control}}
    \label{fig:taxonomy_access_control}
\end{figure*}

\subsubsection{Access Control} %
%
%

\noindent\textit{Access control} covers security features that are concerned with regulating access to protected resources and granting them only to authorized subjects.
\add{For example, in the EHRS from our example, doctors should only be allowed to read sensitive data of a patient they are designated to, while other doctors are only allowed to write to it. However, doctors have the permission to view anonymized statistics, such as past treatment of patients for different diagnoses. This requires the control of all accesses to the system.}
All papers include security features for access control in their ontologies.
The top-level feature comprises two major blocks of features, one grouped under the sub-feature \textit{authentication} and one under \textit{authorization} (Figure~\ref{fig:taxonomy_access_control}).

\textit{Authentication} is presented in the papers both, as a security notion, objective, or a means to achieve data confidentiality, and as a security feature that implements these.
\add{The user must be identified, i.e., whether it is a doctor or not, before they gain access to it.}
\textit{P2}, \textit{P5}, \textit{P12}, \textit{P13}, and \textit{P14} describe authentication as the identification and verification of a party sending a request to a network or application, where the associated security features support the realization of this.
Here, authentication is a security feature that often entails multiple other security features because of its complexity or manifests at multiple places in the code base.
According to \textit{P7}, authentication is used to achieve data confidentiality, while the remaining papers define authentication in terms of user, data, and message integrity/authenticity.
In our taxonomy, authentication classifies into more specific security features (see Fig.~\ref{fig:taxonomy_access_control}).
When authentication is performed using \textit{Credentials}, data objects such as usernames or passwords are used to verify the identity of a user.
\add{In the EHRS from our example, each hospital staff member could receive a set of credentials from an administrator, allowing them to log into the system from an arbitrary device within the hospital.}
\textit{One-time-password} is a method where authentication is performed with randomly generated temporary passwords \citep{HabibaMSN14}. \textit{Certificate authentication} refers to authentication using certificates such as X.509~\citep{DenkerKFPS03}.
\textit{Multifactor authentication} requires the user to provide more than one way of authentication \citep{BhatiaV17}.
\textit{Single sign-on} allows users to securely log in to multiple systems using only one set of credentials.

\textit{Authorization} is also described by most papers as a security objective, requirement, or goal, and as a security feature that realizes these.
Further, it is defined as a means to achieve access control, with an access source, access target, and actions that are permitted to be executed \citep{BuschW15}.
\add{In our example's EHRS, after a user has authenticated their identity, authorization determines what actions this user can perform in the system, i.e., viewing or writing medical records.}
\textit{Access quota limitation} refers to limiting the usage of resources so that high-priority actions that could be security sensitive are not delayed by other relatively low-priority tasks.

\add{Several schemes can be implemented to assign permissions to users for different purposes.}
In \textit{attribute-based access control} (ABAC), the access is determined based on attributes such as requested operation, request parameters, or environmental attributes\,\citep{Chung2019}.
\add{In the EHRS from our example, an attribute decides whether a doctor is designated to a patient.}
\textit{Discretionary access control} (DAC) is where the resources are restricted based on the identity of the users. DAC has complete trust in the users\,\citep{IBMDAC}.
On the contrary, \textit{mandatory access control} (MAC) grants access to resources based on clearance of users, or a predefined hierarchy\,\citep{IBMMAC}.
\textit{Lattice-based access control} determines access to the resources based on a hierarchical lattice structure that represents possible interaction between the resources and the users. This lattice structure is created based on the security levels of the resources and the users\,\citep{Denning1976}.
\textit{Location-based access control}, as the name indicates controls the access based on the location of the user\,\citep{Ardagna2009}.
\textit{Role-based access control} restricts access based on the roles assigned to the users\,\citep{Ferraiolo2009}\add{, such as doctors or hospital staff}.
\textit{Rule-based access control} is established based on a predefined set of access rules.
\textit{Timed access control} enforces permission or access to resources based on time parameters such as schedule or duration of access.
\textit{State-based access control} introduces more fine-grained access decisions than a simple ``allow'' or ``deny'' (\cite{Kamra10_state-based_access_control}).
For example, ``request suspension'' is a decision that requires a further negotiation process before deciding whether to grant access (\cite{Bertino11_access_control}).
\textit{Application mode-based access control} is a special case of state-based access control\,\citep{bosch2000}.
In summary, all different access control schemes are implemented such that some property of the access source and/or access target is checked against specific requirements.
\add{The EHRS from our example uses a simplified role- and attribute-based access control scheme -- a common combination in the literature \citep{Jin2012,Ahmadian2017}.
Since multiple roles share permissions, a role hierarchy is implemented in which permissions for a role are inherited from a parent role (see Fig.~\ref{fig:roles-hierarchy}).
For example, a doctor should have the same permissions as the medical staff.
Therefore, in the role hierarchy, the role \textit{doctor} should inherit all permissions from the role \textit{medical staff}, i.e., writing medical records, and extend it with additional permissions such as writing diagnoses.
Since a patient can choose a doctor to be a designated doctor, an attribute denotes whether they have elevated rights over the access to the patient's data.
}

\begin{figure}
\centering
    \begin{tikzpicture}
        \node (staff) at (1,0) {Staff};
        \node (med) at (-1,-1) {Medical Staff};
        \node (admin)at (2,-1) {Hospital Staff};
        \node (doc)at (-2,-2) {Doctor};
        \node (nurse)at (0,-2) {Nurse};

        \draw[-{Triangle[open]}] (med) -- (staff);
        \draw[-{Triangle[open]}] (admin) -- (staff);
        \draw[-{Triangle[open]}] (doc) -- (med);
        \draw[-{Triangle[open]}] (nurse) -- (med);
    \end{tikzpicture}
    \caption{Simplified role hierarchy in the exemplary EHRS}
    \label{fig:roles-hierarchy}
\end{figure}
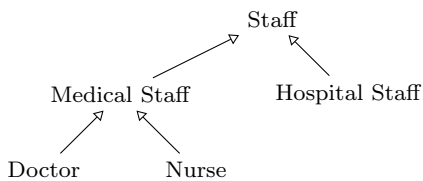

\begin{figure*}
    \centering
    \includegraphics[width = 0.7\textwidth]{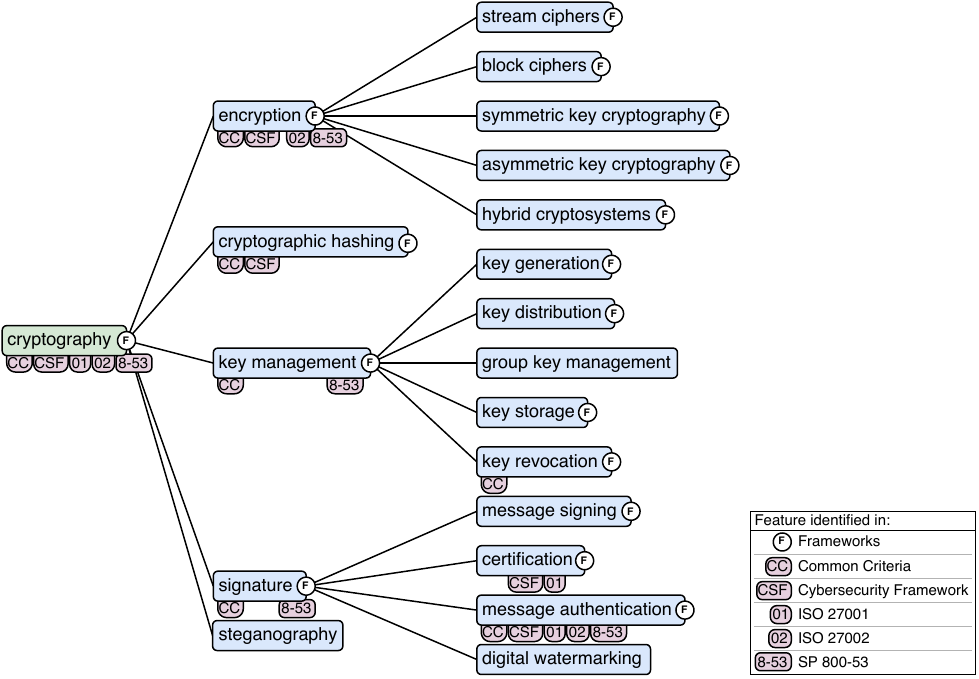}
    \caption{Sub-features of the top-level security feature \textit{cryptography}}
    \label{fig:taxonomy_cryptography}
\end{figure*}

\subsubsection{Cryptography}
\looseness=-1
The feature \textit{Cryptography} aims to ensure secure communication in the presence of adversaries\,\citep{rivest1990}.
The goal is to prevent unauthorized entities from reading a message by binding a key to the message.
A key is a secret consisting of a string of symbols that is used by an algorithm to encrypt or decrypt a message.

Except for \textit{P12}, all investigated papers list cryptography as a security feature, which should be considered when implementing software systems.
In our taxonomy, we separated cryptography into five sub-categories that focus on different aspects of cryptography: \textit{encryption}, \textit{cryptographic hashing}, \textit{key management}, \textit{signature}, and \textit{steganography} (see Fig.~\ref{fig:taxonomy_cryptography}).

\paragraph{Encryption.}
The first sub-category focuses on features for encoding messages in a way that only authorized entity access is able to access its content and protects it from unauthorized modification.
Algorithms used for encryption and decryption purposes are called ciphers and can be divided into groups of \textit{stream ciphers}\,\citep{Jiao20_stream_ciphers} and \textit{block ciphers}\,\citep{Robshaw95_block_ciphers}.
Rivest Cipher 4 (RC4) is one of the most widely used stream ciphers included in various protocols such as TLS.
In contrast, block ciphers encrypt a group of plaintext symbols as one ciphertext block.
The Data Encryption Standard (DES), triple DES (3DES), and the Advanced Encryption Standard (AES) are well-known block ciphers, which are used in modern software systems.

\add{In an EHRS such as the one from our example, patient data must be encrypted before it is stored within the system using any of the described ciphers, in case a third party is able to intercept data transmitted to or within the system.}
Any cipher must define a key that is shared among the multiple parties involved in an encrypted communication.
In \textit{symmetric key cryptography}\,\citep{Bokhari16_symmetric_encryption}, a single key is used for both encryption and decryption, while \textit{asymmetric key cryptography} defines a public key, which is used to encrypt a message, and a private key, which is exchanged between the communication parties and used to decrypt the received message\,\citep{Yassein17_symmetric_asymmetric_encryption}.
A \textit{hybrid cryptosystem} combines the two approaches by using asymmetric key cryptography to encrypt a key from symmetric key cryptography \,\citep{Dent04_hybrid_cryptography}.
While a symmetric key algorithm can either use a block cipher or a stream cipher to handle encryption or decryption, asymmetric cryptosystems rely on specific algorithms such as the Rivest Shamir Adleman (RSA) or the Diffie-Hellman exchange method to securely negotiate keys over a public transmission channel (\cite{Bhanot15_encryption_algorithms}).

\paragraph{Cryptographic hashing.}
This security feature focuses on ensuring that data has not been modified, e.g., a message exchanged between a sender and a receiver, without comparing the entire data. To this end, cryptographic hashing functions irreversibly transform data of arbitrary length into a fixed-length output of enciphered text\,\citep{BuschW15}.
\add{Using cryptographic hashing for medical records in the EHRS from our example ensures that malicious modifications to them can be detected.}
For the same input, the enciphered output is always identical and, therefore, allows a comparison of the calculated value before and after transmission of a message, while also ensuring the confidentiality of the data by making it unreadable for an attacker.
\add{In our exemplary EHRS, passwords in plain text are a major risk to the confidentiality of user data in the system.
Therefore, passwords for each user are stored in the database after state-of-the-art cryptographic hashing is applied.
In case of an incident in which passwords are stolen by an attacker, they are unable to gain access to the stored user accounts, since they can not decipher the hash.
}

\paragraph{Key management.}
Since cryptographic operations rely on secure and confidential keys, features for managing keys are essential\,\citep{Subhabrata23_key_management}.
First, \textit{key generation} must be performed in a secure way, which requires the usage of securely generated random numbers.
Second, to be able to encrypt and decrypt messages, the communicating parties must exchange a key via a \textit{key distribution} scheme (with \textit{group key management} as a special form of key distribution).
Once a key has been negotiated, only authorized users should be allowed access to it, which can be ensured by using a \textit{key storage} method.
Finally, \textit{key revocation} is used to invalidate a key once it is not required anymore, e.g., after a certain timeframe has passed or a criterion has been met.

\paragraph{Signature.}
Receiving a message often raises questions regarding whether its content and sender can be trusted.
To this end, signatures are used to verify the authenticity of a message by binding the identity of the sender to the sent message\,\citep{Katz10_digital_signatures}.
In this process, \textit{message signing} is used to create and bind a signature to a message by associating it with the private key of the sender.
\add{For instance, medical records are signed with a key of a doctor to establish authenticity of the message in our exemplary EHRS.}
Similarly, a \textit{certification} can be realized by a third party to show that a key can be trusted.
Thereafter, the receiver can use \textit{message authentication} to verify the origin and that the received message has not been tampered with in transit.
Here, the corresponding public key is then used to verify the private key bound to the message.
\textit{Digital watermarking} is a method to attach a non-removable signature to data to ensure its origin cannot be tampered with by a third party.

\paragraph{Steganography.}
Steganography can be used to hide a message within another, potentially without the use of an encryption algorithm\,\citep{Kour14_steganography}.
While there is a chance for an unauthorized entity to access the message, the idea is that unknowing entities are not able to notice that secret information is hidden within the message.
One such method hides a message within an image in a way that it cannot be perceived by humans.

\begin{figure*}
    \centering
    \includegraphics[width = 0.6\textwidth]{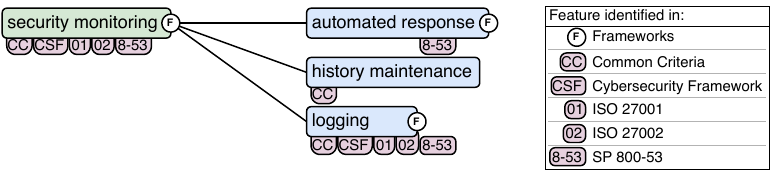}
    \caption{Sub-features of the top-level security feature \textit{security monitoring}}
    \label{fig:taxonomy_security_monitoring}
\end{figure*}

\subsubsection{Security Monitoring}
The top-level feature \textit{security monitoring} (Fig. \ref{fig:taxonomy_security_monitoring}) describes features for monitoring properties of software systems that can indicate the state of security or possible security issues.
For example, monitoring network traffic can be used to detect intrusions or other issues\,\citep{Ghafir16_monitoring_systems}.
In general, monitoring a software system can reveal attempted attacks and help prevent their success\,\citep{mcgraw2004}.
The feature contains \textit{automated response}, \textit{history maintenance}, and \textit{logging} as sub-features.
It is covered by ten of the ontologies (see Table~\ref{tab:ontology-feature}).

\textit{Automated response} refers to responding to incidents that happen in a software system that can lead to potential security violations.
If automated responses independent of human interaction are implemented, the response time to security incidents can be reduced.

\textit{Logging}, while not used as a feature to prevent attacks, can identify and trace back anomalies, such as attacks, within a system.
\add{Whenever a user such as a doctor writes or saves data to the EHRS from our example, the event is logged to a log file, containing the user, time, and action that was performed.
This assures the accountability of certain actions, and helps in reasoning about incidents that may occur.}
Additionally, several considerations regarding the security of the content of the logs must be taken. As such, the secure logging pattern intends to prevent an attacker from gathering sensitive data about a system from its logs\,\citep{dougherty2009}.
\add{In our exemplary EHRS, all logs are encrypted and access to them is restricted to specific users.}
Past work presents logging as a mechanism for providing non-repudiation and ensuring system and data integrity (\textit{P4, P8, P9, P13, P16, P17}), incident management (\textit{P1, P16}), and intrusion detection and prevention (\textit{P10, P14}). 

\textit{History maintenance} preserves the user/system activity logs, enabling the lookup and identification of wrongdoers or unwanted incidents in cases of a security breach.
The storing of relevant information has to be implemented in software systems to allow such investigations after an issue has been detected.
\add{In the EHRS example, changes resulting from system interactions are retained so that, for instance, previous versions of physician reports can be restored.}

\begin{figure*}
    \centering
    \includegraphics[width = 0.7\textwidth]{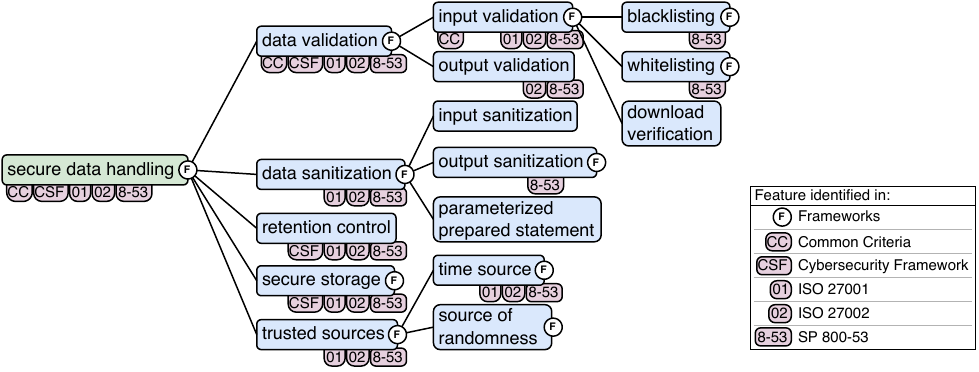}
    \caption{Sub-features of the security feature \textit{Secure Data handling}}
    \label{fig:taxonomy_secure_data_handling}
\end{figure*}

\subsubsection{Secure Data Handling}
This top-level feature is mentioned in 8 out of the 18 papers that we examined. The feature \textit{Secure data handling} covers security features that deal with validation, sanitization, and secure storage and control of the data that is handled in a software system (see Fig.~\ref{fig:taxonomy_secure_data_handling}).
\add{Since data management is a core feature within any EHRS, secure data handling plays a crucial role.}

\textit{Data validation} is characterized by two sub-features, \textit{input validation} and \textit{output validation}.
\add{For example, input validation ensures that entered data complies with a valid data format and does not contain malicious data, such as scripts.}
To this end, \textit{blacklisting} or \textit{whitelisting} can additionally be used to restrict or trust sources from which data can be introduced into the system.

\textit{Download verification} of data that is obtained from external sources can assert that---in addition to not containing any malicious scripts or similar---has not been tampered with and contains the expected content.


\textit{Data sanitization}, which involves \textit{input and output sanitization} (e.g., escaping the user-provided inputs before using them in any kind of \add{database query in the EHRS from our example}), and \textit{parameterized prepared statements} (e.g., pre-compiling an SQL statement before patient data is accessed) form another branch of secure data handling.
These features need to be implemented to ensure that no malicious inputs of a potential attacker are evaluated and that no sensitive data such as passwords are leaked, i.e., they can mitigate attacks such as SQL injections\,\citep{Shar13_sql_injection}.

\textit{Retention control} is another feature under secure data handling that deals with the secure management of data that is no longer needed for any operations but is still maintained in the system.
\add{For instance, personal patient data related to billing will be deleted from our exemplary EHRS after a certain timeframe has passed.}
Specifically, the duration or other indicator for when unused data should be deleted or under which circumstances this should not be the case.
The underlying principle is to reduce the attack surface of a system by minimizing the amount of sensitive data that could be accessed by an unauthorized user or system component if compromised.


\textit{Secure storage} describes storing user and other data in a way that keeps it from being accessed by unauthorized users or software components, preserving the confidentiality and integrity of the stored data\,\citep{Loehr10_secure_storage}.
\add{In the EHRS from our example, cryptographic keys for medical records are not stored in the same database as the medical records themselves.}
For instance, \textit{P11} introduces an encrypted storage feature to securely store data.

\textit{Trusted sources} involve the secure generation of timestamps (\textit{time source}) and random numbers (\textit{randomness}).
A trusted source of time ensures, e.g., that logs can be trusted and used for investigations after a security incident, also between distributed systems that share a trusted source of time.
A trusted source of randomness is vital for many cryptographic operations, e.g., as seeds for encryption protocols or for key generation\,\citep{Schindler09_rng_for_crypto}.


\begin{figure*}
    \centering
    \includegraphics[width = 0.7\textwidth]{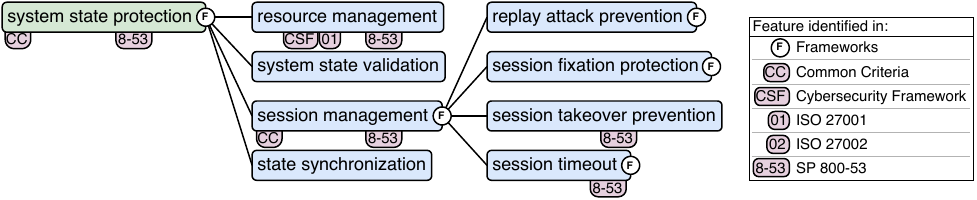}
    \caption{Sub-features of the top-level security feature \textit{system state protection}}
    \label{fig:taxonomy_system_state_protection}
\end{figure*}

\subsubsection{System State Protection.}
The top-level security feature \textit{system state protection} (Figure~\ref{fig:taxonomy_system_state_protection}) describes security features that ensure that the system's operational state is not compromised and that it conforms to defined requirements.
It is mentioned in five of the analyzed ontologies.
\add{In our exemplary EHRS, system state protection is implemented by requiring a doctor to read a patient's data file before they are allowed to prescribe medication via the system to avoid mistreatment.}

\textit{Resource management} refers to implementing control mechanisms for the allocation and access of resources based on a priority level to ensure availability.
As such, resources must be managed to warrant that no attacker can take down a software system by reserving large amounts of resources, e.g., via DDoS attacks\,\citep{Mirkovic04_ddos}\add{, as occurred in the incident at the hospital\,\citep{hospital_example1} described in Section~\ref{subsec:running_eg}}.

\textit{System state validation} can ensure that the system is in a secure state, i.e., that it has not been compromised and that its operational state is correct and secure according to pre-determined rules.
This feature is especially important after events such as the boot-up of the system or recovery after an incident.

\textit{Session management} describes security features mainly to prevent attacks on web applications. 
Four security features to mitigate attacks related to sessions are presented here, \textit{replay attack prevention}, \textit{session fixation protection}\,\citep{Johns11_session_fixation_protection}, \textit{session takeover prevention}\,\citep{Baitha18_session_hijacking}, and \textit{session timeout}.
\add{Session management plays a critical role in our exemplary EHRS by utilizing session timeouts, which closes a session after a period of inactivity, preventing other users in the hospital to gain access to the session.}

\textit{State synchronization} ensures that the states in a system are consistent and synchronized between distributed functions.
This prevents attacks that exploit differences in states between system components.




\subsection{Relation to Security Standards}
\label{sub:comparison_standards}

\begin{figure*}[!ht]
    \centering
    \includegraphics[width = \textwidth]{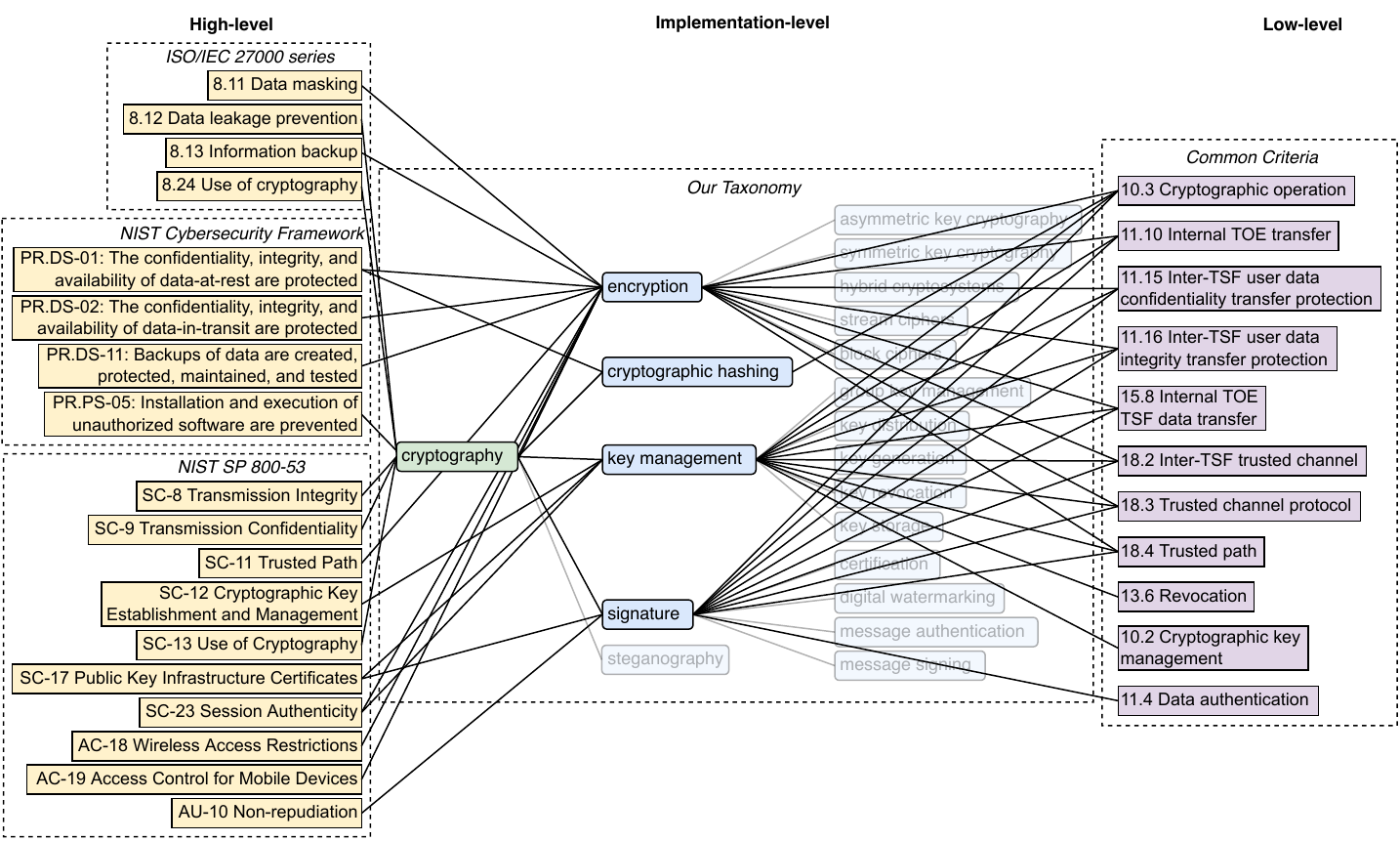}
    \caption{Partial excerpt of the mapping from high-level security standards such as the ISO/IEC 27000 series (left) and from the detailed Common Criteria (CC) (right) to implementation-level security features in our taxonomy (middle). The taxonomy presents security features that can provide the security requirements specified by the high-level standards and generalizes specific low-level details in the CC to actionable advice for developers.}
    \label{fig:mapping_taxonomy_standards}
\end{figure*}

Following the methodology described in Section~\ref{methodology:standard-mapping}, we mapped our taxonomy to security standards, thereby also validating and adapting it.
The mapping and the presented descriptions of the security features can guide developers in adhering to the security requirements of the high-level standards.
In particular, starting from the high-level security standards, developers can use the mappings to select suitable functional security features to realize from our taxonomy.
For the concrete realization, they can then follow the mappings into the relevant detailed aspects of the CC.
In particular, Part 2 of the CC\footnote{\url{https://commoncriteriaportal.org/files/ccfiles/CC2022PART2R1.pdf}} contains detailed descriptions of implementation-level security features that fit the scope of our taxonomy, but the other standards also contain functional security features for which mappings are expected.



\subsubsection{Example Mapping}
As an illustrative example for the mapping, Fig.~\ref{fig:mapping_taxonomy_standards} shows the mapping of the top-level security feature cryptography and its sub-features to the security standards and guidelines used as comparators.
For brevity, we focus on this one feature to illustrate the mapping between our taxonomy and the standards and guidelines.
The complete mapping is made publicly available in our online \cite{Dropbox}.
As shown on the left-hand side in Fig.~\ref{fig:mapping_taxonomy_standards}, multiple parts of the ISO/IEC 27000 series, as well as the NIST Cybersecurity Framework and the NIST SP 800-53, relate to security features in our taxonomy.
On the right-hand side of Figure~\ref{fig:mapping_taxonomy_standards}, the parts of the CC are shown that relate to security features in our taxonomy.
Here, the descriptions in the CC are specific descriptions of certain technologies or techniques to realize the security features they are mapped to.
To this end, the taxonomy provides a more general description of the content of the CC.

\subsubsection{Mapping and Refining of the Taxonomy}

\begin{table}
    \caption{\add{Overlap between security features of standards and taxonomy, showing how many are covered by the taxonomy and how many are missing in the taxonomy and standards}}
    \centering
    \label{tbl:standards_overlap}
    \begin{threeparttable}
    \begin{small}
        \begin{tabular}{lccc}
        \toprule
         & \multicolumn{3}{c}{\textbf{Security Features}} \\
         \cline{2-4}
        \textbf{Standard} & \textbf{\makecell{in \\Standard}} & \textbf{\makecell{not in \\Taxonomy}} & \textbf{\makecell{not in \\Standard}}\\
        \midrule
        CC & 29 (20\tnote{1}\;\,) & 0 (9\tnote{1}\;\,) & 39 \\
        CSF & 18 & 0 & 50 \\
        27001 & 16 & 0 & 52 \\
        27002 & 17 & 0 & 51 \\
        SP800-53 & 27 & 0 & 41\\
        \midrule
        Overall & 37 & 0 & 31\\
        \bottomrule
        \end{tabular}

        \begin{tablenotes}
        \small
            \item[1] Prior to adjusting the taxonomy based on the CC.
        \end{tablenotes}

    \end{small}
    \end{threeparttable}
\end{table}

The following presents the analyzed security standards and describes the process of mapping our taxonomy to the standards and our changes to the taxonomy.
Slight adjustments were made, e.g., to render the taxonomy more generally applicable when we found that the standards described certain security features in a broader scope than their representation in our taxonomy.
Other changes were the addition of further security features that were not contained in the literature and some adjustments to the structure of the taxonomy.
\add{Table~\ref{tbl:standards_overlap} presents the overlap between the taxonomy and the security standards.
Overall, nine security features were identified across the standards that were  not contained in the taxonomy prior to this analysis.
We have adjusted our final taxonomy accordingly.
Out of the 68 security features in the taxonomy, 36 were identified in at least one of the standards, whereas 32 are not mentioned in any of them.
The biggest overlap between an individual standard and the taxonomy was observed for the SP800-53, with 26 security features identified in it that are in the taxonomy.
The ISO27002 showed the smallest overlap with only 12 security features from the taxonomy identified in it.}
The overlaps and performed changes are described in detail in the following.


\paragraph{Common Criteria (CC).}

The CC presents details on low-level security features.
It offers a basis for certifying the security of IT products by listing security properties that need to be fulfilled or for which assurance needs to be provided.
In contrast to the high-level descriptions that are found in many standards, the content of the CC is largely on the implementation level.
Often, specific technologies for achieving a certain security functionality are presented, lacking generality.

The analysis of the CC revealed that 55 of its chapters describe im\-ple\-men\-ta\-tion-level security features.
Of these, 39 (71\%) directly mention security features already contained in the taxonomy.
A mapping between two further chapters and the taxonomy could be achieved by rephrasing the previous feature \textit{escaping user-supplied input} found in the literature to the more general \textit{input sanitization} and \textit{output sanitization}, and rephrasing \textit{log monitoring} to the more general \textit{security monitoring}.
After these changes, we checked the ontologies from the literature that were used to create the taxonomy again and verified that these less specific feature names still fit the descriptions in the literature, which was the case.

The remaining 14 chapters from the CC required the addition of new security features to our taxonomy because they described security features that were too dissimilar to the features already contained in our taxonomy.
Consequently, we added the nine security features \textit{trusted sources} of \textit{time} and \textit{randomness}, \textit{secure storage}, \textit{replay attack prevention}, \textit{resource management}, \textit{retention control}, \textit{system state protection}, \textit{system state validation}, and \textit{state synchronization} to the taxonomy (note that multiple chapters in the CC can relate to the same security feature, therefore the disparity between 14 previously un-mapped chapters and the addition of only nine features).
To validate the updated taxonomy's accordance with the literature, we looked for descriptions of the newly added security features in the ontologies that were the initial sources.
We identified references to all nine of them.
In the ontologies, the descriptions were less concrete than in the CC (for example, paper \textit{P10} presents \textit{system availability} in combination with \textit{DDoS prevention}, which indicates the newly added security feature \textit{resource management}), which is why we did not initially include them in the taxonomy.


The addition of further security features made some adjustments to the structure of the taxonomy necessary to achieve a more coherent grouping.
The new structure better reflects the level of granularity, meaning that now, when comparing two security features in the taxonomy, the lower-level one (i.e., the one more to the right in Fig.~\ref{fig:mapping_taxonomy_standards}) generally addresses a more specific security issue or requirement than the higher-level one (i.e., the one more to the left in Fig.~\ref{fig:mapping_taxonomy_standards}).
One of two changes to the structure concerned the feature \textit{secure data handling}.
Previously, the feature \textit{data validation} had been a top-level feature, which we changed to \textit{secure data handling} being the top-level feature.
%
%
The ontologies in the literature we used as initial sources for the taxonomy have no strict hierarchy, therefore, these changes do not contradict the ontologies.


A second adjustment to the structure of the taxonomy that we performed was prompted by the addition of the feature \textit{system state protection}.
Previously, the feature \textit{session management} had been a top-level feature. 
\textit{System state protection} is an important security feature on a similar level of specificity as the top-level features in our taxonomy.  
Even though it is presented as a security feature in the CC, it is not mentioned in any of the ontologies we examined, which shows a gap in the literature.
We decided that the most coherent structure would be to add it as a top-level feature and organize the features connected to it into the final form shown in Fig.~\ref{fig:taxonomy_system_state_protection}.

\noindent\fbox{%
    \parbox{0.97\linewidth}{%
    In summary, the functional security features described in the CC could be mapped very well to the taxonomy derived from the literature.
    The majority could be mapped directly.
    For the others, we performed slight adjustments to the taxonomy by adding further security features, rephrasing existing ones, and changing the structure.
    The taxonomy was improved by these changes and in its final form (shown in Fig.~\ref{fig:taxonomy_full}) not only represents the literature but also the CC.
    }
}

\paragraph{ISO/IEC 27000 family (27001 and 27002).}
\looseness=-1
This group contains multiple standards that apply to software security in general or to software security of specific domains.
Two standards are specifically related to our work, the \textit{ISO/IEC 27001} and the \textit{ISO/IEC 27002}, which are the two major standards in the family.
ISO/IEC 27001 describes requirements for establishing, maintaining, and improving an Information Security Management System (ISMS).
It presents controls and objectives that organizations might adopt, based on their unique risk landscape.
This standard covers organizational, people, physical, and technological controls.
It primarily offers high-level policies and requirements, making it challenging to identify specific security features that should be implemented.
ISO/IEC 27002 provides guidance and reference for implementing security features to manage information security risks in an ISMS based on ISO/IEC 27001.
It is more explicit than the ISO/IEC 27001 on implementation details of security features. However, most of the standard still consists of high-level descriptions rather than actionable guidance for developers.
In addition, the sheer volume and depth of ISO/IEC 27002 makes identifying implementation-specific features a tedious task, further emphasizing the utility of the taxonomy derived in this~paper.


Analyzing the technological controls presented in the standards revealed that many of their high-level descriptions can be mapped to the implementation-level security features in our taxonomy.
Here, the security features are a way of realizing the technological controls in the standards.
Out of the 34 technological controls that are presented in ISO/IEC 27001 and specified further in ISO/IEC 27002, 14 (41\%) can be mapped in this way to the security features in our taxonomy.
The remaining 20 technological controls can not be realized with implementation-level security features but instead, describe organizational and procedural requirements.
For example, technological controls require that the organization implements hardware redundancy (control 8.14), secure coding principles (control 8.28), or change management processes (control 8.32).
We mapped the 14 technological controls that can be mapped to the taxonomy to the 14 security features \textit{access control}, \textit{cryptography}, \textit{encryption}, \textit{secure storage}, \textit{retention control}, \textit{security monitoring}, \textit{logging}, \textit{trusted source of time}, \textit{data sanitization}, \textit{input validation}, and \textit{output validation} (note that, although 14 controls are mapped to 14 features, the mapping is not injective---some controls are mapped to the same feature and some controls are mapped to more than one feature).
For some mappings, the security features are given as examples of how a technological control can be realized, while others are not explicitly named but the technological controls fit the security features' descriptions (for example, the technological control \textit{Information stored in information systems, devices or in any other storage media shall be deleted when no longer required} is mapped to the security feature \textit{retention control}).

\vspace{1mm}
\noindent\fbox{%
    \parbox{0.97\linewidth}{%
    Overall, no changes to the taxonomy were required to map all implementation-level security features found in the ISO/IEC 27000 family of standards to our taxonomy.
    }
}

\paragraph{NIST SP 800-53 (800-53).}

This standard presents a wide variety of security (and privacy) requirements and related controls.
The descriptions refer to organizational and procedural actions for the most part, only occasionally mentioning implementation-level security features.
In general, the standard calls for an ``organization-wide process to manage risk,'' hence not focusing on technical controls alone.
The description of the faced threats and attacks as ``hostile attacks, human errors, natural disasters, structural failures, foreign intelligence entities, and privacy risks'' further shows the scope of the document and the reason why implementation-level security features are scarce.

Nevertheless, similarly to the ISO/IEC 27000 family, large parts of the NIST SP 800-53 could be mapped to our taxonomy.
Out of the 78 \textit{technical controls} that the standard presents, 40 (51\%) are related to one or more security feature(s) in our taxonomy, meaning that the security features allow the realization of the technical controls.
The remaining 38 technical controls do not require implementation-level security features.
Instead, they describe common software security practices (e.g., \textit{principle of least privilege} (technical control AC-6) or \textit{separation of duties} (technical control AC-5)), user-oriented features (e.g., the display of privacy and security notices (technical control AC-8) or display of information about the last logon (technical control AC-9)), or other information that is non-functional, too specific, or not on the implementation-level.

\vspace{1mm}
\noindent\fbox{%
    \parbox{0.97\linewidth}{%
    All the 40 technical controls presented in the NIST SP 800-53 that describe im\-ple\-men\-ta\-tion-level security features could be mapped to 36 features in our taxonomy without changes.
    }
}




\paragraph{NIST Cybersecurity Framework 2.0 (CSF).}

This resource offers high-level guidelines for organizations to pinpoint risks and threats, along with recommended processes to address them.
It emphasizes the importance of implementing security features as protective measures against potential threats.
However, due to its broad scope (for instance, the detection of and recovery from attacks are also covered), the document generally lacks in-depth details on specific implementation-level security features.

In total, the document is structured into 23 \textit{categories} that describe measures to protect software systems.
Out of these, 14 categories (61\%) could be mapped to security features in our taxonomy.
The other nine categories are non-functional or beyond the scope of our taxonomy for other reasons (e.g., PR.AT-01 and PR.AT-02 are concerned with awareness and training of users, PR.PS-02 and PR.PS-03 ask for the consideration of risks in software and hardware maintenance, replacement, and removal, and PR.IR-02 relates to environmental threats).

\vspace{1mm}
\noindent\fbox{%
    \parbox{0.97\linewidth}{%
    All 14 implementation-level security features identified in the NIST Cybersecurity Framework 2.0 could be mapped to our taxonomy without any changes.
    }
}

\vspace{2mm}


\begin{figure}
    \centering
    \includegraphics[width =0.95\columnwidth]{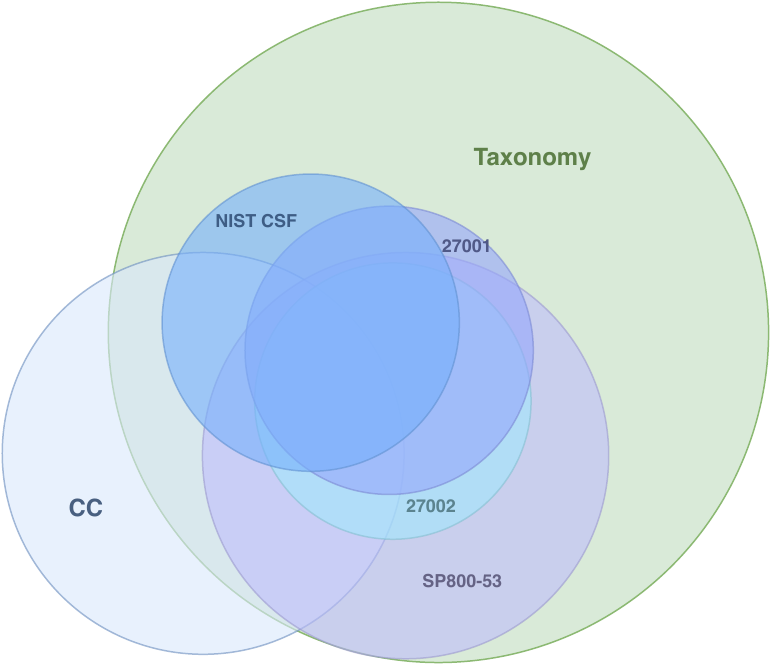}
    \caption{\add{Venn-diagram presenting the overlap between the analyzed standards and the taxonomy concerning security features identified in them.}}
    \label{fig:overlap_venn}
    \vspace{-4mm}
\end{figure}

\add{Figure~\ref{fig:overlap_venn} visualizes the overlap between the standards and our taxonomy in terms of the number of security features (see Table~\ref{tbl:standards_overlap}) before adding the security features that were identified in the Common Criteria that had not been included in the taxonomy at that point.
As shown, the taxonomy covers all other security features mentioned in the analyzed standards and extends this set of functional security features with further features found in the academic literature.
A visualization of the final taxonomy would move the circle representing the Common Criteria also fully inside the circle of the taxonomy.
Based on this depiction and the above descriptions of the mappings between the security standards and guidelines to our taxonomy, we can answer RQ1 as follows.}

\vspace{1mm}
\noindent\fbox{%
    \parbox{0.97\linewidth}{%
    \textbf{RQ1:} We collected a set of 68 implementation-level security features from the literature and security standards. We identified five of them as top-level security features to provide the structure for the taxonomy: \textit{access control}, \textit{cryptography}, \textit{secure data handling}, \textit{security monitoring}, and \textit{system state protection}.
    The taxonomy presented in Figure~\ref{fig:taxonomy_full} provides all security features.
    }
}

\section{Security Features Provided by Security Frameworks (\textbf{\textit{RQ2}})}

\begin{table*}
\centering
\small
\caption{List of selected security frameworks. \# = Number of threads. SO = Stack Overflow. M = Maintained}
\label{tab:frameworks}
\addtolength{\tabcolsep}{-0.3em}

\begin{tabularx}{\textwidth}{lllccc}
    \toprule
    \textbf{ID} & \textbf{Security Framework} & \textbf{URL} & \textbf{\# SO} & \textbf{\# Reddit} & \textbf{M} \\
    \midrule

    01 & Spring Security & \hyperlink{spring.io/projects/spring-security}{spring.io/projects/spring-security} & 91 & 2               & Yes        \\
    02 & Apple Security Framework & \hyperlink{developer.apple.com/documentation/security}{developer.apple.com/documentation/security} & 37 & 0              & Yes        \\
    03 & Apache Shiro & \hyperlink{shiro.apache.org}{shiro.apache.org} & 25 & 0               & Yes        \\
    04 & JAAS & \hyperlink{docs.oracle.com/javase/8/docs/technotes/guides/security/jaas/JAASRefGuide.html}{docs.oracle.com/javase/8/docs} & 12 & 0               & Yes        \\
    05 & Java EE & \hyperlink{oracle.com/java/technologies/java-ee-glance.html}{oracle.com/java/technologies/java-ee-glance.html} & 2      & 0           & Yes        \\
    06 & Java Security Manager & \hyperlink{docs.oracle.com/javase/tutorial/essential/environment/security.html}{docs.oracle.com/javase/tutorial} & 2   & 0              & No         \\
    07 & OpenSSL & \hyperlink{openssl.org}{openssl.org} & 11   & 2             & Yes        \\
    08 & Windows Identity Foundation & \hyperlink{microsoft.com/en-us/download/details.aspx?id=17331}{microsoft.com/en-us/download/details.aspx?id=17331} & 2    & 0             & Yes        \\
    09 & ASP.Net Membership Provider & \hyperlink{learn.microsoft.com/en-us/dotnet/api/system.web.security.membershipprovider}{learn.microsoft.com} & 6   & 0              & Yes        \\
    10 & ASP.Net Role Provider & \hyperlink{learn.microsoft.com/en-us/dotnet/api/system.web.security.roleprovider}{learn.microsoft.com} & 3  & 0               & Yes        \\
    11 & OWASP ESAPI & \hyperlink{owasp.org/www-project-enterprise-security-api}{owasp.org/www-project-enterprise-security-api} & 5        & 0         & Yes        \\
    12 & JBoss Seam Security & \hyperlink{docs.jboss.org/seam/3/security/latest/reference/en-US/html_single}{docs.jboss.org/seam/3/security} & 4    & 0             & No         \\
    13 & Passport & \hyperlink{passportjs.org}{passportjs.org} & 3    & 0             & Yes        \\
    14 & Play Framework Secure Module & \hyperlink{playframework.com/documentation/1.2.5/secure}{playframework.com/documentation/1.2.5/secure} & 2   & 1              & Yes        \\
    15 & OACC & \hyperlink{oaccframework.org}{oaccframework.org} & 2    & 0             & No         \\
    16 & JGuard & \hyperlink{web.archive.org/web/20220619153520/jguard.xwiki.com/xwiki/bin/view/Main/WebHome}{jguard.xwiki.com} & 2        & 0         & No         \\
    17 & Bouncy Castle & \hyperlink{bouncycastle.org}{bouncycastle.org} & 2      & 0           & Yes        \\
    18 & Endpoint Security Framework & \hyperlink{developer.apple.com/documentation/endpointsecurity}{developer.apple.com/documentation/endpointsecurity} & 2    & 0             & Yes        \\
    19 & EveryAuth & \hyperlink{github.com/bnoguchi/everyauth}{github.com/bnoguchi/everyauth} & 2    & 0             & No         \\
    20 & PicketBox & \hyperlink{picketbox.jboss.org}{picketbox.jboss.org} & 2   & 0              & No         \\
    21 & Sureness & \hyperlink{usthe.com/sureness}{usthe.com/sureness} & 2    & 0             & Yes \\
\bottomrule

\end{tabularx}
\end{table*}


We now present our investigation of the security features supported by security frameworks. We discuss the differences between the security features presented in the literature, as documented in our security feature taxonomy (i.e., Section \ref{sec:RQ1}), and those provided to developers by commonly discussed security frameworks. 


\subsection{Identified Security Frameworks}
\noindent We selected 21 security frameworks via our search on Stack Overflow and Reddit.
Table~\ref{tab:frameworks} shows the selected frameworks, sorted by the number of threads on Stack Overflow and Reddit in which each framework was mentioned.
Some frameworks in the table are positioned based on the total number of threads within their parent frameworks, such as ASP.NET, which is a component of the larger .NET framework.
While some framework may be considered outdated when it has not received any updates since 2020, they may still be used by developers, e.g., JGuard was discussed four years after the release of its last stable version, the Java Security Manager is deprecated but still part of maintained JDK versions, and some frameworks still show downloads at the time of writing. 

\begin{table*}
\centering
\setlength{\tabcolsep}{4pt}
\small
\caption{List of security features provided by security frameworks}
\label{tab:secfeatureframeworks}
\begin{tabular}{l|
c>{\columncolor[gray]{0.925}}cc>{\columncolor[gray]{0.925}}cc>{\columncolor[gray]{0.925}}cc>{\columncolor[gray]{0.925}}cc>{\columncolor[gray]{0.925}}cc>{\columncolor[gray]{0.925}}cc>{\columncolor[gray]{0.925}}cc>{\columncolor[gray]{0.925}}cc>{\columncolor[gray]{0.925}}cc>{\columncolor[gray]{0.925}}cc|r}

& \rotatebox{90}{\textbf{01 Spring Security}}
& \rotatebox{90}{\textbf{02 Apple Security Framework}}
& \rotatebox{90}{\textbf{03 Apache Shiro}}
& \rotatebox{90}{\textbf{04 JAAS}}
& \rotatebox{90}{\textbf{05 Java EE / Jakarta EE}}
& \rotatebox{90}{\textbf{06 Java Security Manager}}
& \rotatebox{90}{\textbf{07 OpenSSL}}
& \rotatebox{90}{\parbox{4.25cm}{\textbf{08 Windows .Net Identity Foundation}}}
& \rotatebox{90}{\textbf{09 ASP.Net Membership Provider}}
& \rotatebox{90}{\textbf{10 ASP.Net Role Provider}}
& \rotatebox{90}{\textbf{11 OWASP ESAPI}}
& \rotatebox{90}{\textbf{12 JBoss Seam Security}}
& \rotatebox{90}{\textbf{13 Passport}}
& \rotatebox{90}{\textbf{14 Play Framework Secure Module}}
& \rotatebox{90}{\textbf{15 OACC}}
& \rotatebox{90}{\textbf{16 JGuard}}
& \rotatebox{90}{\textbf{17 Bouncy Castle}}
& \rotatebox{90}{\textbf{18 Endpoint Security Framework}}
& \rotatebox{90}{\textbf{19 EveryAuth}}
& \rotatebox{90}{\textbf{20 PicketBox}}
& \rotatebox{90}{\textbf{21 Sureness}} \\
\hline
\textbf{access control} &  &  &  &  &  &  &  &  &  &  &  &  &  &  &  &  &  &  &  &  & & \\
authentication & \checkmark & \checkmark & \checkmark & \checkmark & \checkmark &  &  &  & \checkmark &  & \checkmark & \checkmark & \checkmark & \checkmark & \checkmark & \checkmark & \checkmark &  & \checkmark & \checkmark & \checkmark & 76.2\%\\
authorization & \checkmark & \checkmark & \checkmark & \checkmark & \checkmark & \checkmark &  & \checkmark &  & \checkmark & \checkmark & \checkmark & \checkmark & \checkmark & \checkmark & \checkmark &  & \checkmark & \checkmark & \checkmark & \checkmark & 85.7\%\\
\hline
\textbf{cryptography} &  &  &  &  &  &  &  &  &  &  &  &  &  &  &  &  &  &  &  &  &  & \\
cryptographic hashing & \checkmark & \checkmark & \checkmark &  & \checkmark &  & \checkmark &  &  &  & \checkmark & \checkmark &  &  &  & \checkmark & \checkmark &  &  &  &  & 47.6\%\\
encryption & \checkmark & \checkmark & \checkmark &  &  &  & \checkmark &  & \checkmark &  & \checkmark &  &  &  & \checkmark &  & \checkmark &  &  &  &  & 38.1\%\\
key management & \checkmark & \checkmark &  &  &  &  & \checkmark &  &  &  &  &  &  &  &  &  & \checkmark &  &  &  &  & 19.0\%\\
signature & \checkmark & \checkmark &  &  &  &  & \checkmark &  &  &  &  &  &  &  &  &  & \checkmark &  &  &  &  & 19.0\%\\
steganography &  &  &  &  &  &  &  &  &  &  &  &  &  &  &  &  &  &  &  &  &  & 0.0\%\\
\hline
\textbf{security monitoring} &  &  &  &  &  &  &  &  &  &  &  &  &  &  &  &  &  &  &  &  &  & \\
automated response &  &  &  &  &  & \checkmark &  &  &  &  &  &  &  &  &  &  &  &  &  &  &  & 4.8\%\\
history maintenance &  &  &  &  &  &  &  &  &  &  &  &  &  &  &  &  &  &  &  &  &  & 0.0\%\\
logging & \checkmark &  &  &  &  &  & \checkmark &  &  &  & \checkmark & \checkmark &  &  &  &  &  & \checkmark & \checkmark & \checkmark &  & 28.6\%\\
\hline
\textbf{system state protection} &  &  &  &  &  &  &  &  &  &  &  &  &  &  &  &  &  &  &  &  &  & \\
resource management &  &  &  &  &  &  &  &  &  &  &  &  &  &  &  &  &  &  &  &  &  & 0.0\%\\
system state validation &  &  &  &  &  &  &  &  &  &  &  &  &  &  &  &  &  &  &  &  &  & 0.0\%\\
session management & \checkmark & \checkmark & \checkmark &  & \checkmark &  &  &  &  &  & \checkmark & \checkmark & \checkmark &  & \checkmark &  &  &  & \checkmark &  &  & 38.1\%\\
state synchronization &  &  &  &  &  &  &  &  &  &  &  &  &  &  &  &  &  &  &  &  &  & 0.0\%\\
\hline
\textbf{secure data handling} &  &  &  &  &  &  &  &  &  &  &  &  &  &  &  &  &  &  &  &  &  & \\
data validation &  &  &  &  &  &  &  &  &  &  & \checkmark &  &  &  &  &  &  &  &  &  &  & 4.8\%\\
data sanitization &  &  &  &  &  &  &  &  &  &  & \checkmark &  &  &  &  &  &  &  &  &  &  & 4.8\%\\
retention control &  &  &  &  &  &  &  &  &  &  &  &  &  &  &  &  &  &  &  &  &  & 0.0\%\\
secure storage &  & \checkmark &  &  &  &  &  &  &  &  &  &  &  &  & \checkmark &  &  &  &  &  &  & 9.5\%\\
trusted sources &  & \checkmark & \checkmark &  &  &  & \checkmark &  &  &  & \checkmark &  &  &  &  &  & \checkmark &  &  &  &  & 19.0\%\\
\hline
\end{tabular}
\end{table*}

In total, we identified 44 security features offered to developers by the selected security frameworks. Terms found in the respective reference guides and documentation were grouped according to our methodology in Section~\ref{subsubsec:methodology_extraction_frameworks}.
The grouping in Table \ref{tab:secfeatureframeworks} is based on the structure of our taxonomy which denotes security features present in each framework.
\add{It should be noted that many security frameworks are tailored towards specific needs or offered for different programming languages, which should not be compared to each other.
Still, some offer similar functionalities for similar use cases, which may be a basis for comparison.
Additionally, some of the frameworks can also be used for different purposes than utilizing them within a software system, such as encrypting messages using a provided command line interface.
However, we only consider these security frameworks at the application-level, i.e., when developers implement software systems.}
In what follows, we summarize the investigated security frameworks, highlighting their coverage in the following areas of security.
We found that the security frameworks can be grouped into two categories that target access control and cryptography.
Besides, some security frameworks offer a wide range of different security features, and can, therefore, be classified as multi-purpose frameworks.

\paragraph{Access Control Frameworks.}
Based on their offered security features, 15 of the selected security frameworks can be mainly used to implement and manage authentication and authorization. Among them, we identified two authentication and authorization middlewares (Passport and EveryAuth) for node.js.
Three modules of the Java standard library implement authentication and authorization (JAAS), enterprise software utilities (Java EE), and security policy enforcement (Java Security Manager).
The Windows .Net Identity Foundation is a security framework for facilitating user authentication in software systems.
JGuard is a security framework based on JAAS for solving access control problems for web applications.
Similarly, JBoss Seam Security and OACC are access control frameworks aiming to provide general functionalities to manage and enforce access control policies.
While the Play Framework Secure Module handles authentication and authorization, in ASP.NET, these features are split among Membership Provider and Role Provider.
The Endpoint Security Framework offered by Apple can be used to monitor and authorize system events.
Finally, the framework Sureness focuses on securing REST APIs.



\paragraph{Cryptography Frameworks.}
OpenSSL and Bouncy Castle offer a large range of \textit{encryption}, \textit{key management}, \textit{hashing}, and \textit{signature} features. While Bouncy Castle is purely a cryptographic library, OpenSSL uses the cryptographic library \textit{libcrypto} for implementing cryptographic features. Cryptographic features are offered by 9 other frameworks as well (Spring Security, Security Framework, Apache Shiro, Java EE, ASP.NET, OWASP ESAPI, JBoss Seam Security, OACC, JGuard).


\paragraph{Multi-Purpose Frameworks.}
\add{Spring Security, Apple Security Framework, Apache Shiro, and OWASP ESAPI offer a large range of security features from our taxonomy, covering both, access control and cryptography features, as well as additional ones such as session management.}
Notably, OWASP ESAPI additionally provides most features for data handling, such as \textit{data validation}, \textit{data sanitization}, and \textit{trusted sources}. 


\begin{table*}
\centering\small
\setlength{\tabcolsep}{3pt}
\caption{\label{tab:features_frameworks}Security features provided by security frameworks via APIs (\texttt{a}), configuration files (\texttt{c}), and annotations (\texttt{n})}
\begin{tabular}{l|
c>{\columncolor[gray]{0.925}}cc>{\columncolor[gray]{0.925}}cc>{\columncolor[gray]{0.925}}cc>{\columncolor[gray]{0.925}}cc>{\columncolor[gray]{0.925}}cc>{\columncolor[gray]{0.925}}cc>{\columncolor[gray]{0.925}}cc>{\columncolor[gray]{0.925}}cc>{\columncolor[gray]{0.925}}cc>{\columncolor[gray]{0.925}}cc}
& \rotatebox{90}{\textbf{01 Spring Security}}
& \rotatebox{90}{\textbf{02 Apple Security Framework}}
& \rotatebox{90}{\textbf{03 Apache Shiro}}
& \rotatebox{90}{\textbf{04 JAAS}}
& \rotatebox{90}{\textbf{05 Java EE  / Jakarta EE}}
& \rotatebox{90}{\textbf{06 Java Security Manager}}
& \rotatebox{90}{\textbf{07 OpenSSL}}
& \rotatebox{90}{\textbf{08 Windows .NET Identity Foundation}}
& \rotatebox{90}{\textbf{09 ASP.NET Membership Provider}}
& \rotatebox{90}{\textbf{10 ASP.NET Role Provider}}
& \rotatebox{90}{\textbf{11 OWASP ESAPI}}
& \rotatebox{90}{\textbf{12 JBoss Seam Security}}
& \rotatebox{90}{\textbf{13 Passport}}
& \rotatebox{90}{\textbf{14 Play Framework Secure Module}}
& \rotatebox{90}{\textbf{15 OACC}}
& \rotatebox{90}{\textbf{16 JGuard}}
& \rotatebox{90}{\textbf{17 Bouncy Castle}}
& \rotatebox{90}{\textbf{18 Endpoint Security Framework}}
& \rotatebox{90}{\textbf{19 EveryAuth}}
& \rotatebox{90}{\textbf{20 PicketBox}}
& \rotatebox{90}{\textbf{21 Sureness}} \\
\hline
\textbf{access control} &  &  &  &  &  &  &  &  &  &  &  &  &  &  &  &  &  &  &  &  &  \\
authentication & a/c & a & a/c & a/c & a/n &  &  &  & a &  & a/c & a & a & a & a & a/c &  & a & a & a/c/n & a \\
authorization & a/c/n & a & a/c/n & a/c & a & a &  & a/c &  & a & a & a/n & a & a/n & a & a/c & a &  & a & a/c/n & c/n \\
\hline
\textbf{cryptography} &  &  &  &  &  &  &  &  &  &  &  &  &  &  &  &  &  &  &  &  & \\
cryptographic hashing & a & a & a &  & a &  & a &  &  &  & a/c & a &  &  &  & a/c & a &  &  &  & \\
encryption & a & a & a &  &  &  & a &  & a &  & a/c &  &  &  & a &  & a &  &  &  & \\
key management & a & a &  &  &  &  & a &  &  &  &  &  &  &  &  &  & a &  &  &  & \\
signature & a & a &  &  &  &  & a &  &  &  &  &  &  &  &  &  & a &  &  &  & \\
steganography &  &  &  &  &  &  &  &  &  &  &  &  &  &  &  &  &  &  &  &  & \\
\hline
\textbf{security monitoring} &  &  &  &  &  &  &  &  &  &  &  &  &  &  &  &  &  &  &  &  & \\
automated response &  &  &  &  &  & a &  &  &  &  &  &  &  &  &  &  &  &  &  &  & \\
history maintenance &  &  &  &  &  &  &  &  &  &  &  &  &  &  &  &  &  &  &  &  & \\
logging & a/c/n &  &  &  &  &  & a &  &  &  & a/c & a/n &  &  &  &  &  & a & a & a & \\
\hline
\textbf{system state protection} &  &  &  &  &  &  &  &  &  &  &  &  &  &  &  &  &  &  &  &  & \\
resource management &  &  &  &  &  &  &  &  &  &  &  &  &  &  &  &  &  &  &  &  & \\
system state validation &  &  &  &  &  &  &  &  &  &  &  &  &  &  &  &  &  &  &  &  & \\
session management & a/c & a & a/c &  & a/c/n &  &  &  &  &  & a/c & a & a & a & a &  &  &  & a &  & \\
state synchronization &  &  &  &  &  &  &  &  &  &  &  &  &  &  &  &  &  &  &  &  & \\
\hline
\textbf{secure data handling} &  &  &  &  &  &  &  &  &  &  &  &  &  &  &  &  &  &  &  &  & \\
data validation &  &  &  &  &  &  &  &  &  &  & a/c &  &  &  &  &  &  &  &  &  & \\
data sanitization &  &  &  &  &  &  &  &  &  &  & a &  &  &  &  &  &  &  &  &  & \\
retention control &  &  &  &  &  &  &  &  &  &  &  &  &  &  &  &  &  &  &  &  & \\
secure storage &  & a &  &  &  &  &  &  &  &  &  &  &  &  & a &  &  &  &  &  & \\
trusted sources &  & a & a &  &  &  & a &  &  &  & a &  &  &  &  &  & a &  &  &  & \\
\hline
\end{tabular}
\end{table*}

\subsection{Provided Security Features and Relation to Security Taxonomy}
To investigate the relationship between the functional security features captured in the taxonomy and those provided to developers, we mapped the frameworks' features to the taxonomy (recall, that we marked them with ``F'' in Fig. ~\ref{fig:taxonomy_access_control}, \ref{fig:taxonomy_cryptography}, \ref{fig:taxonomy_security_monitoring}, \ref{fig:taxonomy_secure_data_handling}, and \ref{fig:taxonomy_system_state_protection}).


\subsubsection{Provided Security Features}
In the following, we present the identified security features in the order of the taxonomy's top-level security features, as shown in Fig.~\ref{fig:taxonomy_full}.

\textit{Access Control.}
As shown in Table~\ref{tab:secfeatureframeworks}, all frameworks except OpenSSL provide features to realize some kind of access control.
Sixteen of the 21 frameworks offer features to realize authentication.
However, from the security features in Fig.~\ref{fig:taxonomy_access_control}, \textit{multifactor authentication} is not provided by any framework.
All frameworks use credentials for authentication but also offer authentication via certificates (Spring Security).
Additionally, Spring Security and Passport provide authentication features via single sign-on. The feature one-time-password is only provided by PicketBox.

\begin{figure*}
\centering
\begin{minipage}{0.95\textwidth}
\begin{lstlisting}[language=Java,label={lst:JAAS-auth},caption={\add{Servlet of our exemplary EHRS login page that implements authentication of a user using JAAS}}]
public class LoginServlet extends HttpServlet {
  // The GET method is called when a user clicks submit on the login page
  protected void doGet(HttpServletRequest request, HttpServletResponse response) throws ServletException, IOException {
    // Get user and password from HTTP response
    final var user = new SimplePrincipal(request.getParameter("user"));
    final var password = request.getParameter("password").toCharArray();

    /* Creating a LoginContext automatically loads the authentication mechanism registered
    with JAAS. User and password are provided via a CallbackHandler. */
    var context = new LoginContext("ehrs", new CallbackHandler(user, password){
      public void handle(Callback[] callbacks){
        // Callbacks used for authentication are provided with user and password
        for(var callback : callbacks) {
          if(callback instanceof NameCallback nameCallback) {
            nameCallback.setName(user);
          } else if(callback instanceof PasswordCallback paswordCallback) {
            passwordCallback.setPassword(password);
          }
        }
      }
    });
    try {
      context.login(); // Authenticating the user throws a LoginException if it fails
    } catch (LoginException e) {
      // In case of a failed login, a simple failure page is returned
      response.getWriter().println("<b>authenticatation failed!</b>");
    }
  }
}
\end{lstlisting}
\end{minipage}
\end{figure*}

\add{
	In our running example, the authentication of users may be realized using JAAS. Listing~\ref{lst:JAAS-auth} shows a possible implementation of this security feature.
	Our exemplary EHRS would be implemented using Java Server Pages (JSP). Whenever a user clicks the login button on the login page, the Servlet shown in Listing~\ref{lst:JAAS-auth} is called.
	In this case, the \texttt{doGet} method is called and handed over an HTTP request containing the data entered into a \emph{user name} and \emph{password} field on the login page.
	This information is retrieved from the request (lines 5 and 6), and then, the JAAS login is instantiated in lines 9--20.
	JAAS automatically loads an authentication mechanism registered with JAAS when instantiating \texttt{LoginContext}.
	The user name and password are handed over via callbacks, which in this case are provided with the values retrieved from the HTTP request in lines 13--18.
	Besides setting login information in code, JAAS supports various callbacks to provide data, e.g., callbacks that show pop-ups to users in a desktop application.
	After instantiating the login context, the provided login information is validated in line 22 by calling \texttt{login()}.
    All of the Java security frameworks examined are compliant with JAAS and extend it by, for example, registering authentication mechanisms or providing easier-to-use wrappers for specific usage scenarios such as Web applications.
}

\begin{figure}
\centering
\begin{minipage}{\linewidth}
\begin{lstlisting}[label = lst:rbac_example, language=Java,
    tabsize=1,caption=\add{An example of implementing role- and attribute-based access control with the Spring Security framework.}]
@RestController
@RequestMapping("/patient-info")
public class PatientInfoController {

    private String patientInfo = "Sensitive Patient Information";

    @PreAuthorize("hasRole('ROLE_DOCTOR') && hasPermission(#patient, 'designated')")
    @GetMapping
    public String getPatientInfo() {
        return patientInfo;
    }

    @PreAuthorize("hasRole('ROLE_DOCTOR')")
    @PostMapping
    public String updatePatientInfo(@RequestBody String newInfo) {
        patientInfo += newInfo;
        return "Patient information updated successfully.";
    }
}
\end{lstlisting}
\end{minipage}
\end{figure}

After a user has been authenticated, it is usually decided whether the user is allowed to perform specific activities.
Even though authorization is one of the most prevalent security features realized in 18 of the 21 frameworks, only \textit{attribute-based} and \textit{role-based access control} are offered.
Fig.~\ref{fig:taxonomy_access_control} emphasizes that the top-level security feature authorization is missing most sub-level features.
\add{Listing~\ref{lst:rbac_example} shows an example of enforcing access control in our exemplary EHRS using Spring Security.
To implement the access control scheme of the EHRS, a combination of role-based and attribute-based access control would be implemented.
First, patient information from the endpoint \texttt{/patient-info} can only be accessed by users with the role \texttt{ROLE\_DOCTOR}. (see lines 7 and 13).
However, to be able to read the patient information (\texttt{getPatientInfo()}), this doctor must also be a designated doctor for the patient, which is expressed as an attribute that is implemented using the 'designated' permission for the concrete patient (see line 7).}

\begin{figure}[tb]
  \setcaptiontype{lstlisting}
        \begin{minipage}[b]{\columnwidth}
            \begin{lstlisting}[language=Java]
public byte[] encryptPatientData(String data) {
  // Pasword and salt for key derivation
String password = "...";  // some random password
String salt = KeyGenerators.string().generateKey();

 /* Create a password-based encryptor
 using 256 bit AES encryption */
 var aes = Encryptors.stronger(password, salt);

 // Encrypt the patient data
 return aes.encrypt(data.getBytes());
}
\end{lstlisting}
            \subcaption{Encryption using Spring Security}
            \label{lst:enc_example_spring}
        \end{minipage}%

        \begin{minipage}[b]{\columnwidth}
            \begin{lstlisting}[language=Java]
Security.addProvider(new BouncyCastleProvider());

public byte[] encryptPatientData(String data) {
  // Pasword and salt for key derivation
  char[] password = "...";  // some random password
  byte[] salt = ...;

  // Generate AES key
  var keyGen = SecretKeyFactory.getInstance("PBKDF2WithHmacSHA1");
  var spec = new javax.crypto.spec.PBEKeySpec(password, salt, 1024, 256);
  var secretKey = keyGen.generateSecret(spec);

  // Create and initialize cipher
  Cipher cipher = Cipher.getInstance("AES/GCM/PKCS5Padding", "BC");
  cipher.init(Cipher.ENCRYPT_MODE, secretKey);

  // Encrypt the patient data
  return cipher.doFinal(data.getBytes());
}
\end{lstlisting}
            \subcaption{Encryption using Bouncy Castle}
            \label{lst:enc_example_bc}
        \end{minipage}
    \caption{Credential-based Encryption using an AES Block Cipher in Spring Security and Bouncy Castle, using the predefined parameters of Spring Security.}
    \label{lst:encryption_examples}
\end{figure}

\textit{Cryptography.}
In addition to the cryptography libraries OpenSSL and Bouncy Castle, the framework Spring Security and the Apple Security Framework offer the most cryptography features.
In total, eleven frameworks offer security features for cryptography.
Excluding Bouncy Castle and OpenSSL, half of the frameworks offer \textit{cryptographic hashing} which is designed explicitly for access control or password encryption (Spring Security, Security Framework, Apache Shiro, OWASP ESAPI, JBoss Seam Security, OACC, JGuard).
Eight frameworks provide support for \textit{encryption}, offering a diverse selection of cryptographic algorithms. These algorithms span across \textit{symmetric cryptography}, including \textit{block and stream ciphers}, and \textit{asymmetric cryptography}. In addition to enabling the generation and verification of \textit{signatures}, the frameworks Spring Security, Apple's Security Framework, OpenSSL, and Bouncy Castle (19.0\%), offer capabilities for \textit{key management}.

\add{An example of how the encryption in the EHRS example  can be implemented using Spring Security is shown in Listing~\ref{lst:enc_example_spring}, while the same encryption using Bouncy Castle is shown in Listing~\ref{lst:enc_example_bc} for comparison.
While Spring Security abstracts most of the configuration details from the user, but therefore allows limited configuration, Bouncy Castle allows for detailed configuration but is more complicated to use.
Both require credentials and a seed for encryption.
While Spring Security comes with a utility function to generate the seed (line 4 in Listing~\ref{lst:enc_example_spring}), the seed for Bouncy Castle must be generated in handwritten code.
In both frameworks, a secret key is generated from a password and salt, but in Spring Security this is hidden from the user.
Spring Security provides some standard configurations via the class \texttt{Encryptors}, of which we initialize the \texttt{stronger} variant in line 8 of Listing~\ref{lst:enc_example_spring}.
In lines 9 to 15 of Listing~\ref{lst:enc_example_bc}, we configure Bouncy Castle in the same way as the selected configuration.
First, we generate a secret key in lines 9 to 11, and then we initialize the cipher used for encryption in lines 14 and 15, in both cases using the predefined configuration parameters of Spring Security.
Finally, in both cases, Spring Security and Bouncy Castle, the data is encrypted (line 10 in Listing~\ref{lst:enc_example_spring} and line 18 in Listing~\ref{lst:enc_example_bc}).
}

The features \textit{steganography}, \textit{group key management}, and \textit{digital watermarking} are not provided by any of the security frameworks considered (see Fig.~\ref{fig:taxonomy_cryptography}).

\textit{Security Monitoring.}
Seven frameworks offer logging features (Table~\ref{tab:secfeatureframeworks}).
Although \textit{logging} is not inherently designed as a proactive defense against attacks, it plays a crucial role in identifying anomalies within a system and retroactively tracing back problems of a system. 
Most frameworks offer some support for integrating external logging frameworks. Finally, Java Security Manager is the only security framework that offers the configuration of \textit{automated responses} to security incidents.

\begin{figure}[tb]
  \setcaptiontype{lstlisting}
        \begin{minipage}[b]{\columnwidth}
\begin{lstlisting}[language=Java]
public class SessionListener implements HttpSessionListener {
   @Override
   public void sessionCreated(HttpSessionEvent e) {
      // Set session timeout in seconds
      e.getSession().setMaxInactiveInterval(300);
   }
}
\end{lstlisting}
            \subcaption{\add{Session timeouts in Spring Security using the SessionListener}}
            \label{lst:timeout_example_spring}
        \end{minipage}%

        \begin{minipage}[b]{\columnwidth}
\begin{lstlisting}[language=Config]
[main]
...
# 300.000 milliseconds = 5 minutes
securityManager.sessionManager.globalSessionTimeout = 300000
\end{lstlisting}
            \subcaption{\add{Session timeouts in Apache Shiro configuration file shiro.ini}}
            \label{lst:timeout_example_shiro}
        \end{minipage}
    \caption{Setting session timeouts in Spring Security and Apache Shiro}
    \label{lst:session_management_examples}
\end{figure}

\textit{System State Protection.}
\textit{Session management} and three of its sub-features are the only \textit{system state protection} features offered by eight of the security frameworks.
As such, many of the \textit{system state protection} features rely on the manual implementation of developers rather than the usage of security frameworks.
\add{Listing~\ref{lst:timeout_example_spring} shows an example of how session timeouts can be implemented in the EHRS using Spring Security, which relies on the implementation of a \texttt{SessionListener} provided by the Java Standard Library.
In contrast, implementing session timeouts in Apache Shiro relies on configuring a session manager provided by the security framework itself as Listing~\ref{lst:timeout_example_shiro} shows.}

\begin{figure}
\centering
\begin{minipage}{0.49\textwidth}
\begin{lstlisting}[label = lst:secure_storage_example, language=Swift, caption=\add{Adding data to a keychain in the Apple Security Framework}]
let stat = SecItemAdd(addquery as CFDictionary, nil)
guard stat == errSecSuccess else {throw <# error #>}
\end{lstlisting}
\end{minipage}
\end{figure}

\looseness=-1
\textit{Secure Data Handling.}
Only OWASP ESAPI offers security features for \textit{data validation} and \textit{data sanitization}, including some of its low-level security features (see Fig.~\ref{fig:taxonomy_secure_data_handling}).
Although OWASP ESAPI provides a set of methods for data validation, the capabilities of the offered validations are limited to basic validations and, therefore, require developers to extend these with validation rules tailored to specific security requirements of their applications.
Only the Apple Security Framework and OACC offer means to realize secure storage.
The former offers a secure storage solution named keychain which assists developers in implementing secure storage, \add{with an example of adding data to it shown in Listing~\ref{lst:secure_storage_example}}.
The latter, OACC offers a comparable solution by providing a fully implemented database specifically designed to manage security-related information. This is achieved through the execution of setup scripts tailored to different databases.
Four frameworks provide features to create trusted sources, such as random number generators and timestamps.
The latter is a foundational security feature that can be used with other features, such as \textit{retention control} or \textit{session takeover prevention}.
Note that some security features (e.g., \textit{parameterized prepared statement}) are provided by standard libraries of programming languages, such as Java.

\subsubsection{Relation to the Taxonomy}
\noindent While the security frameworks support all top-level features from the taxonomy, we observed some noticeable differences to the literature.
The selected security frameworks offer only 64\% of the security features from our taxonomy. 
While nearly all features of \textit{cryptography} are provided.
The frameworks mainly lack features concerning the three of the five top-level features \textit{authorization}, \textit{secure data handling} and \textit{system state protection} as visible in Fig.~\ref{fig:taxonomy_access_control}, \ref{fig:taxonomy_secure_data_handling},~and~\ref{fig:taxonomy_system_state_protection}.

It seems that many access control features might not be used in practice or did not reach practice, yet.
The literature considers 11 access control features for \textit{authorization} (see Fig.~\ref{fig:taxonomy_access_control}), but the security frameworks only offer 2 of these.
In some cases, the selected frameworks might be able to realize security features such as \textit{discretionary access control} or \textit{rule-based access control} by utilizing other offered features.
However, this was not mentioned in any of the documentation. Consequently, we did not label these security features in the taxonomy in Fig.~\ref{fig:taxonomy_access_control} as being provided by the frameworks.
In the case of \textit{secure data handling}, 10 out of the 15 security features are provided by the frameworks, as shown in Fig.~\ref{fig:taxonomy_secure_data_handling}. 
For \textit{system state protection}, the security frameworks offer 4 out of 8 security features we collected in the literature, as depicted in Fig.~\ref{fig:taxonomy_system_state_protection}. 

Additionally, our research suggests that security frameworks sometimes promote the application of various features as novel features, which may not align with our definition of security features according to literature.
For example, Apache Shiro, Java EE, JBoss, and Seam Security offer a security feature called \textit{remember-me} that they advertise as an \textit{authentication} feature signifying an entity as "remembered from a successful authentication during a previous session"~\citep{ApacheShiro}.
Concerning our taxonomy in Fig.~\ref{fig:taxonomy_cryptography}, this is not considered a security feature but implies a specific usage of \textit{session management} features to keep a session open when an application is reopened.
This suggests, that security frameworks have different views on what level of security features should be considered. Note that while some frameworks are not considered security frameworks, they may offer features that support the implementation of security features, or may even directly provide security features, even though these frameworks are not included in our list.

Finally, we found that many security features rely on manual implementation and are not offered by security frameworks.
Most features for access control, some of secure data handling, and system state protection from our taxonomy have to be manually implemented without the use of a framework.
Security monitoring features are offered by a few frameworks and might not be tailored to the needs of every project.



\vspace{1mm}
\noindent\fbox{%
    \parbox{0.97\linewidth}{%
    \textbf{RQ2}: We collected 44 security features from a set of 21 security frameworks identified in discussions on \textit{Stack Overflow and Reddit}. 
    The features overlap with the taxonomy obtained via the literature review. The most significant overlap occurs within the domains of \textit{access control} and \textit{cryptography}. Conversely, the least overlap is found in the domain of \textit{security monitoring}. 
    The relation of frameworks to the taxonomy is indicated in Figure~\ref{fig:taxonomy_access_control}, \ref{fig:taxonomy_cryptography}, \ref{fig:taxonomy_security_monitoring}, \ref{fig:taxonomy_secure_data_handling}, and \ref{fig:taxonomy_system_state_protection} (features marked with \textbf{F}).


    }
}

\section{Manifestation of Functional Security Features in Source Code (\textbf{\textit{RQ3}})}
\label{sec:traceability}

Since the goal of our work is to provide guidance in locating security features in source code, we need to understand how they manifest in codebases as this information can be utilized for creating traceability.
To this end, we captured how each security feature in the analyzed security framework is provided to developers.

\subsection{Security Feature Manifestation}
As indicated in Table~\ref{tab:features_frameworks}, we found that the security frameworks provide security features based on three mechanisms:

\begin{itemize}
    \looseness=-1
    \item \textit{APIs} provide security features directly.
    A framework can define API classes, methods, or fields to invoke or configure security features.
    \add{Listing~\ref{lst:encryption_examples} contains examples of the usage of APIs for encryption}.
    As such, their usage is clearly visible within the codebase, which makes them easy to locate.
    This property can be leveraged to establish traceability, since APIs can serve as an entry point for the location of security features.
    \item \textit{Configuration files} can be used to enable security features, potentially in addition to APIs to provide configurations of used security features.
    \add{Examples of such are given in Listings~\ref{lst:timeout_example_shiro} and \ref{lst:esapi_config}.}
    Developers can change values within a configuration file to modify specific values used by a framework.
    While they are clearly separated from the source code, they are still interacted with by the code base to fetch data that is relevant for security measures.
    Therefore, there is a need to connect the configuration file to the corresponding security features along with the API.
    \item \textit{Annotations} are used by many programming languages, such as Java to extend functionalities of methods or classes within the implementation.
    A developer simply prepends a keyword with a corresponding marker, such as @, to the program element.
    \add{Listing~\ref{lst:rbac_example} shows examples of such annotations in lines 7 and 13.}
    Security annotations can be used to either clearly define a context in which a method or class should be used, e.g., a security level\,\citep{Peldszus2023}, or to provide additional functionalities to methods or classes.
    Like APIs, they are clearly visible within the source code and can be used for tracing security features to locate their implementation.
\end{itemize}

\begin{figure}
    \centering
    \includegraphics[width =\columnwidth]{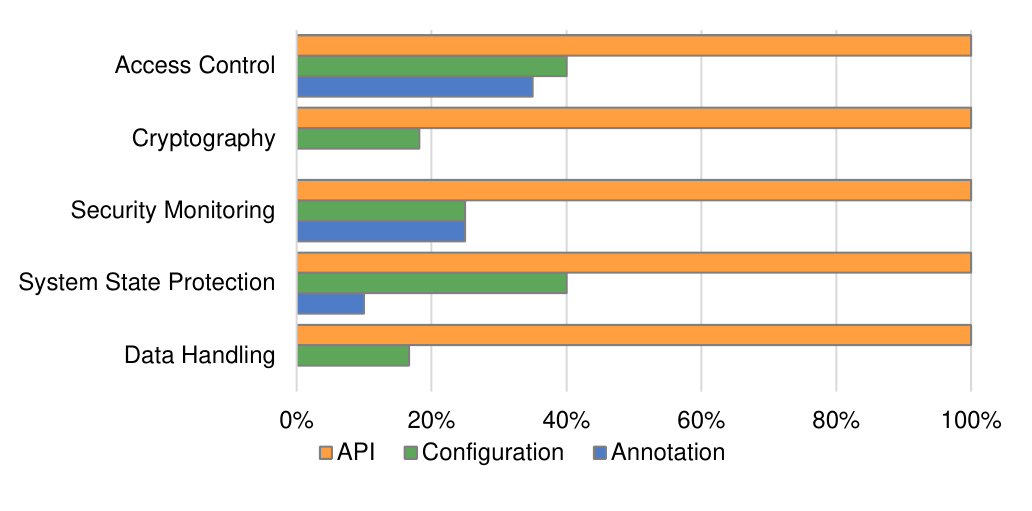}
    \vspace{-8mm}
    \caption{Mechanisms used by security frameworks for providing security features for use in software systems}
    \label{fig:mechanisms}
    \vspace{-4mm}
\end{figure}

For each top-level security feature, Fig.~\ref{fig:mechanisms} shows how often each mechanism is utilized by the security frameworks to provide it.
A security framework can provide the same security feature using multiple mechanisms.

All frameworks provide APIs for each security feature they offer.
While many core functionalities of security features are implemented by using their APIs, configuration files are additionally used by less than half of the frameworks for each feature.
Configurable values within these files are used by the security frameworks to allow simple modification of general parameters such as \textit{timeouts} or used \textit{encryption} algorithms.
Finally, annotations are used for \textit{access control}, \textit{security monitoring} and \textit{system state protection} features.
Typically, classes or methods need to be annotated to extend the functionality of implemented \textit{authentication} mechanisms or to enforce \textit{authorization} rules.


\subsubsection{Access Control}
\label{subsub:ac}
All frameworks offer APIs for access control.
As an example, user \textit{credentials} can be generated in Apache Shiro by instantiating the class \texttt{UsernamePasswordToken} with a username and password.
Then, the method \texttt{hasRole()} can be used on the user accessing a resource to perform a role check.

40\% of access control features include configuration files, which are often used to specify properties of authentication and authorization mechanisms.
Picketbox uses configuration files to select login modules provided by the framework or to define roles that are then specified using Picketbox's API.
In configuration files provided by JGuard, the developer can select authentication schemes and define scopes.
\add{Spring Security also provides several configuration options through configuration files, such as the definition of a role hierarchy, which allows the inheritance of permissions between roles as illustrated in Listing~\ref{lst:spring_xml}.}

\begin{lstlisting}[label = lst:spring_xml, language=xml, caption=\add{XML-configuration file showing the role hierarchy of our exemplary EHRS},float]
<bean id="roleHierarchy" class="org.springframework.security.access.hierarchicalroles.RoleHierarchyImpl">
 <property name="hierarchy">
  <value>
   ROLE_DOCTOR        > ROLE_MEDICSTAFF
   ROLE_NURSE         > ROLE_MEDICSTAFF
   ROLE_MEDICSTAFF    > ROLE_STAFF
   ROLE_HOSPITALSTAFF > ROLE_STAFF
   ROLE_STAFF         > ROLE_UNAUTHENTICATED
  </value>
 </property>
</bean>
\end{lstlisting}

In addition, annotations are typically used to restrict access to methods to a specific group of users.
7 of the selected security frameworks (Spring Security, Apache Shiro, Java EE, Secure Module, JBoss Seam Security, PicketBox, Sureness) make use of Java annotations to handle access control on the method level.
For example, Spring Security allows developers to annotate methods with the annotation \texttt{@PreAuthorize} to restrict method invocations to a specified role given as a parameter\add{, as we have shown in Listing~\ref{lst:rbac_example} for different types of roles}.

We also found that some features build up on a combination of an API and a configuration file.
For example, OWASP ESAPI provides an API to handle login requests as a part of authentication.
In a configuration file, the developer must set additional properties, such as maximum login attempts or timeout duration.
In Apache Shiro, access control can be handled in multiple ways.
For once, as explained before, an API and annotations can be used to perform role checks.
Additionally, Apache Shiro offers a configuration file in which resources and authorization requirements, such as roles, can be defined.

\add{Additionally, our example of the EHRS uses a combination of an API and a configuration file to realize access control in Spring Security.
First, a role hierarchy is defined in a configuration file, as shown in Listing~\ref{lst:spring_xml}.
Then, methods are annotated using Spring Security's defined annotations such as in line 7 and 13 of Listing~\ref{lst:rbac_example}, thereby, referring to these roles.}

\vspace{1mm}
\noindent\fbox{%
    \parbox{0.97\linewidth}{%
In summary, APIs are mainly used to implement the main structure for authentication and authorization mechanisms.
Annotations can be used to limit access for methods within the implementation and to decorate authentication mechanisms.
Configuration files are used to provide additional configuration parameters, such as authentication schemes or properties.
}}

\subsubsection{Cryptography}
Cryptography features are mainly realized by the usage of APIs.
Several cryptographic methods are offered in Apple's Security Framework (iOS and OSX).
The developer can, for example, use the method \texttt{SecKey\-Create\-EncryptedData()} to encrypt blocks of data using a public key and a given algorithm.
\add{In Listing~\ref{lst:enc_example_spring}, we demonstrated how the EHRS from our example uses an API provided by Spring Security for encrypting data.
There, we identified that even though it provides classes and methods for encrypting data, developers still need to write an implementation for some parts of the process, such as the generation of a salt in line 4.
As the comparison between Spring Security and Bouncy Castle in Listing~\ref{lst:encryption_examples} shows, the extent of needed code can vary among different frameworks.}
APIs are used to sign and verify digital signatures in OpenSSL as well.
A signature can be created by using the method \texttt{EVP\_DigestSign} and verified using the method \texttt{EVP\_DigestVerify}.
Several key management features, such as the generation of random keys of a fixed length, are also realized through API usages in Spring Security.
The class \texttt{Bytes\-Key\-Gen\-er\-a\-tor} provides several methods for key generation, such as \texttt{generateKey()} or \texttt{secureRandom()}.

OWASP ESAPI and JGuard are the only security frameworks that provide configuration files for \textit{encryption} and \textit{cryptographic hashing} features.
In the OWASP ESAPIs configuration file, it is possible to choose a cryptography algorithm used to encrypt or hash data.
For example, a default hashing algorithm for passwords can be defined within a configuration file, while the method \texttt{cryptPassword()} can be used to hash passwords using the specified algorithm.
In the same manner, the method \texttt{encrypt()} is used to encrypt plain text using the algorithm specified in the configuration file.

\vspace{1mm}
\noindent\fbox{%
    \parbox{0.97\linewidth}{%
In general, cryptography features are provided via APIs to allow the encryption and signing of data.
Configuration files can be used to set default algorithms for encryption, hashing, and more, as an alternative to setting them each time using method parameters.
In total, 18\% of identified cryptographic features can include configuration files, but these are only used for configuring low-level details of the feature usages.
}}

\subsubsection{Security Monitoring}
\textit{Logging} is the security monitoring feature that is realized the most by the security features.
Logging is typically handled through the usage of an API.
As an example for logging, EveryAuth offers a Boolean variable called \texttt{debug} to turn on and off logging.
Similarly, PicketBox uses the class \texttt{PicketBoxLogger} to log certain predefined events.
To define an \textit{automated response}, the Java Security Manager throws a \texttt{SecurityException} once a security violation has been detected, which can then be used to specify response actions.

In OWASP ESAPI and Spring Security configuration files can be used for configuring a security event logger that is used over an API.
Configuration parameters comprise secure encoding of logged HTML messages and the definition of an upper bound of log file size.
\add{For example, in Spring Security, a configuration parameter must be set to enable the logging of authentication attempts.}

Spring Security and JBoss Seam Security are the only security framework employing an annotation for security monitoring purposes.
For Seam security, the documentation describes that the \texttt{@Logger} annotation can be used to inject a shared instance of a logger, avoiding configuring the logger in every class. This logger could also potentially be a secure logger that has been configured by the developer or follows a security pattern\,\citep{dougherty2009}.
However, no further information is given in the documentation.

\vspace{1mm}
\noindent\fbox{%
    \parbox{0.97\linewidth}{%
To conclude, APIs are used to actively log events and define responses to security violations.
Additionally, configurations are used to specify properties to modify and customize logs for specific purposes.
Annotations only play a minor role in a few frameworks for security monitoring.
}}

\subsubsection{System State Protection}
Only \textit{session management} features are offered by the selected security frameworks for the system state protection category.
API calls are usually used to actively handle functionalities for persisting sessions, such as through the use of cookies.
OWASP ESAPI, for example, offers the method \texttt{getCookie()} to receive a session cookie.

While the general functionality of session management features is realized through the use of APIs, 40\% of the selected security frameworks offer configuration files to configure security features.
As in the case of Apache Shiro and OWASP ESAPI, properties of session management features such as the duration until a session timeout, are often specified in a configuration file\add{, as we have shown in Listing~\ref{lst:timeout_example_shiro}}.
The actual sessions are implemented in further web frameworks such as the Jakarta XML Web Services (JAX-WS) of Java EE.
The security frameworks target to secure the sessions of such web frameworks.

Java EE is the only security framework using annotations for persisting sessions by using the \texttt{@RememberMe} annotation.
After annotating an implemented authentication mechanism class, the login is remembered by the system to keep the session persistent.

\vspace{1mm}
\noindent\fbox{%
    \parbox{0.97\linewidth}{%
In summary, while APIs provide ways to manage sessions actively, configuration files are used to configure specific properties.
Annotations can be used in addition to authentication mechanisms to keep a session open between actions.
}}

\subsubsection{Secure Data Handling}
Along with cryptography, secure data handling features mainly rely on APIs for their realization.
However, OWASP ESAPI is the only framework making use of configuration files for secure data handling to define rules for validation using regular expressions.

Apple's Security Framework provides the \textit{secure storage} of data that is shared among applications over its keychain service.
\add{As shown in Listing~\ref{lst:secure_storage_example},} the class \texttt{SecKeychain} is used to store passwords, cryptographic keys, certificates, and notes.

Additionally, APIs are used to generate \textit{timestamps} and provide a \textit{source of randomness}, which are used by other security features, e.g., for logging or cryptographic purposes.
Apache Shiro offers the method \texttt{getStartTimestamp()} for receiving the starting time of an opened session and a class \texttt{Se\-cure\-Ran\-dom\-Num\-ber\-Gen\-erator} to generate secure random numbers.

APIs can also be used to set rules for regular expressions for validation features.
For example, OWASP ESAPI explicitly offers a \texttt{ValidationRule} interface for specifying rules for data from an untrusted source.
A corresponding configuration file is used to register these rules with the security framework.

\vspace{1mm}
\noindent\fbox{%
    \parbox{0.97\linewidth}{%
In summary, secure data handling features, such as secure storage, generating data from trusted sources, and data validation and sanitation are mainly provided via APIs.
Configuration files are only used in OWASP ESAPI for specifying validation rules through regular expressions.
}}



\subsection{Locating Security Features}
\looseness=-1
The mechanisms used for integrating security features into a system are essential for locating the features.
We observed that a majority of the implementations of security features are via API classes, methods, or fields.
However, we still found a large number of additional functionalities of security frameworks that moved parts of the implementation, e.g., session management parameters and authorization policies, into configuration files and annotations.
To estimate how well the different security features can be located, we investigated how their mechanisms can be used to map them to specific security features.

\subsubsection{Source Code APIs}
\looseness=-1
Every security feature that we identified within the security frameworks is provided through an API.
Since APIs are explicitly used in the implementation, their usages can be located by searching the source code.
As cryptographic features, such as encryption and hashing mostly use APIs, their usages are easy to locate in principle.
However, as also observed in existing works\,\citep{TPS2022}, in some cases, APIs use the same method call for the realization of different security features.
For instance, Bouncy Castle provides engine classes to realize different ciphers based on the method \texttt{init()} with a mode parameter for switching between encryption and decryption.
\add{This process is illustrated in Listing~\ref{lst:enc_example_bc}, line 15, in which \texttt{ENCRYPT\_MODE} denotes that a message should be encrypted.
Respectively, messages can be decrypted by using \texttt{DECRYPT\_MODE} as a parameter.}
The same API call realizes different features, and it becomes difficult to distinguish between them when no additional information is provided.
Furthermore, the usage of API methods can only serve as an entry point for feature location techniques, since the security-critical code for using them also needs to be identified.
\add{Listing~\ref{lst:encryption_examples} shows that the amount of relevant code can range from a few code statements to complex configuration code as for Bouncy Castle.}
In summary, to facilitate the location of security features such as access control, cryptography, etc. from security frameworks, traceability can be established through the security frameworks' API calls.

\begin{figure}[tb]
  \setcaptiontype{lstlisting}
        \begin{minipage}[b]{\columnwidth}
\begin{lstlisting}[label = lst:esapi_config, language=Config]
# ESAPI Encryption
#
# The ESAPI Encryptor provides basic cryptographic
# functions with a simplified API. [...]
Encryptor.MasterKey=tzfztf56ftv
Encryptor.MasterSalt=123456ztrewq
# Provides the default JCE provider that ESAPI will
# "prefer" for its symmetric encryption and hashing.
# [...] Default: Keeps the JCE provider set to
# whatever JVM sets it to.
Encryptor.PreferredJCEProvider=
# By default, ESAPI Java 1.4 uses "PBEWithMD5AndDES"
# and which is very weak.
Encryptor.EncryptionAlgorithm=AES
# For ESAPI Java 2.0 - New encrypt / decrypt methods
Encryptor.CipherTransformation=AES/CBC/PKCS5Padding
\end{lstlisting}
            \subcaption{\add{OWASP ESAPI configuration file showing configuration properties for using the encryption algorithm}}
            \label{lst:esapi_config}
        \end{minipage}%

        \begin{minipage}[b]{\columnwidth}
\begin{lstlisting}[label = lst:esapi_impl, language=Java]
public EncryptedPatientDataObject encryptPatientData(PlainText data) {
    EncryptedPatientDataObject encrypted_data = esapi.encrypt(data);
}
\end{lstlisting}
            \subcaption{\add{API call of OWASP ESAPI encryption feature. Note, that no concrete cipher is specified here, since it is provided through the configuration file in Listing~\ref{lst:esapi_config}}}
            \label{lst:esapi_impl}
        \end{minipage}
    \caption{Encryption of patient data using the OWASP ESAPI encryption feature.}
    \label{lst:encryption_examples_esapi}
\end{figure}

\subsubsection{Configuration Files}
As described above, configuration files are mainly used for access control and system state protection features within the security frameworks.
In principle, configuration files can be related to security features based on knowledge of the security frameworks.
Some values in configuration files are only used for configuring security features that are provided through explicit APIs, whose location we discussed above.
In such cases, no further configuration file-specific tracing is required.
The challenge lies in locating the custom-implemented source code locations that interact with parts of the configuration file since configuration files typically do not require explicit interaction by the developer to provide some kind of functionality.
Often, configuration files affect features simply by including them.
For example, while the security-related parts of the Java EE API allow realizing session management, some lower-level features, such as session timeouts, can be set centrally in a configuration file of a security framework.

Configuration files can also contain references to relevant places in the code.
For example, configuration files can apply features on namespaces or locations within the implementation, which can be used for tracing.
OWASP ESAPI includes a structured configuration file listing configurable properties of features, which can be related to the implementation.
As it is structured in different categories, it is possible to directly relate some of the categories to a specific feature.
For example, the category \texttt{ESAPI Encryption} can be trivially mapped to the security feature encryption, since it contains different parameters for encryption features.
Difficulties arise for developers when mapping lower-level security features to the configuration, as substantial security knowledge is required.
Furthermore, as in the case of source code APIs, some configuration options can be related to different security features as well.
\add{As shown in Listing~\ref{lst:esapi_config},} the option \texttt{Encryptor.CipherTransformation} in the OWASP ESAPI configuration file defines what encryption should be used by default and could potentially point towards both a stream cipher or a block cipher, \add{as the concrete cipher is specified by implementing the interface \texttt{Encryptor}}, which aggravates the relation to the concrete security feature.
\add{When used in different parts of the system, the cipher to be used is not provided as a parameter, and the function call does not reveal the concrete low-level security feature in this case, as shown in Listing~\ref{lst:esapi_impl}.
Therefore, source code APIs need to be considered as well in correctly identifying the concrete low-level security feature.}

\add{A significant challenge in identifying security features using configuration files arises when they need to be parsed.
While APIs and annotations can be identified by parsing the code and traversing the abstract syntax tree using one of the many available parsers for each programming language, configuration files often use custom file formats and follow an individual structure.
Therefore, configuration files require additional information and reasoning to map specific parts of them to security features.}

\subsubsection{Annotations}
As described in Section~\ref{subsub:ac}, annotations can be used to apply security features to classes or methods. Mainly authorization on the method level\add{, such as in Spring Security, shown in line 7 and 13 of Listing~\ref{lst:rbac_example},} and parts of authentication features can be identified by extracting annotations from the implementation.
Logging and session management are further security features that can be realized by providing annotations.
Traceability between source code and features can be achieved by annotating code with identifiers for features\,\citep{Martinson.2021, Bergel.2021, Entekhabi.2019, Andam.2017}.
However, annotating every security-relevant line or block of code can be an exhaustive and tedious task, which creates a lot of overhead during development.
As annotations of frameworks have unique names and correspond to specific security features, they can be seen as information-wise equivalent to feature annotations of tracing frameworks.
These security features could then be located by locating every occurrence of their respective annotations.
As there is also no further code for using the security features, no further steps are required to provide a link between them.
In the example of Spring Security shown \add{in Listing~\ref{lst:rbac_example}}, extracting the \texttt{@PreAuthorize} annotation for a method can clearly reveal which method is accessible by which role, thus providing a simple mapping between access control and the annotation.

\vspace{1mm}
\noindent\fbox{%
    \parbox{0.97\linewidth}{%
    \textbf{RQ3}: \add{When security frameworks are used to integrate security features into a system, the extracted knowledge of how the frameworks provide security features can be used to locate features for all 5 top-level security features we identified. This knowledge encapsulates annotations, code APIs, and configuration files, as well as which interfaces provide which security features. In particular, the methods belonging to the API that provide a framework's cryptographic security features. However, embedding security features into the system requires writing code that actually uses the provided security features. This code may involve more than just calling the API, but may also involve changing data formats or configuring security features such as supported authentication schemes or key lengths. Therefore, feature location techniques must also locate source code that is required to use the security frameworks, e.g., through information flow analysis of the data handed over to an API.}



    }
}

\add{
\section{Application Example}
To demonstrate the utility of our taxonomy, we consider our exemplary EHRS.
As we discussed throughout this work, the system utilizes a wide range of security features to handle the access and protection of patient data in a hospital.
Recall that in the scope of the system, there are multiple roles that have access to the system.
Among them are patient, doctor, and hospital staff.
Through this system, a designated doctor is able to view patient data for which they have been granted permission by the patient.
Other doctors should not be able to view the full extent of the patient's data.
Only hospital staff is allowed to read and write data related to billing.
}

\add{
\parhead{Selecting security features.}
When starting the development of the EHRS, developers have to select the security features required to secure the system.
The taxonomy of functional security features supports the reasoning of required security features by providing an overview of the types of security features that could be implemented.
For example, based on the division of permissions illustrated in Fig.~\ref{fig:use_case} and the covered features shown in Fig.~\ref{fig:taxonomy_access_control}, a combination of role-based access control and attribute-based access control is a suitable choice.

Due to the criticality of the system, it must be developed in compliance with relevant standards \citep{HIPAA1996,MDR2017}. 
For the US, for example, such a system would have to be developed in compliance with standards such as the NIST SP 800-53.
The concrete standards to be followed depend on the concrete system and the target market.
For illustration, we discuss how the mapping between the NIST SP 800-53, which only implicitly considers functional security features, and the security features in the taxonomy supports developers in selecting concrete security features that are required to address the high-level aspects considered in the standard. 
For example, the segment of the standard on system backups shown in Fig.~\ref{fig:encryption-standard} is mapped to the cryptographic security features of the taxonomy (Fig.~\ref{fig:taxonomy_cryptography}).
By following this mapping, the developers discuss possible functional security features and end up with deciding to encrypt backups using a block chiper.

\vspace{1mm}
\noindent\fbox{%
    \parbox{0.97\linewidth}{%
The taxonomy and the mapping to high-level security standards provide a basis to developers for systematic reasoning about and selection of appropriate security features for developing a secure software system.
}}
}

\add{
\parhead{Realization of security features.}
To avoid insecure implementation of security features, developers can follow the best practice of using frameworks that provide ready to use implementations of the planned security features.
To get an overview of possible frameworks, they could follow the mapping between the selected security features from the taxonomy and the security frameworks that provide those features.
For example, they could look up which frameworks provide authentication features by following the mapping between the taxonomy and the frameworks.
Based on this, they can see that the security feature is covered by several security frameworks and could decide to use Spring Security to realize role-based access control using an API, configuration file, and annotations.
}

\add{
The permission system have to be implemented by a developer, who must make a lot of considerations regarding the correct distribution of permissions to the users.
One mistake in assigning permissions to users can have a significant impact, allowing unauthorized users to gain access to sensitive data.
Still, developers have to write code to integrate the security framework into the system, which can be prone to errors.
For example, to restrict access of a method to specific roles, they must annotate the method with the correct roles.
Additionally, they may define a role hierarchy in a configuration file, which allows the inheritance of permissions between roles by granting a child role the same permissions as a parent role.

\vspace{1mm}
\noindent\fbox{%
\parbox{0.97\linewidth}{%
    The mapping between security features in the taxonomy and which of those are provided by popular security frameworks helps developers in selecting appropriate frameworks for realizing the planned security features.
}}
\smallskip
}

\add{
\parhead{Certification of a software system.}
Many critical systems must be compliant with the Common Criteria,\footnote{\url{https://www.commoncriteriaportal.org/products/index.cfm}} a requirement that is likely applicable to the EHRS as well.
Among others, the CC requires the implementation of access control policies and functions, and provides details on how those security features must be realized.
However, the huge size of the standard makes it challenging to identify those aspects that are relevant for the developed system.
To systematically identify those aspects, developers can take the list of security features from the taxonomy that have been selected above and systematically look up the relevant locations in the standard using the mapping between the two.
This way, they can effectively review whether all security features are implemented compliant with the CC.

\vspace{1mm}
\noindent\fbox{%
    \parbox{0.97\linewidth}{%
Using the mapping between the taxonomy and low-level security standards, developers can systematically assess all implementation details needed for a certification of a software system.
}}
}

\add{
\parhead{Locating security features in case of an incident.}
Assuming an incident occurred, in which a nurse was able to access billing data related to the treatment of a patient, developers have to quickly react and recover all locations of the security features.
A developer, who might (not) be familiar with the system, is tasked with resolving the cause of the incident.
To support the investigation, the developer could leverage the taxonomy, which lists and describes security features in a hierarchical order, helping them to reason about security features that might play a role in the incident.
Based on the assumption of an authorization issue as a root cause of the incident, the developer could search for the usages of the authorization functionality of the used security framework.
In combination with these usage locations, they could investigate the source code providing the functionality that the nurse was able to execute. The analysis would show that the method \texttt{getBillingData()} is annotated with \texttt{@PreAuthorize\{ROLE\_HOSPITAL\-STAFF\}}, which allows users with the role hospital staff to access billing information.
Since the method appears to be correctly annotated, there might be an issue with the role hierarchy.

Consequently, they additionally need to investigate the corresponding xml file, which configures the role hierarchy.
Here, it becomes evident that according to the role hierarchy, the role nurse inherits all permissions from the role hospital staff.
According to the use-case diagram shown in Figure~\ref{fig:use_case}, only hospital staff should have access to billing data, not nurses.
Due to the inheritance relation in the configuration, the nurse gained the permissions to read and write billing data.
Therefore, the developer would need to fix the error by introducing two roles, medical staff and administration staff, which splits the granted access permissions to measurements and billing accordingly, and correct the role hierarchy as shown in Listing~\ref{lst:spring_xml_diff}, as well as the annotation.
This change would lead to the correct role hierarchy as shown in Listing~\ref{lst:spring_xml}.
The differences in the distributed permissions are shown in Table~\ref{tab:accessrights}.
}

\begin{lstlisting}[identifierstyle=\color{black},basicstyle=\ttfamily\scriptsize,
    label=lst:spring_xml_diff,
    caption=\add{Diff of a change in the role hierarchy}]
@@ -1,8 +1,10 @@
  <property name="hierarchy">
   <value>
<@{\color{red}-~~~ROLE\_DOCTOR~~~~~~~~~~> ROLE\_HOSPITALSTAFF}@>
<@{\color{red}-~~~ROLE\_NURSE~~~~~~~~~~~> ROLE\_HOSPITALSTAFF}@>
<@{\color{codegreen}+~~~ROLE\_DOCTOR~~~~~~~~~~> ROLE\_MEDICSTAFF}@>
<@{\color{codegreen}+~~~ROLE\_NURSE~~~~~~~~~~~> ROLE\_MEDICSTAFF}@>
<@{\color{codegreen}+~~~ROLE\_MEDICSTAFF~~~~~~> ROLE\_STAFF}@> 
<@{\color{codegreen}+~~~ROLE\_HOSPITALSTAFF~~~> ROLE\_STAFF}@>
   ROLE_STAFF       > ROLE_UNAUTHENTICATED
  </value>
 </property>
\end{lstlisting}

\begin{table}
\centering
\scriptsize
\caption{\add{Role-based access control policy for our EHRS. r = read, w = write, (r) = read if attribute is accepted; Bold permissions are explicitly specified and italic ones inherited.}}
\label{tab:accessrights}
\addtolength{\tabcolsep}{-3pt}
\begin{tabular}{rrl|cccc}
    \toprule
    &&\multirow{2}{*}{\textbf{Role}} & \multirow{2}{*}{\textbf{Diag.}} & \textbf{Health} & \multirow{2}{*}{\textbf{Planning}} & \multirow{2}{*}{\textbf{Billing}} \\
    && & & \textbf{Measurem.} & &  \\
    \midrule
    \multirow{8}{*}{\rotatebox{90}{\textbf{Role Inheritance}}}& \multirow{3}{*}{\rotatebox{90}{{Wrong}}}&
        Doctor & \textbf{w/(r)} & \textit{w\textsuperscript{1}}/\textbf{(r)} & \textit{w/r\textsuperscript{1}} & \textit{w/r\textsuperscript{1}} \\
    &&  Nurse & none & \textbf{w} & \textit{w/r\textsuperscript{1}}  & \textit{w/r\textsuperscript{1}}  \\
    &&  Staff & none & none & \textbf{w/r}  & \textbf{w/r} \\
    \cmidrule{2-7}
    &\multirow{5}{*}{\rotatebox{90}{{Correct}}} & Doctor & \textbf{w/(r)} & \textit{w\textsuperscript{1}}/\textbf{(r)} & \textit{w/r\textsuperscript{1}}  & none \\
    &&Nurse & none & \textit{w\textsuperscript{1}} & \textit{w/r\textsuperscript{1}}  & none  \\
    &&\textit{(Med. Staff)\textsuperscript{2}} & none & \textbf{w} & \textit{w/r\textsuperscript{1}}  & none \\
    &&\textit{(Adm. Staff)\textsuperscript{2}} & none & none & \textit{w/r\textsuperscript{1}}  & \textbf{w/r} \\
    &&Staff & none & none & \textbf{w/r}  & none \\
    \bottomrule
\end{tabular}

\textsuperscript{1} Inherited permissions, \textsuperscript{2} Abstract roles
\end{table}

\add{
\vspace{1mm}
\noindent\fbox{%
    \parbox{0.97\linewidth}{%
As discussed in Sec.~\ref{sec:traceability}, security features manifest in a system via the usages of APIs and annoations in source code as well as configuration files, which can be automatically detected and help developers in locating security features.
}}
\smallskip
}

\add{
\parhead{Reasoning about related security features.}
After the role hierarchy has been changed to fix the vulnerability discussed above, it is unclear what other parts of the system are impacted by this change.
Here, developers may also need to investigate other related security features.
For instance, while attribute-based access control and role-based access control may now be correctly implemented, developers should still check what other security features related to storing or retrieving data, such as secure storage, might be affected.
Another entry point for further investigation could be the parent feature Authorization, which may be implemented either through custom code or other frameworks, which should all be checked.
Recovering this code requires feature location techniques which must take into account API code, configurations, and annotations that security frameworks use to realize these security features.

\vspace{1mm}
\noindent\fbox{%
\parbox{0.97\linewidth}{%
    The taxonomy provides developers with an concise overview of functional security features, which helps in reasoning about security features related with each other, i.e., since multiple variants of access control are combined.
}}
}

\section{Discussion and Implications}
\label{sec:discussion}

\noindent The results of our study suggest the following implications for practitioners and research directions.

\subsection{Practitioners}
\label{subsec:implications}
Practitioners can use our results to better understand security features, their coverage by security frameworks as well as their relation to security standards.
\add{
   Because the taxonomy provides an overview of security features, concise explanations, and references to more detailed literature, it is also a good starting point for developers new to IT security.
}
Our derived taxonomy indicated that for each functional security feature, there are many different sub-features relevant to practitioners, which need to be selected appropriately for each software system.
The taxonomy offers a selection of needed security features on a high abstraction level, which are linked to multiple security standards.
This facilitates security feature selection when working with security standards.
Thereafter, as the taxonomy shows for which security feature a security framework exists, it is possible to choose an appropriate security framework based on the selection.
For almost all of these security features, libraries and frameworks should be used to minimize risks for security issues through custom implementations.

Functional security features provided by libraries and frameworks can also be used as an entry point when performing code reviews.
Our results show that configuration files are used by many security frameworks and play an essential role when realizing certain functional security features. Therefore, they should be reviewed as well.
Aside from security features based on annotations, the use of security frameworks and APIs still requires a substantial amount of security-critical code, which is prone to insecurity and requires careful scrutiny.
\add{
    By identifying relevant code locations and configuration files, either on the entire project or when corresponding locations are changed, reviewers can be immediately pointed toward those locations. 
    Thereby, identified code locations can be automatically related to the security features that are realized there instead of only low-level code statements that must be put into context manually.
}

Many more sophisticated security concepts of the taxonomy, e.g., multi-factor authentication, have to be realized by developers in terms of combining other security features. 
Here, the academic literature describes multiple helpful implementation-level security features by which frameworks should be extended to provide more straightforward use and reduce the probability of insecure implementations.

\subsection{Researchers}
\looseness=-1
As our results reveal that multiple security features are not offered by security frameworks, our taxonomy implies several research directions.
In our SLR, we captured five kinds of functional imp\-le\-men\-ta\-tion-level security features.
While most features are provided by security frameworks, despite best practices that advise otherwise\,\citep{JO2016}, developers are likely to implement some of them on their own.
Based on what the security frameworks provide, we assume \textit{logging} and \textit{data validation} features to be most likely not used as provided by security frameworks but to be mainly based on custom code.
Additionally, we found multiple security features, such as \textit{retention control} and \textit{resource management}, to not be offered by any of the security frameworks.
As such, it is essential to consider the tendency toward custom implementation of such security features in research on security compliance checks.

\add{The reasons on why some security features are not included in security frameworks demand further investigation in future work.
As security frameworks, such as Bouncy Castle or OpenSSL, are usually tailored towards specific use cases, some security frameworks implement some subfeatures of our taxonomy, but not all of them.
Researchers should therefore investigate, whether security framework developers are not aware of these security features, or if there are other reasons for their exclusion.}

\add{Further, while our investigation provides a structured overview of security features available within security frameworks, future research should delve deeper into understanding their actual implementation and usage in practice. Specifically, this could involve conducting developer studies to understand how practitioners implement and adapt these features in practice or mining public repositories, such as GitHub, to identify security features in codebases. Our taxonomy provides a foundation for such investigations, enabling researchers to analyze security features in actual implementations.}

The use of established security frameworks not only lowers the risk of security issues in a software system by avoiding custom insecure implementations of security features but also provides easy-to-locate entry points for feature location techniques.
To improve the location of security-critical code, additional annotations for labeling source code can help identify relevant code portions.
Still, developers should not be overwhelmed by too many additional annotations. 
Instead, information relevant to feature location can be gathered from security frameworks, as discussed in Section~\ref{sec:traceability}.

In the context of security audits or security compliance checks, the location of the source code portions corresponding to security features is essential. 
Our findings showed that concrete implementation-level security features might be relatively simple to locate.
However, when looking at the literature on design-time security requirements\,\citep{Juerjens05, Peldszus2022}, we notice a divergence in abstraction between the security requirements, e.g., declaration of what is sensitive information, and the concrete security features identified that will be used for implementing such security requirements. \add{Following security by design techniques, security features are usually planned very abstractly but must be implemented taking a number of aspects into account to ensure that they are used securely and cannot be bypassed. This gap in abstraction is comparable to the differences observed above between high-level and low-level security standards and is a significant obstacle to checking the implementation for compliance with its security design \citep{PTS+2019,Peldszus2023,TPS2022}. Since our taxonomy of functional security features resides in between those two abstractions and effectively maps between them}, our findings can be used as a basis for novel security feature traceability methods.

\section{Threats to Validity}
\label{sec:threats_to_validity}
\looseness=-1
We now discuss threats to validity.

\subsection{Internal Validity}
\looseness=-1
Internal validity might be threatened by author bias.
For once, the keywords that were used for the systematic literature research were chosen by the authors.
This might additionally have an impact on the selection of the security standards, which were chosen based on the expert knowledge of the authors.
To minimize the bias, we employed several authors from different research areas, such as the software engineering, security, and human factors domains, who held frequent discussions.
Through this process, the paper selection revealed a large sample of security features considered in the literature.
Therefore, relating the selected standards to our taxonomy still revealed a strong overlap while also providing more general terms for some categories within our taxonomy, confirming the representativeness of the sample.
The same bias could also threaten the validity of the mapping between the standards and our taxonomy.
To ensure the validity of the mapping, the first five authors held discussions on matches and discrepancies.
Discrepancies that lead to changes were resolved by the same authors as well.

A bias in selecting the security frameworks might be introduced by the Stack Overflow and Reddit security framework selection. 
\add{Security frameworks discussed on these platforms may not accurately represent those widely used in industry, introducing a potential bias in our selection. These discussions may highlight frameworks with greater usability challenges or ones associated with popular programming languages. Additionally, thread recency could skew the results, favoring frameworks with active recent discussions while potentially excluding those still relevant but less frequently discussed. To address these potential biases, we expanded our search to include multiple developer communities, using Reddit alongside Stack Overflow to confirm that selected frameworks are relevant across different communities.}
\add{Further}, we might have wrongly excluded security frameworks based on our interpretation of the discussion in the threads.
Similarly, the investigation of the homepage, reference guide, and the API documentation of the corresponding security framework for the framework security feature extraction could bias the resulting security features, as some security features might not have been considered as security features in the analysis process.
To minimize the threat of wrongly excluding a security framework or security feature, two of the authors independently participated in the framework selection process, extracted the security features from the security frameworks, and held frequent discussions on the inclusion and exclusion of extracted features of the security frameworks.

A final threat to the correctness and completeness might be imposed by the sources used for the security feature search.
The security features that can be extracted from the homepage, reference guide, and the API documentation of the corresponding security frameworks may not reveal a complete set of security features of each framework, as it might offer security features that are not well documented.
Therefore, there may be a few security features offered by the security frameworks that we did not consider in this work.
Still, with the selected security frameworks and sources, we were able to provide a rich comparison of their features to our taxonomy.
Additionally, we were able to thoroughly reason about the realization of security features within commonly discussed security frameworks.

\subsection{External Validity}
\looseness=-1
Multiple external factors threaten the generalizability of our results.
The systematic literature research and Stack Overflow and Reddit search might not allow us to capture a representative sample of literature and security frameworks relevant to our study.
Still, the chosen general keywords provide an extensive collection of literature and security frameworks, emphasized by the strong overlap between the taxonomy and security features from the security frameworks, as well as the security standards.

Generalizability might also be threatened by the content of the selected papers we investigated.
A large overlap between the investigated security standards and the literature in the validation of the taxonomy revealed that this concern is not significant to our work.
We can, therefore, conclude that our taxonomy covers a wide range of security features to be considered when implementing software systems.

Another threat to the generalizability of our results is introduced by utilizing Stack Overflow and Reddit as a source for the security framework investigation.
The selection criteria for security frameworks (being mentioned in two or more threads on Stack Overflow and Reddit) might not reveal all popular frameworks.
Since developers on Stack Overflow and Reddit mainly discuss frameworks available to everyone, we could have missed frameworks that are closed for public usage. 
Nonetheless, the investigated security frameworks contained a large number of security features, which are included in our taxonomy.

\section{Conclusion}
\noindent
In this paper, we present a taxonomy of functional implementation-level security features based on an SLR of the literature, their mapping to widely used security standards, and their relation to popular security frameworks.
Following an empirical approach, we aim to improve the understanding of the requirements for light-weight security feature location support.
Our taxonomy contains 68 security features
with the top-level features \textit{access control}, \textit{cryptography}, \textit{security monitoring}, \textit{system state protection}, and \textit{secure data handling}.
To examine which security features are contained in security frameworks commonly discussed on Stack Overflow and Reddit, we investigated existing security frameworks and related the provided security features to our taxonomy.
While most functional security features considered in the literature are provided by security frameworks, there are still many that need substantial implementation effort by developers.

Finally, as a first step towards light-weight security feature approaches, we investigated how security frameworks provide security features to developers and discussed strategies for locating security features to reduce the manual location effort to a minimum. 
We found, that security features provided by security frameworks mostly utilize manifest in forms such as API calls, which are easy to identify in the codebase. 
As such, traceability techniques are able to leverage this information to enable the quick location of security features.

The practical implications show how developers can use our taxonomy to choose security features required to adhere by popular security standards and select appropriate security frameworks.
We focused on the literature and security frameworks as reliable sources, constituting a self-contained study, still, follow-up work should investigate more data sources.
A logical next step is an empirical investigation of the security features presented in the taxonomy with practitioners to identify challenges and best practices in implementing them.
\add{Future work should examine how these security features are applied in real software systems, either through developer studies or by mining public repositories such as GitHub. Our taxonomy serves as a basis for this investigation, allowing researchers to assess the practical usage of both framework-provided and custom-implemented security features.}
An affirmation of its quality and usability would further support the claim of the practical implications.

Finally, we call for action to improve the location of security features while lowering additional development-time effort.
Our findings build a foundation for this objective by providing a deeper understanding of im\-ple\-men\-ta\-tion-level security features and which indicators could be used as entry points for their location.
We hope that other researchers complement our taxonomy.
Based on that foundation, we aim to develop methods that can be used to establish traceability between security feature models and their implementation in code.

\section{Data Availability}
\noindent
A replication package of all the data of our systematic reviews of the literature, security standards and security frameworks is publicly available at Dropbox\,\citep{Dropbox}. Upon acceptance of this article, we will move our replication package to Zenodo.

\section*{Acknowledgments}
\noindent
Supported by the Deutsche Forschungsgemeinschaft (DFG, German Research Foundation) under Germany's Excellence Strategy - EXC 2092 CASA - 390781972.

\balance
\bibliographystyle{spbasic}
\bibliography{IEEEabrv,doc}

\begin{thebibliography}{117}
\providecommand{\natexlab}[1]{#1}
\providecommand{\url}[1]{{#1}}
\providecommand{\urlprefix}{URL }
\expandafter\ifx\csname urlstyle\endcsname\relax
  \providecommand{\doi}[1]{DOI~\discretionary{}{}{}#1}\else
  \providecommand{\doi}{DOI~\discretionary{}{}{}\begingroup
  \urlstyle{rm}\Url}\fi
\providecommand{\eprint}[2][]{\url{#2}}

\bibitem[{Abbas et~al.(2005)Abbas, Saddik, and Miri}]{Abbas2005ASO}
Abbas A, Saddik AE, Miri A (2005) {A} {S}tate {O}f {T}he {A}rt {S}ecurity
  {T}axonomy {O}f {I}nternet {S}ecurity: {T}hreats {A}nd {C}ountermeasures.
  Computer Science

\bibitem[{Abukwaik et~al.(2018)Abukwaik, Burger, Andam, and
  Berger}]{Abukwaik.2018}
Abukwaik H, Burger A, Andam BK, Berger T (2018) {S}emi-{A}utomated {F}eature
  {T}raceability {W}ith {E}mbedded {A}nnotations. In: International Conference
  on Software Maintenance and Evolution (ICSME), IEEE, pp 529--533,
  \doi{10.1109/ICSME.2018.00049}

\bibitem[{Acar et~al.(2017)Acar, Backes, Fahl, Garfinkel, Kim, Mazurek, and
  Stransky}]{Acar17}
Acar Y, Backes M, Fahl S, Garfinkel S, Kim D, Mazurek ML, Stransky C (2017)
  {C}omparing {T}he {U}sability {O}f {C}ryptographic {A}{P}{I}{s}. In: 2017
  IEEE Symposium on Security and Privacy (S\&P), pp 154--171,
  \doi{10.1109/SP.2017.52}

\bibitem[{Adat and Gupta(2018)}]{AdatG18}
Adat V, Gupta BB (2018) {S}ecurity {I}n {I}nternet {O}f {T}hings: {I}ssues,
  {C}hallenges, {T}axonomy, {A}nd {A}rchitecture. Telecommunication Systems
  67(3):423--441, \doi{10.1007/s11235-017-0345-9}

\bibitem[{Ahmadian et~al.(2017)Ahmadian, Peldszus, Ramadan, and
  J\"{u}rjens}]{Ahmadian2017}
Ahmadian AS, Peldszus S, Ramadan Q, J\"{u}rjens J (2017) {M}odel-{B}ased
  {P}rivacy {A}nd {S}ecurity {A}nalysis {W}ith {C}{A}{R}{i}{S}{M}{A}. In: Joint
  Meeting on Foundations of Software Engineering, ACM, ESEC/FSE 2017, p
  989–993, \doi{10.1145/3106237.3122823}

\bibitem[{Andam et~al.(2017)Andam, Burger, Berger, and Chaudron}]{Andam.2017}
Andam B, Burger A, Berger T, Chaudron MRV (2017) {FLOrIDA: Feature LOcatIon
  DAshboard for extracting and visualizing feature traces}. In: International
  Workshop on Variability Modelling of Software-intensive Systems (VaMoS), ACM,
  pp 100--107, \doi{10.1145/3023956.3023967}

\bibitem[{Ardagna et~al.(2009)Ardagna, Cremonini, De~Capitani~di Vimercati, and
  Samarati}]{Ardagna2009}
Ardagna CA, Cremonini M, De~Capitani~di Vimercati S, Samarati P (2009) Access
  Control in Location-Based Services, Springer, pp 106--126.
  \doi{10.1007/978-3-642-03511-1_5}

\bibitem[{Baitha and Vinod(2018)}]{Baitha18_session_hijacking}
Baitha AK, Vinod S (2018) {S}ession {H}ijacking {A}nd {P}revention {T}echnique.
  International Journal of Engineering \& Technology 7(2.6):193--198

\bibitem[{Batory et~al.(2004)Batory, Sarvela, and Rauschmayer}]{Batory:2004bw}
Batory D, Sarvela JN, Rauschmayer A (2004) {Scaling Step-Wise Refinement}. IEEE
  Transactions on Software Engineering 30(6):355--371

\bibitem[{Bau et~al.(2012)Bau, Wang, Bursztein, Mutchler, and
  Mitchell}]{bau2012vulnerability}
Bau J, Wang F, Bursztein E, Mutchler P, Mitchell JC (2012) {V}ulnerability
  {F}actors {I}n {N}ew {W}eb {A}pplications: {A}udit {T}ools, {D}eveloper
  {S}election \& {L}anguages. Stanford, Tech Rep

\bibitem[{{BBC}(2020)}]{hospital_example1}
{BBC} (2020) {{P}olice {L}aunch {H}omicide {I}nquiry {A}fter {G}erman
  {H}ospital {H}ack}. \url{https://www.bbc.com/news/technology-54204356/},
  [Online; accessed 04-December-2024]

\bibitem[{{ben Othmane} et~al.(2015){ben Othmane}, Chehrazi, Bodden, Tsalovski,
  Brucker, and Miseldine}]{benOthmane.2015}
{ben Othmane} L, Chehrazi G, Bodden E, Tsalovski P, Brucker AD, Miseldine P
  (2015) {F}actors {I}mpacting {T}he {E}ffort {R}equired {T}o {F}ix {S}ecurity
  {V}ulnerabilities. In: Information Security, Lecture Notes in Computer
  Science, vol 9290, {Springer}, pp 102--119,
  \doi{10.1007/978-3-319-23318-5\_6}

\bibitem[{{ben Othmane} et~al.(2017){ben Othmane}, Chehrazi, Bodden, Tsalovski,
  and Brucker}]{benOthmane.2017}
{ben Othmane} L, Chehrazi G, Bodden E, Tsalovski P, Brucker AD (2017) {T}ime
  {F}or {A}ddressing {S}oftware {S}ecurity {I}ssues: {P}rediction {M}odels
  {A}nd {I}mpacting {F}actors. Data Science and Engineering 2(2):107--124,
  \doi{10.1007/s41019-016-0019-8}

\bibitem[{Bergel et~al.(2021)Bergel, Ghzouli, Berger, and
  Chaudron}]{Bergel.2021}
Bergel A, Ghzouli R, Berger T, Chaudron MRV (2021) {FeatureVista: interactive
  feature visualization}. In: ACM International Systems and Software Product
  Line Conference - Volume A, ACM, pp 196--201, \doi{10.1145/3461001.3471154}

\bibitem[{Berger et~al.(2015)Berger, Lettner, Rubin, Gr{\"u}nbacher, Silva,
  Becker, Chechik, and Czarnecki}]{berger2015feature}
Berger T, Lettner D, Rubin J, Gr{\"u}nbacher P, Silva A, Becker M, Chechik M,
  Czarnecki K (2015) {What Is a Feature? A Qualitative Study of Features in
  Industrial Software Product Lines}. In: Systems and Software Product Line
  Conference

\bibitem[{Bertino et~al.(2011)Bertino, Ghinita, and
  Kamra}]{Bertino11_access_control}
Bertino E, Ghinita G, Kamra A (2011) Access Control for Databases: Concepts and
  Systems. Now Foundations and Trends

\bibitem[{Bhanot and Hans(2015)}]{Bhanot15_encryption_algorithms}
Bhanot R, Hans R (2015) {A Review and Comparative Analysis of Various
  Encryption Algorithms}. International Journal of Security and Its
  Applications 9(4):289--306

\bibitem[{Bhatia and Verma(2017)}]{BhatiaV17}
Bhatia T, Verma AK (2017) {Data Security in Mobile Cloud Computing Paradigm: A
  Survey, Taxonomy and Open Research Issues}. Journal of Supercomputing
  73(6):2558--2631, \doi{10.1007/s11227-016-1945-y}

\bibitem[{Biggerstaff et~al.(1994)Biggerstaff, Mitbander, and
  Webster}]{biggerstaff1994program}
Biggerstaff TJ, Mitbander BG, Webster DE (1994) {Program Understanding and the
  Concept Assignment Problem}. Communications of the ACM 37(5):72--82

\bibitem[{Blythe et~al.(2019)Blythe, Sombatruang, and Johnson}]{Blythe2019}
Blythe JM, Sombatruang N, Johnson SD (2019) {{What Security Features and Crime
  Prevention Advice Is Communicated in Consumer Iot Device Manuals and Support
  Pages?}} Journal of Cybersecurity 5(1), \doi{10.1093/cybsec/tyz005}

\bibitem[{Bokhari and Shallal(2016)}]{Bokhari16_symmetric_encryption}
Bokhari MU, Shallal QM (2016) {A Review on Symmetric Key Encryption Techniques
  in Cryptography}. International journal of computer applications 147(10)

\bibitem[{Bosch(2000)}]{bosch2000}
Bosch J (2000) Design \& Use of Software Architectures—Adopting and Evolving
  a Product Line Approach. Pearson Education Ltd.

\bibitem[{Busch and Wirsing(2015)}]{BuschW15}
Busch M, Wirsing M (2015) {An Ontology for Secure Web Applications}.
  International Journal of Software and Informatics 9(2):233--258

\bibitem[{Chen et~al.(2005)Chen, Zhang, Zhao, and Mei}]{chen2005}
Chen K, Zhang W, Zhao H, Mei H (2005) {An Approach to Constructing Feature
  Models Based on Requirements Clustering}. In: International Conference on
  Requirements Engineering, IEEE, p 31–40

\bibitem[{Chung et~al.(2019)Chung, Ferraiolo, Kuhn, Schnitzer, Sandlin, Miller,
  and Scarfone}]{Chung2019}
Chung, Ferraiolo D, Kuhn D, Schnitzer A, Sandlin K, Miller R, Scarfone K (2019)
  {Guide to Attribute Based Access Control (ABAC) Definition and
  Considerations}. \doi{https://doi.org/10.6028/NIST.SP.800-162}

\bibitem[{Cornell(2012)}]{cornell2012}
Cornell D (2012) {Remediation Statistics: What Does Fixing Application
  Vulnerabilities Cost}. In: RSA Conference

\bibitem[{Denker et~al.(2003)Denker, Kagal, Finin, Paolucci, and
  Sycara}]{DenkerKFPS03}
Denker G, Kagal L, Finin TW, Paolucci M, Sycara KP (2003) Security for {DAML}
  web services: Annotation and matchmaking. In: International Semantic Web
  Conference, Springer, Lecture Notes in Computer Science, vol 2870, pp
  335--350

\bibitem[{Denning(1976)}]{Denning1976}
Denning DE (1976) {A Lattice Model of Secure Information Flow}. Communications
  of the ACM 19(5):236–243, \doi{10.1145/360051.360056}

\bibitem[{Dent(2004)}]{Dent04_hybrid_cryptography}
Dent AW (2004) Hybrid cryptography. Cryptology ePrint Archive, Paper 2004/210,
  \url{https://eprint.iacr.org/2004/210}

\bibitem[{Dit et~al.(2013)Dit, Revelle, Gethers, and Poshyvanyk}]{dit2013}
Dit B, Revelle M, Gethers M, Poshyvanyk D (2013) {Feature Location in Source
  Code: A Taxonomy and Survey}. Journal of Software Maintenance and Evolution:
  Research and Practice 25, \doi{10.1002/smr.567}

\bibitem[{Dougherty et~al.(2009)Dougherty, Sayre, Seacord, Svoboda, and
  Togashi}]{dougherty2009}
Dougherty C, Sayre K, Seacord R, Svoboda D, Togashi K (2009) {Secure Design
  Patterns}. Tech. Rep. CMU/SEI-2009-TR-010, Carnegie Mellon University,
  Software Engineering Institute's Digital Library, \doi{10.1184/R1/6583640.v1}

\bibitem[{Egele et~al.(2013)Egele, Brumley, Fratantonio, and
  Kruegel}]{egele2013empirical}
Egele M, Brumley D, Fratantonio Y, Kruegel C (2013) {An Empirical Study of
  Cryptographic Misuse in Android Applications}. In: ACM SIGSAC conference on
  Computer \& communications security, ACM, pp 73--84

\bibitem[{Entekhabi et~al.(2019)Entekhabi, Solback, Stegh{\"o}fer, and
  Berger}]{Entekhabi.2019}
Entekhabi S, Solback A, Stegh{\"o}fer JP, Berger T (2019) {Visualization of
  Feature Locations With the Tool FeatureDashboard}. In: International Systems
  and Software Product Line Conference volume B, {ACM}, pp 1--4,
  \doi{10.1145/3307630.3342392}

\bibitem[{{European Parliament and Council of the European
  Union}(2017)}]{MDR2017}
{European Parliament and Council of the European Union} (2017) {Regulation (EU)
  2017/745 of the European Parliament and of the Council of 5 April 2017 on
  medical devices, amending Directive 2001/83/EC, Regulation (EC) No 178/2002
  and Regulation (EC) No 1223/2009 and repealing Council Directives 90/385/EEC
  and 93/42/EEC}.
  \urlprefix\url{https://eur-lex.europa.eu/eli/reg/2017/745/oj}, [Online;
  accessed 19-December-2024]

\bibitem[{Fahl et~al.(2013)Fahl, Harbach, Perl, Koetter, and
  Smith}]{fahl2013rethinking}
Fahl S, Harbach M, Perl H, Koetter M, Smith M (2013) {Rethinking SSL
  development in an appified world}. In: ACM SIGSAC conference on Computer \&
  communications security, ACM, pp 49--60

\bibitem[{Fang et~al.(2012)Fang, Miller, and Kupsch}]{FangMK12}
Fang W, Miller BP, Kupsch JA (2012) {Automated Tracing and Visualization of
  Software Security Structure and Properties}. In: 9th International Symposium
  on Visualization for Cyber Security (VizSec), {ACM}, pp 9--16,
  \doi{10.1145/2379690.2379692}

\bibitem[{Ferraiolo and Kuhn(2009)}]{Ferraiolo2009}
Ferraiolo DF, Kuhn DR (2009) {Role-Based Access Controls}. ArXiv abs/0903.2171

\bibitem[{Ghafir et~al.(2016)Ghafir, Prenosil, Svoboda, and
  Hammoudeh}]{Ghafir16_monitoring_systems}
Ghafir I, Prenosil V, Svoboda J, Hammoudeh M (2016) {A Survey on Network
  Security Monitoring Systems}. In: 2016 IEEE 4th International Conference on
  Future Internet of Things and Cloud Workshops (FiCloudW), pp 77--82,
  \doi{10.1109/W-FiCloud.2016.30}

\bibitem[{Glaser(1978)}]{glaser1978}
Glaser B (1978) Theoretical Sensitivity: Advances in the Methodology of
  Grounded Theory. Advances in the methodology of grounded theory, Sociology
  Press

\bibitem[{Habiba et~al.(2014)Habiba, Masood, Shibli, and Niazi}]{HabibaMSN14}
Habiba U, Masood R, Shibli MA, Niazi MA (2014) Cloud identity management
  security issues {\&} solutions: a taxonomy. Complex Adaptive Systems Modeling
  2:5, \doi{10.1186/s40294-014-0005-9}

\bibitem[{Hakeem and Shah(2004)}]{hakeem2004ontology}
Hakeem A, Shah M (2004) Ontology and taxonomy collaborated framework for
  meeting classification. In: International Conference on Pattern Recognition,
  IEEE, vol~4, pp 219--222

\bibitem[{Harbi et~al.(2019)Harbi, Aliouat, Harous, Bentaleb, and
  Refoufi}]{HarbiAHBR19}
Harbi Y, Aliouat Z, Harous S, Bentaleb A, Refoufi A (2019) {A Review of
  Security in Internet of Things}. Wireless Personal Communications
  108(1):325--344, \doi{10.1007/s11277-019-06405-y}

\bibitem[{Harzing(2007)}]{Harzing07}
Harzing A (2007) Publish or perish.
  \urlprefix\url{https://harzing.com/resources/publish-or-perish}, online;
  accessed 20-December-2023

\bibitem[{Hendre and Joshi(2015)}]{HendreJ15}
Hendre A, Joshi KP (2015) {A Semantic Approach to Cloud Security and
  Compliance}. In: Pu C, Mohindra A (eds) 8th {IEEE} International Conference
  on Cloud Computing (CLOUD), IEEE, pp 1081--1084, \doi{10.1109/CLOUD.2015.157}

\bibitem[{Herzog et~al.(2007)Herzog, Shahmehri, and Duma}]{HerzogSD07}
Herzog A, Shahmehri N, Duma C (2007) {An Ontology of Information Security}.
  International Journal of Information Security and Privacy 1(4):1--23,
  \doi{10.4018/jisp.2007100101}

\bibitem[{Hewett and Kijsanayothin(2009)}]{hewett2009}
Hewett R, Kijsanayothin P (2009) {On modeling software defect repair time}.
  Empirical Software Engineering 14:165--186, \doi{10.1007/s10664-008-9064-x}

\bibitem[{Houmb et~al.(2010)Houmb, Islam, Knauss, Jürjens, and
  Schneider}]{Houmb2010}
Houmb S, Islam S, Knauss E, Jürjens J, Schneider K (2010) {Eliciting security
  requirements and tracing them to design: An integration of Common Criteria,
  heuristics, and UMLsec}. Requirements Engineering 15:63--93,
  \doi{10.1007/s00766-009-0093-9}

\bibitem[{IBM(2023)}]{IBMDAC}
IBM (2023) {Discretionary access control (MAC)}.
  \urlprefix\url{https://www.ibm.com/docs/en/zos/3.1.0?topic=controls-discretionary-access-control-dac},
  accessed: 2023-Dec-20

\bibitem[{{IBM}({2023})}]{DOORS}
{IBM} ({2023}) {IBM Engineering Requirements Management DOORS Family}.
  \urlprefix\url{https://www.ibm.com/docs/en/engineering-lifecycle-management-suite/doors/9.7.2},
  online; accessed 20-December-2023

\bibitem[{IBM(2023)}]{IBMMAC}
IBM (2023) {Mandatory access control (MAC)}.
  \urlprefix\url{https://www.ibm.com/docs/en/zos/3.1.0?topic=environment-mandatory-access-control-mac},
  accessed: 2023-Dec-20

\bibitem[{Islam et~al.(2011)Islam, Mouratidis, and Jürjens}]{Islam2011}
Islam S, Mouratidis H, Jürjens J (2011) A framework to support alignment of
  secure software engineering with legal regulations. Software and System
  Modeling 10:369--394, \doi{10.1007/s10270-010-0154-z}

\bibitem[{{ISO/IEC JTC 1/SC 27}(2009)}]{CC}
{ISO/IEC JTC 1/SC 27} (2009) {Common Criteria for Information Technology
  Security Evaluation}. International Standard ISO/IEC 15408, {International
  Organization for Standardization (ISO)}

\bibitem[{{ISO/TC 22/SC 32 }(2021)}]{ISO21434}
{ISO/TC 22/SC 32 } (2021) {Road vehicles -- Cybersecurity engineering}.
  International Standard ISO/SAE 21434, {International Organization for
  Standardization (ISO)}

\bibitem[{Jakobsen and Orlandi(2016)}]{JO2016}
Jakobsen J, Orlandi C (2016) {On the CCA (in)Security of MTProto}. In: Workshop
  on Security and Privacy in Smartphones and Mobile Devices, p 113–116,
  \doi{10.1145/2994459.2994468}

\bibitem[{Ji et~al.(2015)Ji, Berger, Antkiewicz, and Czarnecki}]{Ji2015}
Ji W, Berger T, Antkiewicz M, Czarnecki K (2015) {Maintaining Feature
  Traceability with Embedded Annotations}. In: International Conference on
  Software Product Line, ACM, pp 61--70, \doi{10.1145/2791060.2791107}

\bibitem[{Jiao et~al.(2020)Jiao, Hao, and Feng}]{Jiao20_stream_ciphers}
Jiao L, Hao Y, Feng D (2020) Stream cipher designs: a review. Science China
  Information Sciences 63:1--25

\bibitem[{Jin et~al.(2012)Jin, Sandhu, and Krishnan}]{Jin2012}
Jin X, Sandhu R, Krishnan R (2012) {RABAC: Role-Centric Attribute-Based Access
  Control}. In: International Conference on Mathematical Methods, Models, and
  Architectures for Computer Network Security (MMM-ACNS),
  \doi{10.1007/978-3-642-33704-8_8}

\bibitem[{Johns et~al.(2011)Johns, Braun, Schrank, and
  Posegga}]{Johns11_session_fixation_protection}
Johns M, Braun B, Schrank M, Posegga J (2011) {Reliable Protection against
  Session Fixation Attacks}. In: ACM Symposium on Applied Computing, ACM, SAC
  '11, p 1531–1537, \doi{10.1145/1982185.1982511}

\bibitem[{J{\"{u}}rjens(2005)}]{Juerjens05}
J{\"{u}}rjens J (2005) {Secure Systems Development with UML}. Springer

\bibitem[{Kamra and Bertino(2010)}]{Kamra10_state-based_access_control}
Kamra A, Bertino E (2010) {Privilege States Based Access Control for
  Fine-Grained Intrusion Response}. In: International Conference on Recent
  Advances in Intrusion Detection, Springer-Verlag, Berlin, Heidelberg,
  RAID'10, p 402–421

\bibitem[{Kang et~al.(1990)Kang, Cohen, Hess, Novak, and Peterson}]{kang1990}
Kang K, Cohen S, Hess J, Novak W, Peterson A (1990) {Feature-Oriented Domain
  Analysis (FODA) Feasibility Study}. Tech. Rep. CMU/SEI-90-TR-021, Software
  Engineering Institute, Carnegie Mellon University, Pittsburgh, PA

\bibitem[{Kang and Liang(2013)}]{KangL13}
Kang W, Liang Y (2013) A security ontology with {MDA} for software development.
  In: International Conference on Cyber-Enabled Distributed Computing and
  Knowledge Discovery (CyberC), IEEE, pp 67--74, \doi{10.1109/CyberC.2013.20}

\bibitem[{Katz(2010)}]{Katz10_digital_signatures}
Katz J (2010) Digital signatures, vol~1. Springer

\bibitem[{Kaur et~al.(2016)Kaur, Singh, Singh, and Sharma}]{Kaur2016}
Kaur R, Singh A, Singh S, Sharma S (2016) {Security of software defined
  networks: Taxonomic modeling, key components and open research area}. In:
  2016 International Conference on Electrical, Electronics, and Optimization
  Techniques (ICEEOT), pp 2832--2839, \doi{10.1109/ICEEOT.2016.7755214}

\bibitem[{Khanam et~al.(2020)Khanam, Ahmedy, Idris, Jaward, and
  Sabri}]{KhanamAIJS20}
Khanam S, Ahmedy IB, Idris MYIB, Jaward MH, Sabri AQM (2020) {A Survey of
  Security Challenges, Attacks Taxonomy and Advanced Countermeasures in the
  Internet of Things}. {IEEE} Access 8:219709--219743,
  \doi{10.1109/ACCESS.2020.3037359}

\bibitem[{Kim et~al.(2007)Kim, Luo, and Kang}]{KimLK07}
Kim A, Luo J, Kang MH (2007) {Security Ontology to Facilitate Web Service
  Description and Discovery}. Journal on Data Semantics 9:167--195,
  \doi{10.1007/978-3-540-74987-5 \_6}

\bibitem[{Kour and Verma(2014)}]{Kour14_steganography}
Kour J, Verma D (2014) {Steganography techniques--A review paper}.
  International Journal of Emerging Research in Management \&Technology ISSN pp
  2278--9359

\bibitem[{Krombholz et~al.(2017)Krombholz, Mayer, Schmiedecker, and
  Weippl}]{krombholz2017have}
Krombholz K, Mayer W, Schmiedecker M, Weippl E (2017) {"I Have No Idea What
  I{\textquoteright}m Doing" - On the Usability of Deploying {HTTPS}}. In: 26th
  USENIX Security Symposium, USENIX Association, pp 1339--1356

\bibitem[{Krueger et~al.(2019)Krueger, Berger, and Leich}]{krueger2019}
Krueger J, Berger T, Leich T (2019) {Software Engineering for Variability
  Intensive Systems}, Auerbach Publications, pp 153--172.
  \doi{10.1201/9780429022067-7}

\bibitem[{Kumar and Goyal(2019)}]{KumarG19}
Kumar R, Goyal R (2019) On cloud security requirements, threats,
  vulnerabilities and countermeasures: {A} survey. Computer Science Reviews
  33:1--48, \doi{10.1016/j.cosrev.2019.05.002}

\bibitem[{Lazar et~al.(2014)Lazar, Chen, Wang, and Zeldovich}]{lazar2014does}
Lazar D, Chen H, Wang X, Zeldovich N (2014) {W}hy {D}oes {C}ryptographic
  {S}oftware {F}ail?: {A} {C}ase {S}tudy {A}nd {O}pen {P}roblems. In:
  Asia-Pacific Workshop on Systems, ACM, p~7

\bibitem[{Löhr et~al.(2010)Löhr, Sadeghi, and
  Winandy}]{Loehr10_secure_storage}
Löhr H, Sadeghi AR, Winandy M (2010) {Patterns for Secure Boot and Secure
  Storage in Computer Systems}. In: 2010 International Conference on
  Availability, Reliability and Security (ARES), pp 569--573,
  \doi{10.1109/ARES.2010.110}

\bibitem[{Mahapatra et~al.(2020)Mahapatra, Singh, and Kumar}]{Mahapatra2020}
Mahapatra S, Singh B, Kumar V (2020) A survey on secure transmission in
  internet of things: Taxonomy, recent techniques, research requirements, and
  challenges. Arab Journal for Science and Engineering p 6211–6240

\bibitem[{Martinson et~al.(2021)Martinson, Jansson, Mukelabai, Berger, Bergel,
  and Ho-Quang}]{Martinson.2021}
Martinson J, Jansson H, Mukelabai M, Berger T, Bergel A, Ho-Quang T (2021)
  {HAnS: IDE-based editing support for embedded feature annotations}. In: ACM
  International Systems and Software Product Line Conference - Volume B, ACM,
  pp 28--31, \doi{10.1145/3461002.3473072}

\bibitem[{McDonald et~al.(2019)McDonald, Schoenebeck, and
  Forte}]{McDonald2019Reliability}
McDonald N, Schoenebeck S, Forte A (2019) {Reliability and Inter-rater
  Reliability in Qualitative Research: Norms and Guidelines for CSCW and HCI
  Practice}. ACM on Human-Computer Interaction 3(CSCW), \doi{10.1145/3359174}

\bibitem[{McGraw(2004)}]{mcgraw2004}
McGraw G (2004) Software security. IEEE Security \& Privacy 2(2):80--83,
  \doi{10.1109/MSECP.2004.1281254}

\bibitem[{Mirkovic and Reiher(2004)}]{Mirkovic04_ddos}
Mirkovic J, Reiher P (2004) {A Taxonomy of DDoS Attack and DDoS Defense
  Mechanisms}. SIGCOMM Computer Communincation Reviews 34(2):39–53,
  \doi{10.1145/997150.997156}

\bibitem[{Mukelabai et~al.(2023)Mukelabai, Hermann, Berger, and
  Steghöfer}]{Mukelabai2023}
Mukelabai M, Hermann K, Berger T, Steghöfer JP (2023) {FeatRacer: Locating
  Features Through Assisted Traceability}. IEEE Transactions on Software
  Engineering 49(12):5060--5083, \doi{10.1109/TSE.2023.3324719}

\bibitem[{Nadi et~al.(2016)Nadi, Kr{\"u}ger, Mezini, and
  Bodden}]{nadi2016jumping}
Nadi S, Kr{\"u}ger S, Mezini M, Bodden E (2016) Jumping {T}hrough {H}oops:
  {W}hy {D}o {J}ava {D}evelopers {S}truggle {W}ith {C}ryptography {A}{P}{I}s?
  In: International Conference on Software Engineering, ACM, pp 935--946

\bibitem[{Oyetoyan et~al.(2016)Oyetoyan, Cruzes, and Jaatun}]{Oyetoyan2016}
Oyetoyan TD, Cruzes DS, Jaatun MG (2016) {An Empirical Study on the
  Relationship between Software Security Skills, Usage and Training Needs in
  Agile Settings}. In: 2016 11th International Conference on Availability,
  Reliability and Security (ARES), pp 548--555, \doi{10.1109/ARES.2016.103}

\bibitem[{Oyetoyan et~al.(2019)Oyetoyan, Jaatun, and Cruzes}]{Oyetoyan2019}
Oyetoyan TD, Jaatun MG, Cruzes DS (2019) {Measuring Developers' Software
  Security Skills, Usage, and Training Needs}. In: Exploring Security in
  Software Architecture and Design, pp 260--286,
  \doi{10.4018/978-1-5225-6313-6.ch011}

\bibitem[{Patnaik et~al.(2019)Patnaik, Hallett, and Rashid}]{Patnaik2019}
Patnaik N, Hallett J, Rashid A (2019) Usability smells: An analysis of
  {Developers{\textquoteright}} struggle with crypto libraries. In: Fifteenth
  Symposium on Usable Privacy and Security (SOUPS 2019), USENIX Association, pp
  245--257

\bibitem[{Peldszus(2020)}]{Peldszus2020}
Peldszus S (2020) {Development of Secure Software with GRaViTY}. In: Workshop
  on Software-Reengineering \& -Evolution

\bibitem[{Peldszus(2022)}]{Peldszus2022}
Peldszus S (2022) {Security Compliance in Model-driven Development of Software
  Systems in Presence of Long-Term Evolution and Variants}. Springer,
  \doi{10.1007/978-3-658-37665-9_6}

\bibitem[{Peldszus et~al.(2019)Peldszus, Tuma, Str{\"{u}}ber, J{\"{u}}rjens,
  and Scandariato}]{PTS+2019}
Peldszus S, Tuma K, Str{\"{u}}ber D, J{\"{u}}rjens J, Scandariato R (2019)
  {Secure Data-Flow Compliance Checks between Models and Code based on
  Automated Mappings}. In: International Conference on Model-driven Engineering
  Languages and Systems (MODELS), IEEE, pp 23--33,
  \doi{10.1109/MODELS.2019.00-18}

\bibitem[{Peldszus et~al.(2021)Peldszus, B{\"{u}}rger, Kehrer, and
  J{\"{u}}rjens}]{PBKJ2021}
Peldszus S, B{\"{u}}rger J, Kehrer T, J{\"{u}}rjens J (2021) {Ontology-Driven
  Evolution of Software Security}. Data \& Knowledge Engineering (DKE) 134,
  \doi{10.1016/j. datak.2021.101907}

\bibitem[{Peldszus et~al.(2024)Peldszus, Burger, and Jurjens}]{Peldszus2023}
Peldszus S, Burger J, Jurjens J (2024) {UMLsecRT: Reactive Security Monitoring
  of Java Applications with Round-Trip Engineering}. IEEE Transactions on
  Software Engineering (01):1--31, \doi{10.1109/TSE.2023.3326366}

\bibitem[{Potter and McGraw(2004)}]{Potter2004}
Potter B, McGraw G (2004) Software security testing. IEEE Security \& Privacy
  2(5):81--85, \doi{10.1109/MSP.2004.84}

\bibitem[{Ralph et~al.(2021)Ralph, bin Ali, Baltes, Bianculli, Diaz, Dittrich,
  Ernst, Felderer, Feldt, Filieri, de~França, Furia, Gay, Gold, Graziotin, He,
  Hoda, Juristo, Kitchenham, Lenarduzzi, Martínez, Melegati, Mendez, Menzies,
  Molleri, Pfahl, Robbes, Russo, Saarimäki, Sarro, Taibi, Siegmund, Spinellis,
  Staron, Stol, Storey, Taibi, Tamburri, Torchiano, Treude, Turhan, Wang, and
  Vegas}]{Ralph21_empirical_standards}
Ralph P, bin Ali N, Baltes S, Bianculli D, Diaz J, Dittrich Y, Ernst N,
  Felderer M, Feldt R, Filieri A, de~França BBN, Furia CA, Gay G, Gold N,
  Graziotin D, He P, Hoda R, Juristo N, Kitchenham B, Lenarduzzi V, Martínez
  J, Melegati J, Mendez D, Menzies T, Molleri J, Pfahl D, Robbes R, Russo D,
  Saarimäki N, Sarro F, Taibi D, Siegmund J, Spinellis D, Staron M, Stol K,
  Storey MA, Taibi D, Tamburri D, Torchiano M, Treude C, Turhan B, Wang X,
  Vegas S (2021) {Empirical Standards for Software Engineering Research}.
  \eprint{2010.03525}

\bibitem[{Rana et~al.(2023)Rana, Parast, Kelly, Wang, and
  Kent}]{Subhabrata23_key_management}
Rana S, Parast FK, Kelly B, Wang Y, Kent KB (2023) A comprehensive survey of
  cryptography key management systems. Journal of Information Security and
  Applications 78:103607, \doi{https://doi.org/10.1016/j.jisa.2023.103607}

\bibitem[{{Replication Package}(2023)}]{Dropbox}
{Replication Package} (2023) {Replication Package.}
  \url{https://www.dropbox.com/sh/4p4k2swm8ija64z/AAAr_oakU09SirMEja_yxiUDa?dl=0},
  [Online; accessed 20-December-2023]

\bibitem[{Revelle et~al.(2005)Revelle, Broadbent, and Coppit}]{RevelleBC05}
Revelle M, Broadbent T, Coppit D (2005) {Understanding Concerns in Software:
  Insights Gained from Two Case Studies}. In: 13th International Workshop on
  Program Comprehension (IWPC), {IEEE}, pp 23--32, \doi{10.1109/WPC.2005.43}

\bibitem[{Riebisch(2003)}]{riebisch2003}
Riebisch M (2003) {Towards a More Precise Definition of Feature Models}. In:
  Modeling Variability for Object-Oriented Product Lines

\bibitem[{Rivest(1990)}]{rivest1990}
Rivest RL (1990) {CHAPTER 13 - Cryptography}. In: {van Leeuwen} J (ed)
  Algorithms and Complexity, Handbook of Theoretical Computer Science,
  Elsevier, pp 717--755,
  \doi{https://doi.org/10.1016/B978-0-444-88071-0.50018-7}

\bibitem[{Robillard and Murphy(2007)}]{Robillard2007}
Robillard MP, Murphy GC (2007) Representing concerns in source code. ACM
  Transactions on Software Engineering and Methodology 16(1):3,
  \doi{10.1145/1189748.1189751}

\bibitem[{Robshaw(1995)}]{Robshaw95_block_ciphers}
Robshaw M (1995) Block ciphers

\bibitem[{Roth et~al.(2021)Roth, Gr\"{o}ber, Backes, Krombholz, and
  Stock}]{roth202112}
Roth S, Gr\"{o}ber L, Backes M, Krombholz K, Stock B (2021) {12 Angry
  Developers - A Qualitative Study on Developers' Struggles with CSP}. In:
  Conference on Computer and Communications Security (CCS), ACM, p 3085–3103,
  \doi{10.1145/3460120.3484780}

\bibitem[{Rubin and Chechik(2013)}]{Rubin2013ASO}
Rubin J, Chechik M (2013) {A Survey of Feature Location Techniques}. In: Domain
  Engineering, Product Lines, Languages, and Conceptual Models

\bibitem[{Russo et~al.(2019)Russo, Di~Sorbo, Visaggio, and Canfora}]{RDVC2019}
Russo ER, Di~Sorbo A, Visaggio CA, Canfora G (2019) {Summarizing
  Vulnerabilities’ Descriptions to Support Experts during Vulnerability
  Assessment Activities}. Journal of Systems and Software 156(C):84–99,
  \doi{10.1016/j. jss.2019.06.001}

\bibitem[{Santos et~al.(2017)Santos, Tarrit, and Mirakhorli}]{santos2017}
Santos JC, Tarrit K, Mirakhorli M (2017) {A Catalog of Security Architecture
  Weaknesses}. In: 2017 IEEE International Conference on Software Architecture
  Workshops (ICSAW), IEEE, pp 220--223

\bibitem[{Santos et~al.(2019)Santos, Tarrit, Sejfia, Mirakhorli, and
  Galster}]{santos2019}
Santos JC, Tarrit K, Sejfia A, Mirakhorli M, Galster M (2019) {An Empirical
  Study of Tactical Vulnerabilities}. Journal of Systems and Software
  149:263--284, \doi{https://doi.org/10.1016/j.jss.2018.10.030}

\bibitem[{Schindler(2009)}]{Schindler09_rng_for_crypto}
Schindler W (2009) Random Number Generators for Cryptographic Applications,
  Springer, pp 5--23. \doi{10.1007/978-0-387-71817-0_2}

\bibitem[{Schwarz et~al.(2020)Schwarz, Mahmood, and Berger}]{Schwarz.2020}
Schwarz T, Mahmood W, Berger T (2020) {A Common Notation and Tool Support for
  Embedded Feature Annotations}. In: ACM International Systems and Software
  Product Line Conference - Volume B, ACM, pp 5--8,
  \doi{10.1145/3382026.3431253}

\bibitem[{Seiler and Paech(2017)}]{seiler2017}
Seiler M, Paech B (2017) {Using Tags to Support Feature Management Across Issue
  Tracking Systems and Version Control Systems}. In: Requirements Engineering:
  Foundation for Software Quality, Springer, pp 174--180

\bibitem[{Shar and Tan(2013)}]{Shar13_sql_injection}
Shar LK, Tan HBK (2013) {Defeating SQL Injection}. Computer 46(3):69--77,
  \doi{10.1109/MC.2012.283}

\bibitem[{Sparxsystems(2023)}]{enterprisearchitect}
Sparxsystems (2023) {Enterprise Architect}.
  \urlprefix\url{https://www.sparxsystems.eu/}, accessed: 2023-Dec-20

\bibitem[{Stack~Exchange(2022)}]{stackexchange}
Stack~Exchange I (2022) {Stack Exchange API}.
  \urlprefix\url{https://api.stackexchange.com/}, online; accessed
  20-December-2023

\bibitem[{Talooki et~al.(2015)Talooki, Bassoli, Lucani, Rodriguez, Fitzek,
  Marques, and Tafazolli}]{TalookiBLRFMT15}
Talooki VN, Bassoli R, Lucani DE, Rodriguez J, Fitzek FHP, Marques H, Tafazolli
  R (2015) Security concerns and countermeasures in network coding based
  communication systems: {A} survey. Computer Networks 83:422--445,
  \doi{10.1016/j.comnet.2015.03.010}

\bibitem[{{The Apache Software Foundation}(2010)}]{ApacheShiro}
{The Apache Software Foundation} (2010) {Apache Shiro - Simple. Java.
  Security.} \url{https://shiro.apache.org/}, [Online; accessed
  20-December-2023]

\bibitem[{Tsipenyuk et~al.(2005)Tsipenyuk, Chess, and McGraw}]{Tsipenyuk2005}
Tsipenyuk K, Chess B, McGraw G (2005) {Seven Pernicious Kingdoms: A Taxonomy of
  Software Security Errors}. IEEE Security \& Privacy 3(6):81--84,
  \doi{10.1109/MSP.2005.159}

\bibitem[{Tuma et~al.(2022)Tuma, Peldszus, Strüber, Scandariato, and
  Jürjens}]{TPS2022}
Tuma K, Peldszus S, Strüber D, Scandariato R, Jürjens J (2022) {Checking
  Security Compliance between Models and Code}. International Journal on
  Software and Systems Modeling \doi{10.1007/s10270-022-00991-5}

\bibitem[{{United States Congress}(1996)}]{HIPAA1996}
{United States Congress} (1996) Health insurance portability and accountability
  act of 1996 (public law 104-191).
  \urlprefix\url{https://www.govinfo.gov/content/pkg/PLAW-104publ191/pdf/PLAW-104publ191.pdf},
  [Online; accessed 19-December-2024]

\bibitem[{Valente et~al.(2022)Valente, Holanda, Mariano, Furuta, and
  Da~Silva}]{Valente22_academic_databases_for_slrs}
Valente A, Holanda M, Mariano AM, Furuta R, Da~Silva D (2022) {Analysis of
  Academic Databases for Literature Review in the Computer Science Education
  Field}. In: 2022 IEEE Frontiers in Education Conference (FIE), pp 1--7,
  \doi{10.1109/FIE56618.2022.9962393}

\bibitem[{Venter and Eloff(2003)}]{VenterE03}
Venter HS, Eloff JHP (2003) A taxonomy for information security technologies.
  Computers \% Security 22(4):299--307, \doi{10.1016/S0167-4048(03)00406-1}

\bibitem[{Vorobiev and Bekmamedova(2010)}]{VorobievB10}
Vorobiev A, Bekmamedova N (2010) {An Ontology-Driven Approach Applied to
  Information Security}. Journal of Research and Practice in Information
  Technology 42

\bibitem[{Xia et~al.(2017)Xia, Bao, Lo, Kochhar, Hassan, and
  Xing}]{Xia17_developers_searches}
Xia X, Bao L, Lo D, Kochhar PS, Hassan AE, Xing Z (2017) What do developers
  search for on the web? Empirical Software Engineering 22(6):3149--3185,
  \doi{10.1007/s10664-017-9514-4}

\bibitem[{Yassein et~al.(2017)Yassein, Aljawarneh, Qawasmeh, Mardini, and
  Khamayseh}]{Yassein17_symmetric_asymmetric_encryption}
Yassein MB, Aljawarneh S, Qawasmeh E, Mardini W, Khamayseh Y (2017)
  {Comprehensive Study of Symmetric Key and Asymmetric Key Encryption
  Algorithms}. In: 2017 International Conference on Engineering and Technology
  (ICET), pp 1--7, \doi{10.1109/ICEngTechnol.2017.8308215}

\end{thebibliography}

\end{document}